\documentclass[10pt,twocolumn,showpacs,preprintnumbers,amsmath,amssymb,aps,prb,longbibliography,superscriptaddress]{revtex4-2}
\usepackage[utf8]{inputenc}
\usepackage{color,array,verbatim,multirow,dsfont,graphicx,esint,titlesec,longtable,bbm,listings,braket,siunitx,tikz,mathtools}
\usepackage[shortlabels]{enumitem}
\usepackage[normalem]{ulem}
\usetikzlibrary{positioning}
\usepackage{bm}

\usepackage[unicode=true,
 bookmarks=true,bookmarksnumbered=true,bookmarksopen=false,
 breaklinks=true,pdfborder={0 0 0},backref=false,colorlinks=true]{hyperref}
\hypersetup{citecolor={blue},urlcolor={magenta}}

\usepackage{cleveref}
\crefname{appendix}{App.}{Apps.}
\crefname{equation}{Eq.}{Eqs.}
\crefname{figure}{Fig.}{Figs.}
\crefname{table}{Tab.}{Tabs.}
\crefname{section}{Sec.}{Secs.}

\newcommand{\Z}{\mathbb{Z}}

\newcommand{\br}{\mathbf{r}}

\newcommand{\ri}{\mathrm{i}}

\newcommand{\I}{\mathcal{I}}

\usepackage{lipsum} % Lorem ipsum package
\usepackage[caption=false]{subfig}
\usepackage{float}
\usepackage{picture}

\begin{document}
\title{Entanglement Entropy of Free Fermions in Timelike Slices}
\author{Bowei Liu}
\affiliation{Department of Physics, Princeton University, Princeton, New Jersey 08544, USA}
\author{Hao Chen}
\affiliation{Department of Physics, Princeton University, Princeton, New Jersey 08544, USA}
\affiliation{Department of Electrical and Computer Engineering, Princeton University, Princeton, New Jersey 08544, USA}
\author{Biao Lian}
\affiliation{Department of Physics, Princeton University, Princeton, New Jersey 08544, USA}
\date{\today}

\begin{abstract}
We define the entanglement entropy of free fermion quantum states in an arbitrary spacetime slice of a discrete set of points, and particularly investigate timelike (causal) slices. For 1D lattice free fermions with an energy bandwidth $E_0$, we calculate the time-direction entanglement entropy $S_A$ in a time-direction slice of a set of times $t_n=n\tau$ ($1\le n\le K$) spanning a time length $t$ on the same site. For zero temperature ground states, we find that $S_A$ shows volume law when $\tau\gg\tau_0=2\pi/E_0$; in contrast, $S_A\sim \frac{1}{3}\ln t$ when $\tau=\tau_0$, and $S_A\sim\frac{1}{6}\ln t$ when $\tau<\tau_0$, resembling the Calabrese-Cardy formula for one flavor of nonchiral and chiral fermion, respectively. For finite temperature thermal states, the mutual information also saturates when $\tau<\tau_0$. For non-eigenstates, volume law in $t$ and signatures of the Lieb-Robinson bound velocity can be observed in $S_A$. For generic spacetime slices with one point per site, the zero temperature entanglement entropy shows a clear transition from area law to volume law when the slice varies from spacelike to timelike. 
\end{abstract}

\maketitle

\section{INTRODUCTION}
The entanglement entropy of a subsystem in a bipartite system \cite{entropy1,entropy2,entropy3,entangle_rev} characterizes the spatial entanglement information of quantum states \cite{epr,epr_bohr,bell}. For instance, the entanglement entropy of gapped (gapless) ground states satisfies the area law \cite{bombelli1986,entropy_area_prl,hastings2007,entropy_rev} (area law with a logarithmic factor \cite{Cardy_Peschel_1988,calabrese_qft_2004,calabrese_cft_2009,gioev2006,wolf2006,cramer2007,liweifei2006,barthel2006}), while topological orders can be detected via correction to the area law known as the topological entanglement entropy \cite{topological_entropy,levin2006}.
Moreover, the subsystem entanglement entropy of excited states can exhibit either area law or volume law, indicating (many-body) localization or quantum chaos, respectively \cite{entropy_rev,bauer2013,huse2013}.

The spatial subregion entanglement entropy, although related to thermodynamics in many aspects including the eigenstate thermalization hypothesis \cite{ETH1,ETH2,ETH3} and black holes \cite{bombelli1986,entropy_area_prl,blackhole_1,blackhole_2,blackhole_3,bousso2002}, is a nonlocal quantity.
Therefore, we ask if entanglement entropy of quantum states can be defined in a timelike slice, such as an observer's worldline, which would be locally observable. For free fermion models, we show that the entanglement entropy of a quantum state can be explicitly defined for any spacetime slice consisting of a discrete set of points, which has not been discussed in previous studies of spacetime quantum entanglement \cite{temporal_bell_1,temporal_bell_2,temporal_bell_3,oreshkov2012,fitzsimons2013,brukner2014,ried2015,brukner2015,jia2017,cotler2018,cotler2019,quantum_cellular_automata,nebabu2023bulk}.
It is also different from the temporal entanglement entropy for influence matrix \cite{abanin_prx,abanin_1,abanin_2,abanin_3,feynman-vernon} or similar generalizations in tensor networks \cite{Muller-Hermes_2012,Hasting_connecting_ent_2015}, which is defined for an effective time-direction ``quantum state", instead of for the physical quantum state (see \cref{sec:discussion}, and supplementary material (SM) \cite{suppl} Sec. II).

We particularly investigate the entanglement entropy of one-dimensional (1D) free lattice fermions in an on-site time-direction slice of discrete times separated by $\tau$ (\cref{fig1b}). For lattice fermions with an energy spectrum of range $E_0$, we find the time-direction entanglement entropy of thermal states is maximal and temperature independent 
when $\tau\gg\tau_0=2\pi/E_0$, while stabilizes to the Calabrese-Cardy formula of 1D chiral fermions \cite{Cardy_Peschel_1988,calabrese_qft_2004,calabrese_cft_2009} when $\tau<\tau_0$ at zero temperature, indicating the existence of a continuous time limit ($\tau\rightarrow 0$). The mutual information at finite temperature also stabilizes when $\tau<\tau_0$. For non-eigenstates, the time-direction entanglement entropy exhibits volume law and can probe the Lieb-Robinson bound velocity. Moreover, we show that the entanglement entropy exhibits a crossover of behaviors between spacelike and timelike slices.

This paper is organized as follows. In Sec. II we formulate the generic definition of spacetime slice entanglement entropy of free fermion lattice models, from both operator and path integral formalisms. Then for the example of 1D free fermions, we study the entanglement entropy in the time-direction slice in Sec. III, and in generic spacelike and timelike linear spacetime slices in Sec. IV. We then discuss possible future developments in Sec. V.

\section{ENTANGLEMENT ENTROPY OF FREE FERMIONS IN AN ARBITRARY SPACETIME SLICE}%\label{sec:def}
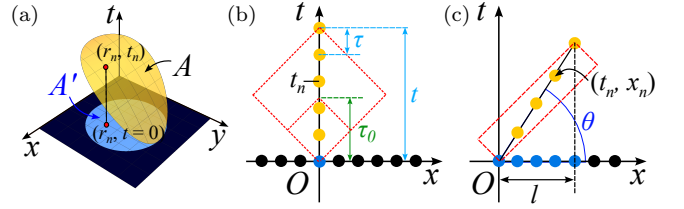
\begin{figure}[tbp]
\centering
\vspace{-5mm}
\subfloat{\label{fig1a}
\begin{picture}(0.32\columnwidth,0.32\columnwidth)
\put(0,0){\includegraphics[width=0.32\columnwidth]{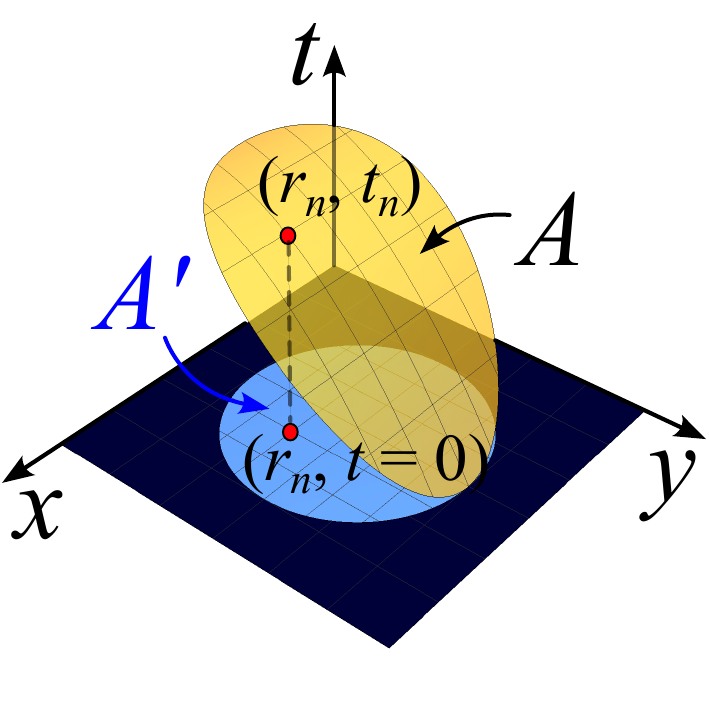}}
\put(-4,70){\footnotesize (a)}
\end{picture}}%
\subfloat{\label{fig1b}
\begin{picture}(0.32\columnwidth,0.32\columnwidth)
\put(0,0){\includegraphics[width=0.32\columnwidth]{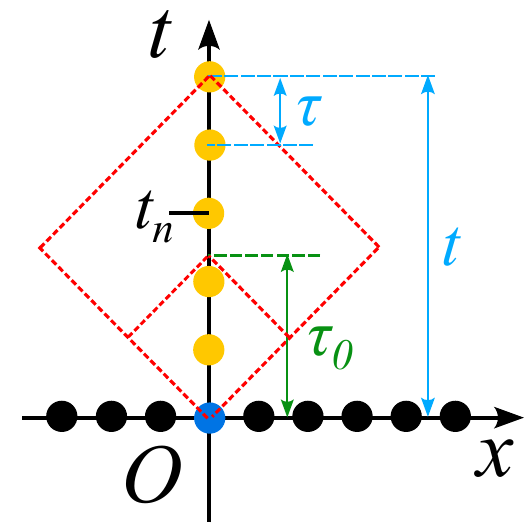}}
\put(-4,70){\footnotesize (b)}
\end{picture}}%
\subfloat{\label{fig1c}
\begin{picture}(0.32\columnwidth,0.32\columnwidth)
\put(0,0){\includegraphics[width=0.32\columnwidth]{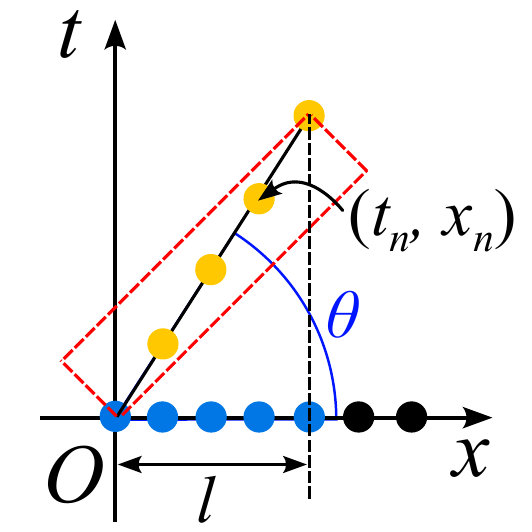}}
\put(-4,70){\footnotesize (c)}
\end{picture}}
\caption{
(a) An arbitrary spacetime slice $A$ (yellow) with discrete spacetime points $(\br_n,t_n)$ at lattice sites $\br_n$ and times $t_n$. In a 1D lattice fermion model, (b) shows a time-direction slice (yellow) of length $t$ containing $K$ points at times $t_n$ ($1\le n\le K$) on the same site $m=0$; (c) shows a linear slice (yellow) containing points at $t_n=v_{\text{max}}^{-1}x_n \tan(\theta)$ on the $n$-th site ($0\le n\le \ell-1$), with $v_{\text{max}}$ defined in \cref{eq:tau0}.
}
\label{fig1}
\end{figure}

Consider a system with a total Hilbert space $h_{\text{tot}}$. For an arbitrary spacetime slice $A$ (\cref{fig1a}), if a sub-Hilbert space $h_A$ can be identified for it, we can define its reduced density matrix $\rho_A$ and entanglement entropy $S_A$ as
\begin{equation}\label{eq:density-S}
\rho_A=\text{tr}_{h_{A^c}}(\rho_{\text{tot}})\ ,\quad S_A=-\text{tr}(\rho_A\ln \rho_A)\ ,
\end{equation}
where $h_{A^c}$ is the complement of $h_A$ ($h_{\text{tot}}=h_A\otimes h_{A^c}$). To do so, we impose the Heisenberg picture, such that quantum states are time-independent, while operators at spacetime point $(\br,t)$ are defined by 
\begin{equation}
O_{\br,j}(t)=e^{iHt}O_{\br,j}e^{-iHt}\ , 
\end{equation}
where $H$ is the Hamiltonian. We then define $h_A$ as the \emph{minimal} sub-Hilbert space such that the correlations of any operators $O_{\br_n,j}(t_n)$ at spacetime points $(\br_n,t_n)$ in slice $A$ are calculable from the reduced density matrix $\rho_A$ in $h_A$, namely, \begin{equation}\label{eq:n-point}
\langle \prod_{\alpha\in A}O_{\br_{n_\alpha},j_\alpha}(t_{n_\alpha})\rangle =\text{tr}\left[\rho_A\prod_{\alpha\in A}O_{\br_{n_\alpha},j_\alpha}(t_{n_\alpha})\right]\ .    
\end{equation}
Such a sub-Hilbert space is generically difficult to identify, but as we shall show, it can be identified straightforwardly for free fermion models.

\subsection{Definition in the Heisenberg picture}\label{sec:def}
We now show how sub-Hilbert space $h_A$ can be identified for free fermion models. Consider a lattice model (in any dimension) where each site $\br_m$ has one fermion degree of freedom with annihilation operator $c_{\br_m}$ and creation operator $c_{\br_m}^\dag$ at time $t=0$, with anti-commutators $\{c_{\br_m}, c_{\br_n}^\dag\}=\delta_{mn}$. Assume the model has a free fermion Hamiltonian of the fermion bilinear form
\begin{equation}\label{eq:H-general}
H=\sum_{i,j}c^\dag_{\br_i} h_{ij}c_{\br_j}\ ,
\end{equation}
which has charge conservation (the case without charge conservation will be discussed in \cref{sec:bogoliubov}). In an arbitrary slice $A$ with $K$ discrete spacetime points $(\br_n,t_n)$, each annihilation operator $c_{\br_n}(t_n)=e^{iHt_n}c_{\br_n}e^{-iHt_n}{ = \sum_{j} \xi_{n j} c_{\br_j}} $ at point $(\br_n,t_n)$ is the linear combination of $c_{\br_m}$ at time $t=0$. Accordingly, the anti-commutations between $c_{\br_n}(t_n)$ and $c_{\br_m}^\dag(t_m)$ are {not delta functions}, but are still numbers:
\begin{equation}\label{eq:anti-comm-matrix}
\{c_{\br_m} (t_m), c_{\br_n}^\dagger (t_n)\} = B_{m n}\ , 
\quad (m, n \in A)\ .
\end{equation}
$B_{m n}$ gives a $K\times K$ non-negative Hermitian commutation matrix $B$ for the $K$ points in slice $A$. By rewriting $B=Q_1 \Lambda Q_1^\dag$ with $\Lambda$ diagonal and $Q_1$ unitary, we can define a matrix $M=Q_2\Lambda^{-1/2}Q_1^\dag$ where $Q_2$ can be an arbitrary unitary matrix, such that $M$ satisfies  
\begin{equation}\label{eq:M-matrix}
M^\dag M=B^{-1}. 
\end{equation}
Particularly, by QL decomposition, we choose $Q_2$ such that $M$ is a lower triangular matrix for later purposes of path integral formalism. We can then write down an orthonormal fermion basis (in terms of the zero-time on-site basis $c_{\br_j}$):
\begin{equation}\label{eq:orthonormal}
d_m=\sum_{n\in A}M_{mn}c_{\br_n}(t_n)  =\sum_{n\in A}\sum_j M_{mn} \xi_{n j} c_{\br_j}\ ,
\end{equation}
and their Hermitian conjugates $d_m^\dag$ ($1\le m\le K$). It is straightforward to see \cite{suppl} that they satisfy 
\begin{equation}
\{d_m, d_n^\dagger\} = \delta_{mn}, 
\end{equation}
and thus form an orthonormal fermion basis. $h_A$ is then the Hilbert space of the $K$ fermion degrees of freedom $d_m$, which has a dimension $\text{dim}[h_A]=2^K$. { More explicitly, the basis of the sub-Hilbert space $h_A$ is given by the $2^K$ Fock states $\prod_m (d_m^\dag)^{\zeta_m} |0\rangle$ for $\zeta_m \in \{0, 1\}$, $1 \leq m \leq K$. In contrast, the total Hilbert space of the system $h_{\text{tot}}$ has a basis given by the $2^L$ Fock states $\prod_m (c_{\br_n}^\dag)^{\xi_n} |0\rangle$ for $\xi_n \in \{0, 1\}$, where $L$ is the total number of sites of the system.} This allows one to explicitly decompose the total Hilbert space into the subsystem of slice $A$, and its complement $A^c$ spanned by the Hilbert space of fermion operators orthogonal to $d_m$ in \cref{eq:orthonormal}:
\begin{equation}
h_{\text{tot}}=h_A\otimes h_{A^c}\ .
\end{equation}
Any correlations among $c_{\br_n}(t_n)$ and $c_{\br_m}^\dag(t_m)$ are completely determined within sub-Hilbert space $h_A$. We note that our method for identifying sub-Hilbert space $h_A$ here is similar to the bulk Hilbert space reconstruction most recently studied in Ref. \cite{nebabu2023bulk}.

As an illustrative example, consider a lattice of 3 sites with corresponding fermion annihilation operators $c_1,c_2,c_3$, respectively, and assume spacetime slice $A$ contains two spacetime points: $(x_1,t_1)=(x_1,0)$ and $(x_2,t_2)$, with $t_1=0$ and $t_2\neq 0$. For concreteness, assume the free Hamiltonian is $H=\sum_{m,n}c_m^\dag \mathcal{H}_{mn}c_n$, and the matrix $\mathcal{H}$ and time $t_2$ are such that the second column of unitary matrix $e^{-i\mathcal{H}t_2}$ has matrix elements $\left(e^{-i\mathcal{H}t_2}\right)_{m2}=\frac{1}{\sqrt{3}}$, we can derive the fermion operators at these two points as
\begin{equation}
c_1(t_1)=c_1\ ,\quad c_2(t_2)=\frac{1}{\sqrt{3}} (c_1+c_2+c_3)\ .
\end{equation}
Obviously, the above two fermion operators are not orthogonal, with their anti-commutation relation given by
\begin{equation}
\{c_m(t_m),c_n^\dag(t_n)\}=B_{mn}\ ,\qquad 
B=\begin{pmatrix}
1 & \frac{1}{\sqrt{3}} \\
\frac{1}{\sqrt{3}} & 1
\end{pmatrix}\ .
\end{equation}
To find the Hilbert space spanned by the two fermion operators $c_1(t_1)$ and $c_2(t_2)$, namely the sub-Hilbert space $h_A$ of slice $A$, we need to find an orthonormal basis from linear superposition of $c_1(t_1)$ and $c_2(t_2)$, one choice of which is
\begin{equation}
\begin{split}
&d_1=c_1=c_1(t_1)\ ,\\ &d_2=\frac{1}{\sqrt{2}}(c_2+c_3)=-\frac{1}{\sqrt{2}}c_1(t_1)+\sqrt{\frac{3}{2}}c_2(t_2)\ ,
\end{split}
\end{equation}
or in matrix form,
\begin{equation}
d_m=\sum_{n}M_{mn}c_n(t_n)\ ,\qquad
M=\begin{pmatrix}
1 & 0 \\
-\frac{1}{\sqrt{2}} & \sqrt{\frac{3}{2}}
\end{pmatrix}\ .
\end{equation}
This is the $M$ matrix in \cref{eq:orthonormal}, which satisfies
\begin{equation}
M^\dag M=B^{-1}\ ,\qquad \rightarrow \qquad \{d_m,d_n^\dag\}=\delta_{mn}\ .
\end{equation}
The sub-Hilbert space $h_A$ spanned by $c_1(t_1)$ and $c_2(t_2)$ is thus the direct product of the fermion Hilbert space of $d_1$ and $d_2$, which has dimension $\text{dim}[h_A]=2^2=4$.

For pure Fock states or thermal mixed states (or generically Gaussian states) with density matrix $\rho_{\text{tot}}$ in the full system (which are time-independent in the Heisenberg picture here), the Wick's theorem holds, and thus the entanglement entropy $S_A$ can be calculated from the $K\times K$ two-point correlation matrix $D$ in the orthonormal fermion basis $d_m$ \cite{peschel_calculation_2003}, with matrix elements:
\begin{equation}\label{eq:correlation-matrix}
D_{mn}=\text{tr}(\rho_{\text{tot}}d^\dag_m d_n)=(M^* CM^T)_{mn}\ ,\\
\end{equation}
where $C$ is the two-point correlation matrix in the spacetime basis with matrix elements 
\begin{equation}
C_{mn}=\text{tr}\left[\rho_{\text{tot}}c_{\br_m}^\dag(t_m) c_{\br_n}(t_n)\right]\ . 
\end{equation}
By Wick's theorem, the entanglement entropy $S_A$ in sub-Hilbert space $h_A$ of slice $A$ is determined by the two-point correlation matrix $D$ as \cite{peschel_calculation_2003}
\begin{equation}\label{eq:S-Dexpression}
\begin{split}
&S_A=-\text{tr}\left[D\ln D+(I-D)\ln(I-D)\right]\ ,
\end{split}
\end{equation}
where $I$ is the identity matrix.

We note that the choice of $d_m$ fermion basis is not unique, but is up to a further unitary transformation (namely, the unitary matrix $Q_2$ in defining the $M$ matrix can be arbitrary). However, the choice of $d_m$ fermion operators does not change the entanglement entropy $S_A$, which is invariant under fermion basis unitary transformations.

%{\color{red}The referee has raised concern about mutual information.}
When $\rho_{\text{tot}}$ is a mixed state, we can also define the mutual information between sub-Hilbert space $h_A$ of slice $A$ and its complement $h_{{A^c}}$:
%it is known that the entanglement entropy $S_A$ of sub-Hilbert space $h_A$ (\cref{eq:S-Dexpression}) is not a measure of the entanglement between $h_A$ and $h_{A^c}$, but is only a von-Neumann entropy. Instead, the entanglement between sub-Hilbert space $h_A$ of slice $A$ and its complement $h_{{A^c}}$ can be characterized by the mutual information:
\begin{equation}\label{eq:mutual}
\I=S_A+S_{A^c}-S_{\text{tot}}\ ,
\end{equation}
where $S_{A^c}=-\text{tr}(\rho_{A^c}\ln \rho_{A^c})$ is the entanglement entropy of subsystem $A^c$, and $S_\text{tot}=-\text{tr}(\rho_\text{tot}\ln \rho_\text{tot})$ is the entanglement entropy of the entire system. For a pure state $\rho_{\text{tot}}$, one has $S_{\text{tot}} = 0$ and $\I=2S_A$. We note that however, for mixed states, neither entanglement entropy nor mutual information is a good measure of the entanglement between $h_A$ and $h_{{A^c}}$. (In fact, for mixed states it is more proper to call $S_A$ the von-Neumann entropy, but we retain the name entanglement entropy for convenience, as is adopted in many literature.) The entanglement of mixed states can be for instance characterized by the entanglement negativity \cite{shapourian2019,ghosh2020}, which we leave for future studies.

\subsection{Path integral formalism}
To provide an understanding of the above definition of spacetime slice entanglement entropy in \cref{eq:S-Dexpression} from a spacetime local perspective, in this subsection we rewrite it in terms of the coherent state path integral formalism. The readers who are solely interested in the physical outcomes of the spacetime slice entanglement entropy in \cref{eq:S-Dexpression} may skip this subsection.

\begin{figure}[tbp]
\centering
\subfloat{
\begin{picture}(0.6\columnwidth,0.6\columnwidth)
\put(0,0){\includegraphics[width=0.5\columnwidth]{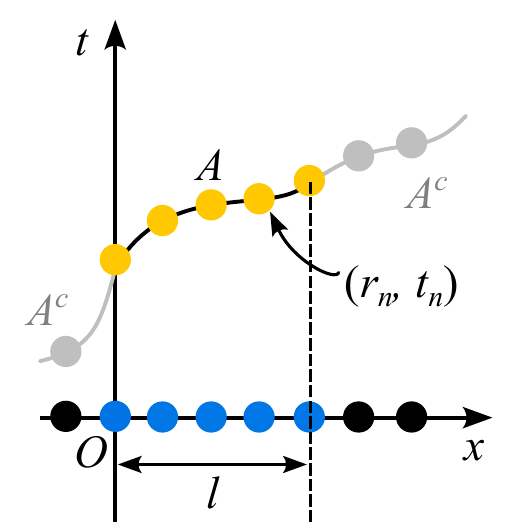}}
%\put(-4,70){\footnotesize (c)}
\end{picture}}
\caption{
Non-constant time slice $A$ is part of a hypersurface $S=A\cup A^c$ that covers the entire space of the system.
}
\label{fig-pathint}
\end{figure}

Conventionally, for a constant time $t$ slice covering the full space, the coherent state basis for path integral is given by
\begin{equation}
|\bm{\eta}(t)\rangle=\prod_j\left[1{-}\eta_{\br_j} (t) c_{\br_j}^{\dag}(t)\right]|0\rangle
\end{equation}
where $\eta_{\br_j}(t)$ are anti-commuting Grassmann numbers (which also anti-commutes with fermion operators), and $\bm{\eta}(t)=(\eta_{\br_1}(t),\eta_{\br_2}(t),\cdots)^T$ is the vector formed by all $\eta_{\br_j}(t)$. They give the values of the local fermion field $c_{\br_j}(t)$ at spacetime coordinate $(\br_j,t)$, namely:
\begin{equation}
c_{\br_j} (t)|\bm\eta(t)\rangle=\eta_{\br_j} (t)|\bm\eta(t)\rangle\ .
\end{equation}
The coherent states also satisfy the completeness condition:
\begin{equation}
\int D\bm{\eta}(t)D\bar{\bm\eta}(t) |\bm{\eta}(t)\rangle\langle\bm{\eta}(t)|=1\ .
\end{equation}
For a free fermion model with Hamiltonian as shown in \cref{eq:H-general}, the conventional path integral between two states at $t_I$ and $t_F$ is done by inserting the coherent state basis of constant time slices
\begin{equation}
\begin{aligned}
&\langle\bm{\eta}_F(t_F)|e^{-i H (t_F-t_I)}| \bm\eta_I(t_I)\rangle 
= \int_{\bm\eta\left(t_I\right)=\bm\eta_I} ^{\bm\eta{\left(t_{F}\right)}=\bm\eta_F} D \bm\eta D \bar{\bm\eta} \\
& \times e^{i \int_{t_I}^{t_F} dt \left[i \sum_j \bar{\eta}_{\br_j} (t) \partial_t \eta_{\br_j} (t)- \sum_{j l} h_{j l} \bar{\eta}_{\br_j} (t) \eta_{\br_l} (t)\right]}\ .
\end{aligned}
\end{equation}

Now for a non-constant time slice $A$ shown in \cref{fig-pathint}, we explain how to rewrite the path integral beginning or ending on slice $A$, which would be useful for defining the reduced density matrix $\rho_A$ for slice $A$. 

Suppose slice $A$ is part of a hypersurface $S = A \cup A^c$ that covers the entire space of the system, where $A^c$ is the complement of $A$ satisfying $A \cap A^c=0$, as shown in \cref{fig-pathint}. We note that the choice of $A_c$ is arbitrary, and the entire hypersurface do not need to be ``spacelike". For a space with $L$ sites, we assume hypersurface $S$ contains $L$ spacetime points $\left(\br_{j}, t_{j}\right)$ with fermion operators $c_{\br_{j}}(t_{j})$, $c_{\br_{j}}^\dag(t_{j})$ defined in the Heisenberg picture. For convenience, we assume slice $A$ contains $K$ spacetime points, and we sort the spacetime points $\left(\br_{j}, t_{j}\right)$ such that the first $K$ points ($1\le j\le K$) are those in slice $A$. 

Our goal is to define a coherent state basis on hypersurface $S$ satisfying 
\begin{equation}
c_{\br_j}(t_j) \left|\bm\eta_S\right\rangle =\eta_{\br_j}(t_j)\left|\bm\eta_S\right\rangle\ ,
\label{eq:coh}\end{equation}
where $\eta_{\br_j}(t_j)$ are Grassmann numbers and is grouped into a vector $\bm{\eta}_S=(\eta_{\br_1}(t_1),\eta_{\br_2}(t_2),\cdots)^T$. To do this, we follow the same prescription as in \cref{sec:def}: we define the anticommutation matrix in hypersurface $S$ to be
\begin{equation}
\{c_{\br_j} (t_j), c_{\br_l}^\dagger (t_l)\} = B^S_{j l}\ ,\quad (j, l \in S) 
\end{equation}
where it is easy to see that $B^S$ is a Hermitian matrix. After diagonalizing it into $B^S=Q_1^S \Lambda^S Q_1^{S\dag}$ with $\Lambda^S$ diagonal and $Q_1^S$ unitary, we define $N^S=\left(\Lambda^S\right)^{-\frac{1}{2}} Q_1^{S\dag}$. We can then find a QL decomposition $N^S = Q_2^S M^S$ where $Q_2^S$ is unitary and 
\begin{equation}
M^S=Q_2^{S\dag}\left(\Lambda^S\right)^{-\frac{1}{2}} Q_1^{S\dag}=\left(\begin{array}{ll}
M & 0 \\
M^{\prime} & M_c
\end{array}\right)
\end{equation}
is a lower triangular matrix. It is easy to see that $M^{S \dag} M^S=\left(B^S\right)^{-1}$. We have written $M^S$ here into a block form with $M$ being a $K\times K$ matrix and $M_c$ a $(L-K)\times (L-K)$ matrix. Accordingly, both $M$ and $M_c$ are lower triangular. Also, $(M^S)^{-1}$ will be lower triangular.

We then define a new fermion complete basis $d_j$ via
\begin{equation}\label{eq:orthonormal-path}
d_j=\sum_{l\in S}M_{jl}^Sc_{\br_l}(t_l)\ ,\ \rightarrow\ c_{\br_j}(t_j)=\sum_{l\in S}(M^{S})^{-1}_{jl}d_l\ ,
\end{equation}
which satisfies 
\begin{equation}
\{d_j,d_l^\dag\}=\delta_{jl}\ . 
\end{equation}
Since $(M^S)$ and $(M^S)^{-1}$ are both lower triangular, the first $K$ fermion operators $d_j$ are only linear combinations of the first $K$ fermion operators $c_{\br_j}(t_j)$ in slice $A$, and vice versa, reducing to the definition of fermion basis within slice $A$ in \cref{eq:orthonormal}. In the new basis of \cref{eq:orthonormal-path} on hypersurface $S$, we can define coherent states
\begin{equation}
\left|\bm\eta_S\right\rangle=\prod_{j \in S}\left[1-\sum_k \eta_{\br_k}\left(t_k\right)M^S_{j k} d_j^\dag\right]|0\rangle\ ,
\end{equation}
which satisfies the definition of coherent states in \cref{eq:coh}: 
\begin{equation}
\begin{split}
&c_{\br_j}(t_j) \left|\bm\eta_S\right\rangle= \sum_{l\in S}(M^{S})^{-1}_{jl}d_l \left|\bm\eta_S\right\rangle \\
&=\sum_{l,k\in S}(M^{S})^{-1}_{jl}M^S_{l k}\eta_{\br_k}\left(t_k\right)\left|\bm\eta_S\right\rangle =\eta_{\br_j}(t_j)\left|\bm\eta_S\right\rangle\ .
\end{split}
\end{equation}
For fixed hypersurface $S$, the completeness condition is still satisfied:
\begin{equation}
|\det(M^S)|^2 \int D\bm{\eta}_S D\bar{\bm\eta}_S |\bm{\eta}_S\rangle\langle\bm{\eta}_S|=1\ ,
\end{equation}
with $|\det(M^S)|^2$ being the Jacobian, which is a constant factor that can be absorbed into the measure of path integral. 

We then note that such a coherent state can be decomposed as a direct product
\begin{equation}
\left| \bm\eta_S \right\rangle=\left|\bm\eta_A\right\rangle \otimes \left|\bm\eta_{A^c}\right\rangle\ , 
\end{equation}
where 
\begin{equation}
\left|\bm\eta_A\right\rangle=\prod_{j \in A}\left[1-\sum_{k\in A} \eta_{\br_k} \left(t_k\right) M_{j k} d_j^{\dag}\right]|0\rangle
\end{equation}
only depends on the values of $\eta_{\br_j} (t_j)$ in subregion $A$, due to the lower triangular form of matrix $M^S$. The other part $\left|\bm\eta_{A^c}\right\rangle$ will depend on the values of $\eta_{\br_j} (t_j)$ in both $A$ and $A^c$ though.

In Schrodinger picture, consider a state with the density matrix $\rho_{\text{tot}}(0)$ at $t=0$.
In terms of path integral, we then define the density matrix on hypersurface $S$ in the coherent state basis $\left|\bm\eta_{S}\right\rangle$ as 
\begin{equation}
\begin{aligned}
&\left\langle\bm \eta_{S}^{\prime}\left|\rho_{\text {tot}}(S)\right|\bm \eta_{S}\right\rangle
= \int D \bm\eta D \bm \bar{\bm\eta} \\
& \times e^{i \int_{0}^{S} dt (i \bar{\bm\eta}^T \partial_{t} \bm\eta-\bar{\bm\eta}^T h \bm\eta)} \rho_{\text {tot}}(0) e^{i \int_{S}^{0} dt (i \bar{\bm\eta}^T \partial_{t} \bm\eta-\bar{\bm\eta}^T h \bm\eta)}
\end{aligned}
\end{equation}
where initial and final states in the path integral are given by $\bm\eta_{I}=\bm\eta_{S}$ and $\bm\eta_{F}=\bm\eta_{S}^{\prime}$.

The reduced density matrix $\rho_A$ in slice $A$ is then defined by integrating out $\bm\eta_{A^{C}}=\bm\eta_{A^{C}}^{\prime}$ in the path integral while fixing $\bm\eta_{A}$ and $\bm\eta_{A}^{\prime}$, which is equivalent to tracing over the Hilbert space generated by $d_{j}$ with $j>K$. After this partial trace, $\rho_A$ only depends on $\eta_{\br_{j}}(t_{j}) \in A$, which is equivalent to our formalism in \cref{sec:def}. In this formalism, the path integral is spacetime local.

\subsection{Definition for free fermions with pairing}\label{sec:bogoliubov}

We can generalize our definition of spacetime entanglement entropy for free fermion Bogoliubov-de Gennes (BdG) models with pairing of the form
\begin{equation}
H=\sum_{i,j}\left[h_{ij}c^\dag_{\br_i} c_{\br_j}+\frac{1}{2}\left(\Delta_{ij}c^\dag_{\br_i} c^\dag_{\br_j}+h.c.\right)\right]\ ,
\end{equation}
where $\Delta_{ij}=-\Delta_{ji}$ is the pairing amplitude. 

In the slice $A$ with $K$ points $(\br_n,t_n)$, the annihilation operator $c_{\br_n}(t_n)=e^{iHt_n}c_{\br_n}e^{-iHt_n}$ at point $(\br_n,t_n)$ will be a Bogoliubov fermion mode which is the linear combination of both $c_{\br_m}$ and $c^\dag_{\br_m}$ at time $t=0$. This leads to anti-commutation relations
\begin{equation}\label{eq:anti-comm-matrix-BdG}
\begin{split}
&\{c_{\br_m} (t_m), c_{\br_n}^\dagger (t_n)\} = B_{m n}\ ,\\
&\{c_{\br_m} (t_m), c_{\br_n} (t_n)\} = V_{m n}\ ,
\quad (m, n \in A)\ ,
\end{split}
\end{equation}
where by definition $B_{mn}$ is a Hermitian matrix, and $V_{mn}$ is a symmetric matrix (not necessarily real). By defining the Nambu basis $\Phi_A=(c_{\br_1} (t_1),\cdots,c_{\br_K} (t_K),c_{\br_1}^\dag (t_1),\cdots, c_{\br_K}^\dag (t_K))^T$ of slice $A$, we can rewrite the above anti-commutation relations as
\begin{equation}\label{eq:anti-comm-Nambu}
\{\Phi_A,\Phi_A^\dag\}=\left(
\begin{array}{cc}
B&V\\
V^* &B^*
\end{array}
\right)=R\ ,
\end{equation}
where $B^*$ is the complex conjugation of matrix $B$, and $V^\dag$ is the Hermitian conjugation of matrix $V$. In the absence of pairing $\Delta_{ij}$, one would have $V=0$. By definition, the entire matrix $R$ is non-negative, Hermitian and respects the particle-hole symmetry $PRP^{-1}=R^*$, as ensured by $P\Phi_A=(\Phi_A^\dag)^T$, where the particle-hole transformation matrix $P=I_K\otimes \sigma_x$, with $I_K$ being the $K\times K$ identity matrix and $\sigma_x$ the Pauli-$x$ matrix.

Then, one can show that there exists a transformation matrix $L$ (not unique) such that
\begin{equation}\label{eq:pairing-L}
L^\dag L=R^{-1}\ ,\qquad PLP^{-1}=L^{*}\ ,
\end{equation}
namely, preserves the particle-hole symmetry $P$. This allows us to transform the fermion operators into a new Nambu basis $\Psi_A=(d_1,d_2,\cdots, d_K,d_1^\dag,d_2^\dag,\cdots, d_K^\dag)^T$ given by
\begin{equation}
\Psi_A=L\Phi_A\ ,
\end{equation}
where the fermion annihilation and creation operators $d_j$, $d_j^\dag$ are orthogonal:
\begin{equation}
\{d_i,d_j^\dag\}=\delta_{ij}\ ,\qquad \{d_i,d_j\}=\{d_i^\dag,d_j^\dag\}=0\ .
\end{equation}
A systematical way of obtaining the transformation matrix $L$ satisfying \cref{eq:pairing-L} can be found in the SM \cite{suppl}. Such a set of $K$ fermion modes $d_j$ spans the $2^K$ dimensional subHilbert space $h_A$ of spacetime slice $A$, similar to the fermion number conserving case. 

Once the fermion operators $d_j$, $d_j^\dag$ in \cref{eq:pairing-L} are obtained, which identifies the degrees of freedom of the subsystem of spacetime slice $A$, we can calculate the entanglement entropy $S_A$ following Ref. \cite{peschel_calculation_2003}. For any given quantum state $\rho_{\text{tot}}$ which is a Fock state or generically Gaussian state, one can calculate the entanglement entropy $S_A$ inside the spacetime slice $A$ by calculating both the normal and pairing correlation matrices:
\begin{equation}
\begin{split}
&D_{ij}=\text{tr}\left[\rho_{\text{tot}}d_i^\dag d_j\right]\ ,\quad F_{ij}=\text{tr}\left[\rho_{\text{tot}}d_j^\dag d_i^\dag\right]\ ,
\end{split}
\end{equation}
and then calculate the reduced density matrix $\rho_A$ by following the standard formula derived in Ref. \cite{peschel_calculation_2003}, which we will not repeat here due to its relatively complicated form.

\section{ENTANGLEMENT ENTROPY IN TIME-DIRECTION SLICE}\label{sec:time} 

\subsection{Model and definitions}

We note that our definition of entanglement entropy in generic spacetime slice above holds for any free fermion models in any spacetime dimensions. For simplicity,  hereafter, we investigate the spacetime slice entanglement entropy for the 1D tight-binding free fermion model with fermion number conservation (namely, without pairing). 

We consider the following model with $L$ sites and periodic boundary condition:
\begin{equation}\label{eq:H-model}
H=-u \sum_{m=0}^{L-1}\left(c_{m+1}^{\dagger} c_{m}+c_{m}^{\dagger} c_{m+1}\right)\ ,
\end{equation}
where the hopping $u>0$. The fermion energy at quasi-momentum $k$ is $E(k)=-2u\cos(ka_0)$, where $a_0=1$ is the lattice constant. The energy bandwidth is $E_0=4u$. Accordingly, we define the characteristic time $\tau_0$ and the maximal Fermi velocity (the Lieb-Robinson bound velocity \cite{lieb1972}) $v_{\text{max}}$ of the model as
\begin{equation}\label{eq:tau0}
\tau_0=\frac{2\pi}{E_0}\ ,\qquad v_{\text{max}}=\text{max}\left|\frac{dE(k)}{dk}\right|=2u a_0\ .
\end{equation}
Hereafter we set $u=\frac{1}{4}$. Besides, we denote the filling (average fermion number per site) as $\nu\in[0,1]$.

We now consider the case of slice $A$ being a time-direction line segment of total time $t$ on a fixed site $\br_n=0$, containing $K$ equally spaced points at time $t_n=(n-1)\tau$ with coordinates 
\begin{equation}
(\br_n,t_n)=(0,t_n)=(0,(n-1)\tau)\ ,\quad (1\le n\le K)
\end{equation}
and $\tau=\frac{t}{K-1}$ is the spacing between adjacent times, as shown in \cref{fig1b}.

We first study the entanglement entropy $S_A$ of thermal states $\rho_{\text{tot}}$ at temperature $T$ and filling $\nu$ in slice $A$, and then study that for non-eigenstates. We set $L\gg K$, for which $L$ can be approximately viewed as infinite (the thermodynamic limit), and $S_A$ is independent of $L$.

\subsection{Zero temperature} 
\begin{figure}[tbp]
\centering
\vspace{-12mm}
\subfloat{\label{fig2a}
\begin{picture}(0.48\linewidth,0.48\linewidth)
\put(0,0){\includegraphics[width=0.48\linewidth]{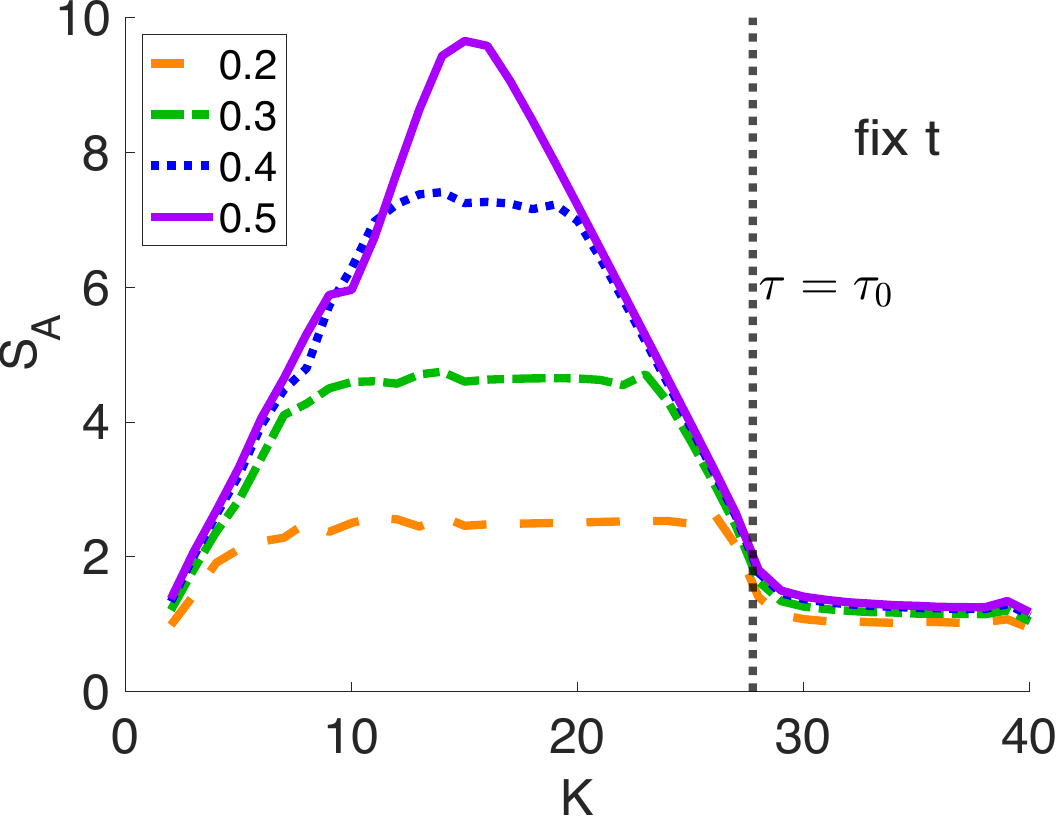}}
\put(-5,90){\footnotesize (a)}
\end{picture}}
\subfloat{\label{fig2b}
\begin{picture}(0.48\linewidth,0.48\linewidth)
\put(0,0){\includegraphics[width=0.48\linewidth]{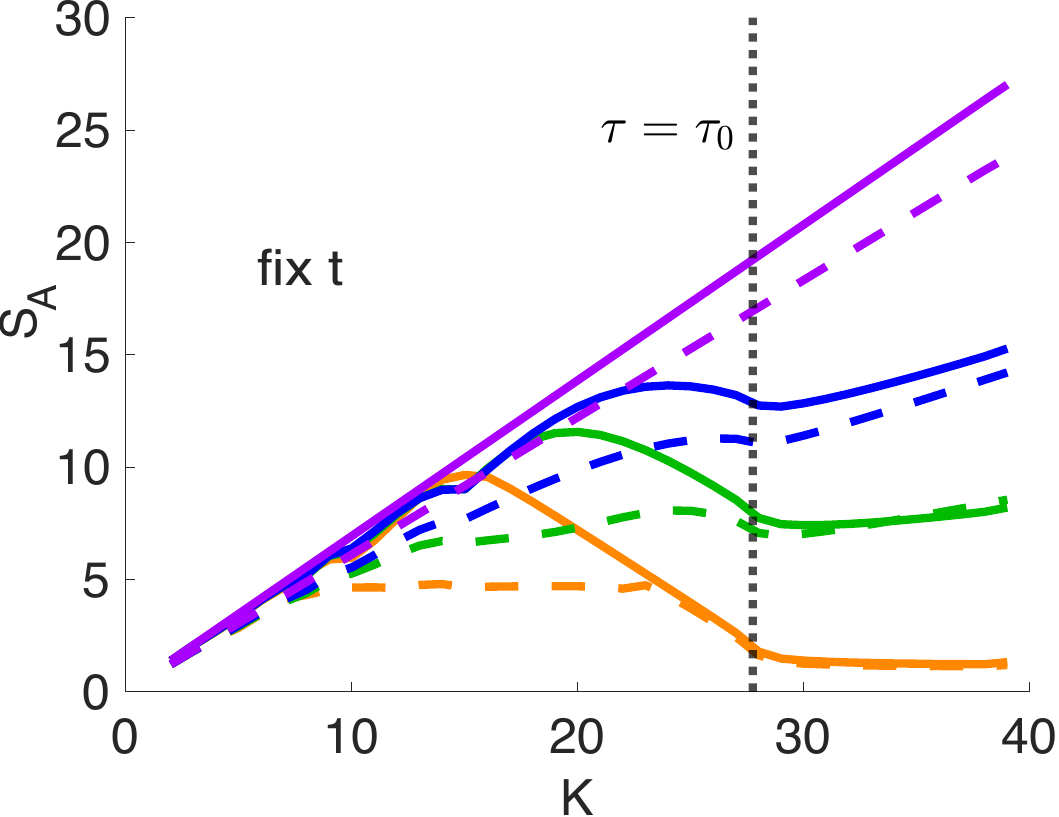}}
\put(-5,90){\footnotesize (b)}
\end{picture}}

\vspace{-12mm}
\subfloat{\label{fig2c}
\begin{picture}(0.48\linewidth,0.48\linewidth)
\put(0,0){\includegraphics[width=0.48\linewidth]{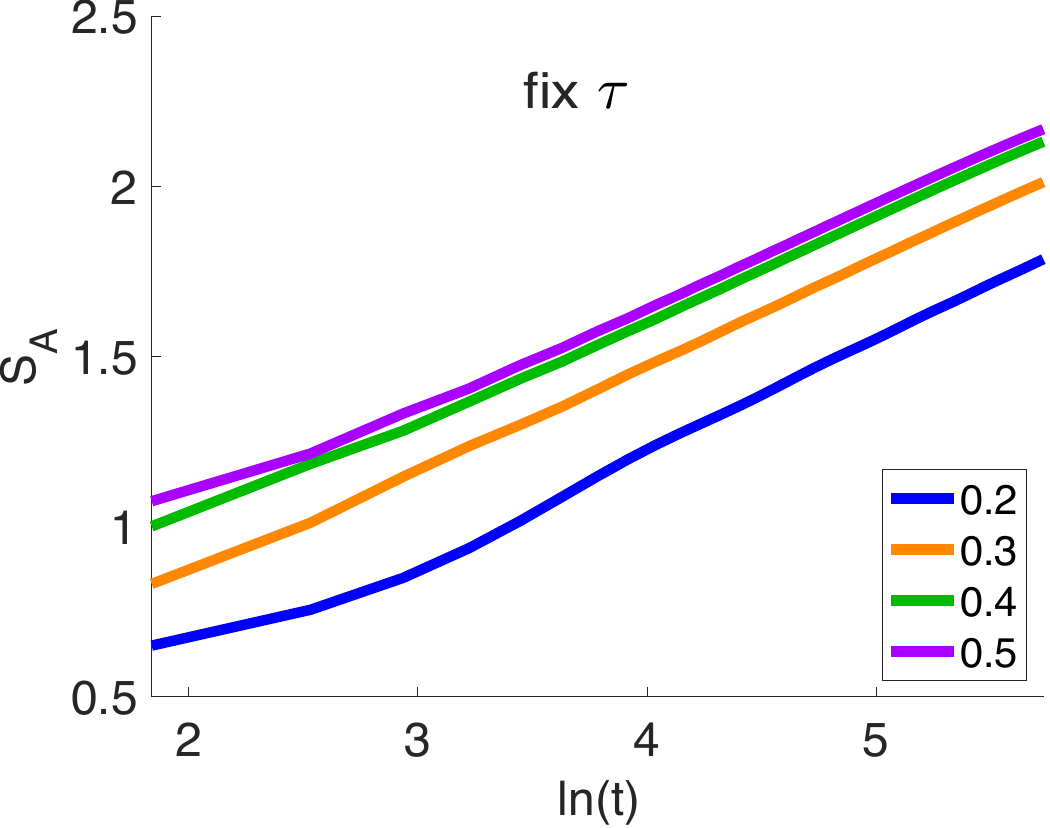}}
\put(-5,90){\footnotesize (c)}
\end{picture}}
\subfloat{\label{fig2d}
\begin{picture}(0.48\linewidth,0.48\linewidth)
\put(0,0){\includegraphics[width=0.48\linewidth]{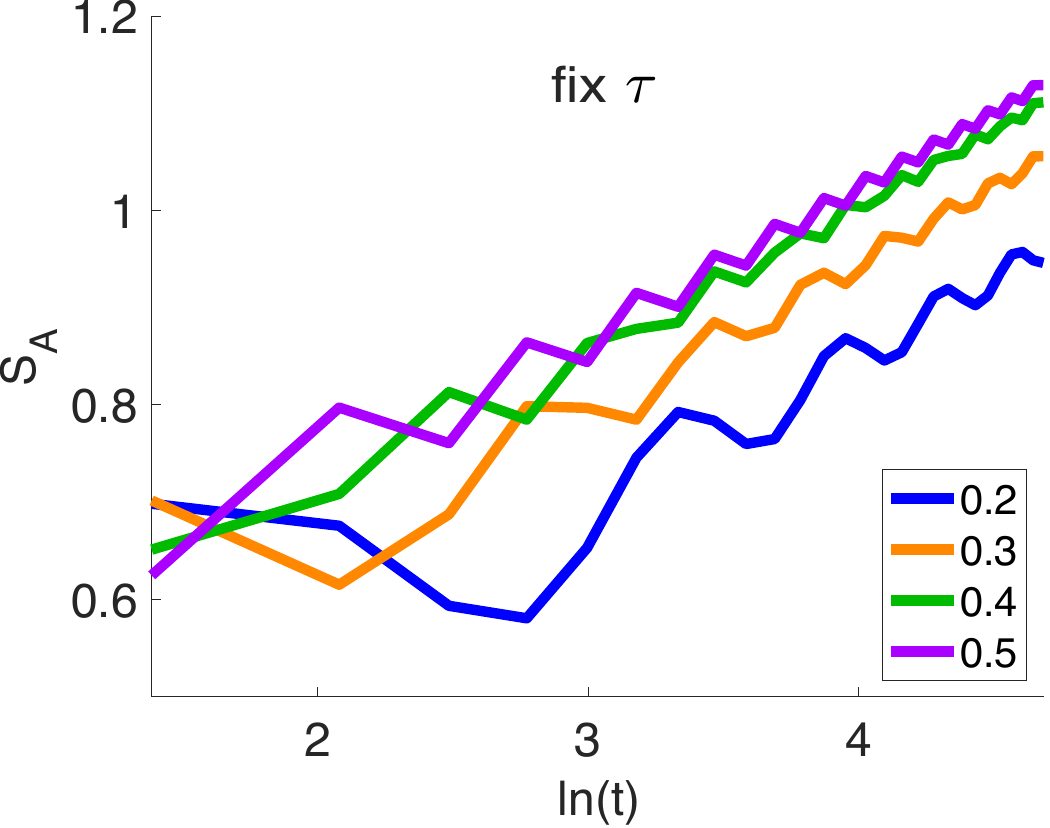}}
\put(-5,90){\footnotesize (d)}
\end{picture}}

\vspace{-12mm}
\subfloat{\label{fig2e}
\begin{picture}(0.48\linewidth,0.48\linewidth)
\put(0,0){\includegraphics[width=0.48\linewidth]{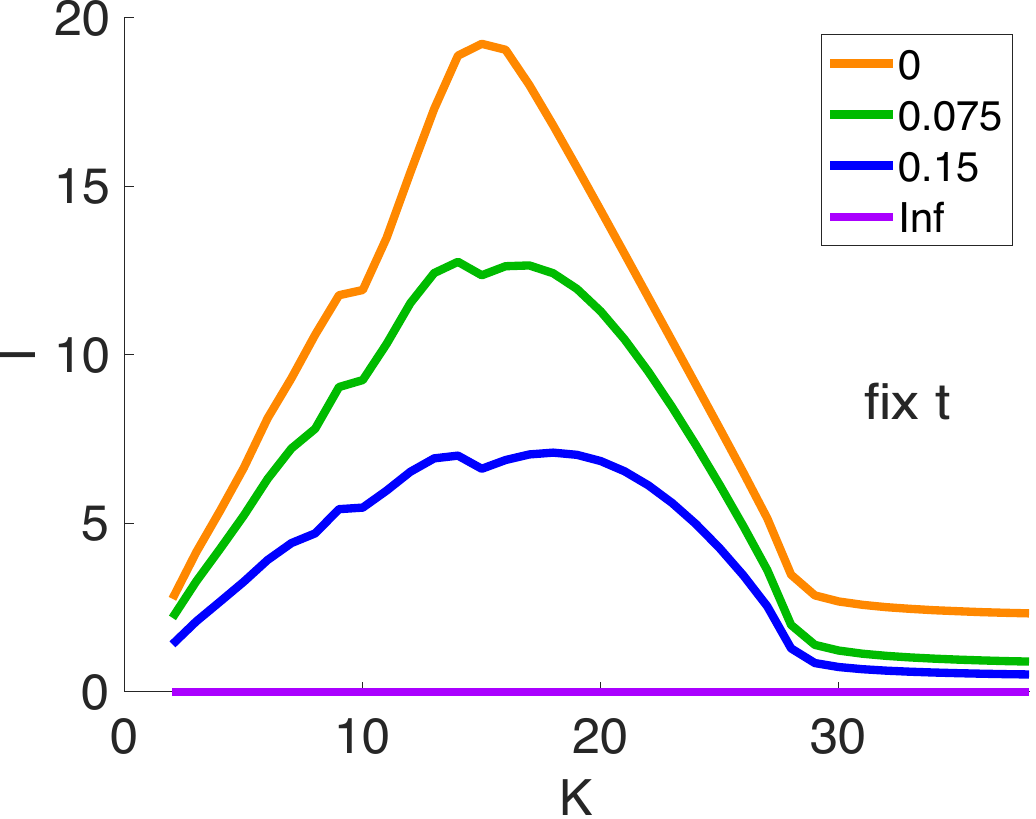}}
\put(-5,90){\footnotesize (e)}
\end{picture}}
\subfloat{\label{fig2f}
\begin{picture}(0.48\linewidth,0.48\linewidth)
\put(0,0){\includegraphics[width=0.48\linewidth]{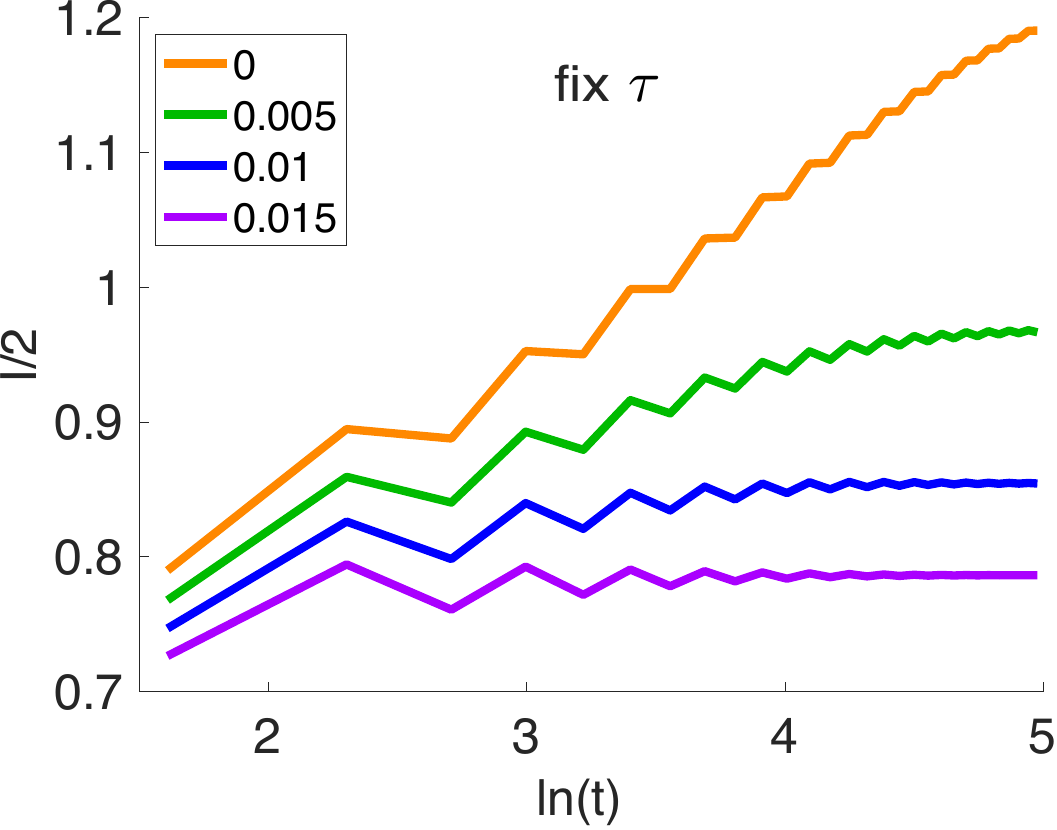}}
\put(-5,90){\footnotesize (f)}
\end{picture}}
\caption{(a)-(d) Entanglement entropy $S_A$ of model \cref{eq:H-model} with $L=500$ in time-direction slice of length $t$ with $K$ points separated by $\tau=t/(K-1)$, for (a) different filling $\nu$ given in the legend at temperature $T=0$ and fixed $2\pi t/\tau_0 = 168$; (b) finite temperature $S_A$ with fixed $2\pi t/\tau_0=168$, at fillings $\nu=0.5$ (solid) and $\nu=0.3$ (dashed). For each $\nu$, the four curves from low to high have $T/E_0=0,0.075, 0.15, \infty$, respectively. (c)-(d) $S_A$ versus $\ln t$ for different fillings $\nu$ (see legend) at $T=0$ and fixed (c) $\tau=\tau_0$ and (d)  $\tau=\frac{2}{\pi} \tau_0$. (e)-(f) Mutual information $\I$ with $L=200$ at $\nu = 0.5$ and temperatures $T/E_0$ given in the legends while fixing (e) $2\pi t/\tau_0 = 168$ and (f) $\tau = \frac{5}{2\pi} \tau_0$, respectively.
}
\label{fig2}
\end{figure}

At temperature $T=0$, $\rho_{\text{tot}}$ is a pure state (ground state with a Fermi sea). For fixed total time $t$, as shown in \cref{fig2a}, within the range $K<t/\tau_0$, or equivalently $\tau>\tau_0$, $S_A$ shows a triangular (trapezoid) shape curve with respect to $K$ for filling $\nu=0.5$ ($\nu\neq 0.5$); while in the range $K>t/\tau_0$, or $\tau<\tau_0$, $S_A$ approaches a stabilized value. 

We further examine the total time $t$ dependence of the stabilized value of $S_A$ by fixing the time separation $\tau$ and changing the number of points $K=\frac{t}{\tau}+1$. Interestingly, we find $S_A$ is approximately linear in $\ln t$, where the slope $\left(\frac{\partial S_A}{\partial\ln K}\right)_\tau\approx\left(\frac{\partial S_A}{\partial\ln t}\right)_\tau$ is $\frac{1}{3}$ at $\tau=\tau_0$ (\cref{fig2c}), and quickly saturates to $\frac{1}{6}$ when $\tau<\tau_0$ (\cref{fig2d}). In addition, $S_A$ shows an oscillation  of period $\text{max}(\frac{1}{\nu},\frac{1}{1-\nu})$ with respect to $K$ \cite{suppl} at filling $\nu$. These conclusions are insensitive to the energy dispersion \cite{suppl}.

The zero temperature $S_A$ observed above can be theoretically understood in three regimes as follows. Since here we are considering the infinite system size limit $L\rightarrow\infty$, the single-particle energy $E$ takes continuous values. Assume $\Omega(E)\ge 0$ denotes the normalized density of states of the fermion band satisfying $\int \Omega(E) dE=1$. For the time-direction slice $A$, the matrix in \cref{eq:anti-comm-matrix} reads 
\begin{equation}
B_{m n} = \int e^{-\ri E (t_m - t_n)} \Omega(E) dE\ ,
\end{equation}
while the spacetime correlation matrix in \cref{eq:correlation-matrix} is 
\begin{equation}
C_{m n}=\int \Omega(E)n_F(E) e^{\ri E \left(t_{m}-t_{n}\right)}dE
\end{equation}
where $n_F(E)$ is the zero temperature Fermi-Dirac distribution function at filling $\nu$. The above expressions are generically valid in the infinite system size limit $L\rightarrow\infty$.

\subsubsection{The $\tau\gg \tau_0$ case: volume law}

When $\tau\gg \tau_0$, or $K\ll t/\tau_0$, one has $|t_m-t_n|\gg \tau_0=2\pi/E_0$ for $m\neq n$. This leads to
\begin{equation}\label{eq:BC-largetau}
\begin{split}
&B_{mn}\approx\delta_{mn}\int \Omega(E) dE=\delta_{mn}\ ,\\
&C_{mn}\approx \delta_{mn}\int \Omega(E)n_F(E) dE=\nu\delta_{mn}\ ,
\end{split}
\end{equation}
since the off-diagonal elements with $m\neq n$ are suppressed by their fast oscillating integrands proportional to $e^{\ri E \left(t_{m}-t_{n}\right)}$. As a result, the $B$ matrix is simply the identity matrix, and thus by \cref{eq:M-matrix} one can choose the matrix $M=I$. This yields $D=M^*CM^T\approx\nu I$, and by \cref{eq:S-Dexpression} one has
\begin{equation}\label{eq:S-smallK}
S_A \approx K[-\nu\ln \nu - (1-\nu) \ln(1-\nu)]\ . \quad (\tau\gg \tau_0)
\end{equation}
This fits well with the slope $\left(\frac{\partial S_A}{\partial K}\right)_t$ at small $K$ in \cref{fig2a} \cite{suppl}. Note that this result also holds for any finite temperature, since the finite temperature Fermi-Dirac distribution function $n_F(E)$ will not change \cref{eq:BC-largetau}. 

In particular, if one approximates $\Omega(E)$ as uniform within the range $[-\frac{E_0}{2},\frac{E_0}{2}]$:
\begin{equation}\label{eq:DOS}
\Omega(E)= E_0^{-1}\Theta\left(\frac{E_0}{2}-|E|\right)\ ,
\end{equation}
where $\Theta(x)=1$ ($=0$) when $x>0$ ($x<0$), 
\cref{eq:S-smallK} becomes exact when $\nu=\frac{p}{q}$ ($p,q\in\mathbb{N}^+$ coprime) and $\tau=jq\tau_0$ ($j\in\mathbb{N}^+$). 

Since $t=(K-1)\tau$, \cref{eq:S-smallK} implies that for fixed $\tau\gg\tau_0$, the time-direction entanglement entropy has a time volume law (in the infinite system size limit $L\rightarrow\infty$): 
\begin{equation}
S_A\propto t\ ,\qquad (\tau\gg\tau_0) \,
\end{equation}
%{\color{blue} in the infinite system size limit ($L\rightarrow\infty$).} 
Intuitively, this is because the discrete points $t_m$ in the time-direction slice $A$ are too far from each other when $\tau\gg \tau_0$, and thus they are nearly uncorrelated (as seen from $C\propto I$) and have almost orthogonal fermion basis (as seen from $B=I$). As a result of lack of correlations, the time-direction entanglement entropy $S_A$ is approximately the sum of the entropy of each time point $t_m$, and thus obey the volume law when $\tau\gg \tau_0$.

We note that $S_A$ in this limit is similar to the tensor network temporal entanglement entropy in \cite{Muller-Hermes_2012,Hasting_connecting_ent_2015}, although the definitions differ significantly \cite{suppl}.

\subsubsection{The $\tau=\tau_0$ case: duality to the spatial case}

When $\tau=\tau_0$, or $K=\frac{t}{\tau_0}+1$, the time direction entanglement entropy shows a similar behavior as the conventional spatial entanglement entropy. This can be most easily seen by taking the uniform $\Omega(E)$ approximation in \cref{eq:DOS}, for which one has exactly 
\begin{equation}
B_{mn} = E_0^{-1}\int_{-E_0/2}^{E_0/2} e^{-\ri (m - n) E \tau_0} dE=\delta_{mn}\ ,
\end{equation}
namely, $B=I$ is simply the identity matrix. This indicates that the fermion operators $c_{0}(t_n)$ at different time $t_n$ already form an orthogonal basis, and thus we can choose the matrix $M=I$. 
By \cref{eq:correlation-matrix}, this gives 
\begin{equation}
D_{mn}=C_{mn}=\frac{\tau_0}{2\pi}\int_{-\pi/\tau_0}^{\pi/\tau_0} n_F(E)e^{\ri (m - n) E \tau_0} dE\ .
\end{equation}

Note that this $D_{mn}$ can be viewed as the \emph{spatial} correlation matrix of a ``lattice fermion model" with a ``lattice constant" $\tau_0$, where $E$ is the ``quasi-momentum" in the ``Brillouin zone" $[-\frac{\pi}{\tau_0},\frac{\pi}{\tau_0}]$, and $n_F(E)=\Theta(E_F-E)$ for some Fermi energy $E_F$. Therefore, the time-direction correlation matrix $D_{mn}$ is equivalent to the spatial correlation matrix of a ``zero temperature Fermi sea" state of a 1D free fermion lattice model occupying the interval of ``quasi-momentum" $-\frac{E_0}{2}\le E\le E_F$. In other words, we see that the time-direction entanglement entropy $S_A$ at $\tau=\tau_0$, which is a function of correlation matrix $D$, is dual to the conventional spatial entanglement entropy of a free fermion lattice model.

By the Calabrese-Cardy formula \cite{Cardy_Peschel_1988,calabrese_qft_2004,calabrese_cft_2009}, the zero-temperature entanglement entropy of gapless states of infinite 1D space in a spatial subregion of length $\ell$ is $S_A =\frac{c_L+c_R}{6}\ln(\ell/\ell_c)$ with some ultraviolet (UV) cutoff constant $\ell_c$, where $c_L$ and $c_R$ are the left and right central charges, respectively. Therefore, the above duality indicates that our zero-temperature time-direction entanglement entropy $S_A$ at $\tau=\tau_0$ should resemble that of the 1D gapless lattice fermion (which is nonchiral with $c_L=c_R=1$) in a length $t$ subregion (in the limit $L\rightarrow\infty$):
\begin{equation}\label{eq:S-tau0}
S_A \approx\frac{1}{3}\ln \left(\frac{t}{t_c}\right)\approx\frac{1}{3}\ln \left(\frac{K\tau}{t_{c}}\right),\quad (\tau=\tau_0,T=0)
\end{equation}
where $t_c$ is some UV cutoff constant. This agrees with our observations in \cref{fig2c}. The similarity between time-direction entanglement entropy in \cref{eq:S-tau0} and the conventional spatial entanglement entropy of free lattice fermions also suggests that this $\tau=\tau_0$ case here might be similar to the dual unitary circuits \cite{Prosen_2020,Kasim_2023,Logaric_2024}.

\subsubsection{The $\tau\ll\tau_0$ case: the continuum time limit}

When $\tau\ll\tau_0$, or $K\gg t/\tau_0$, we are approaching the continuum time limit where the neighboring points are close in time. 

\begin{figure}[tbp]
\centering
\subfloat{\label{fig16a}
\begin{picture}(0.48\linewidth,0.48\linewidth)
\put(0,0){\includegraphics[width=0.48\linewidth]{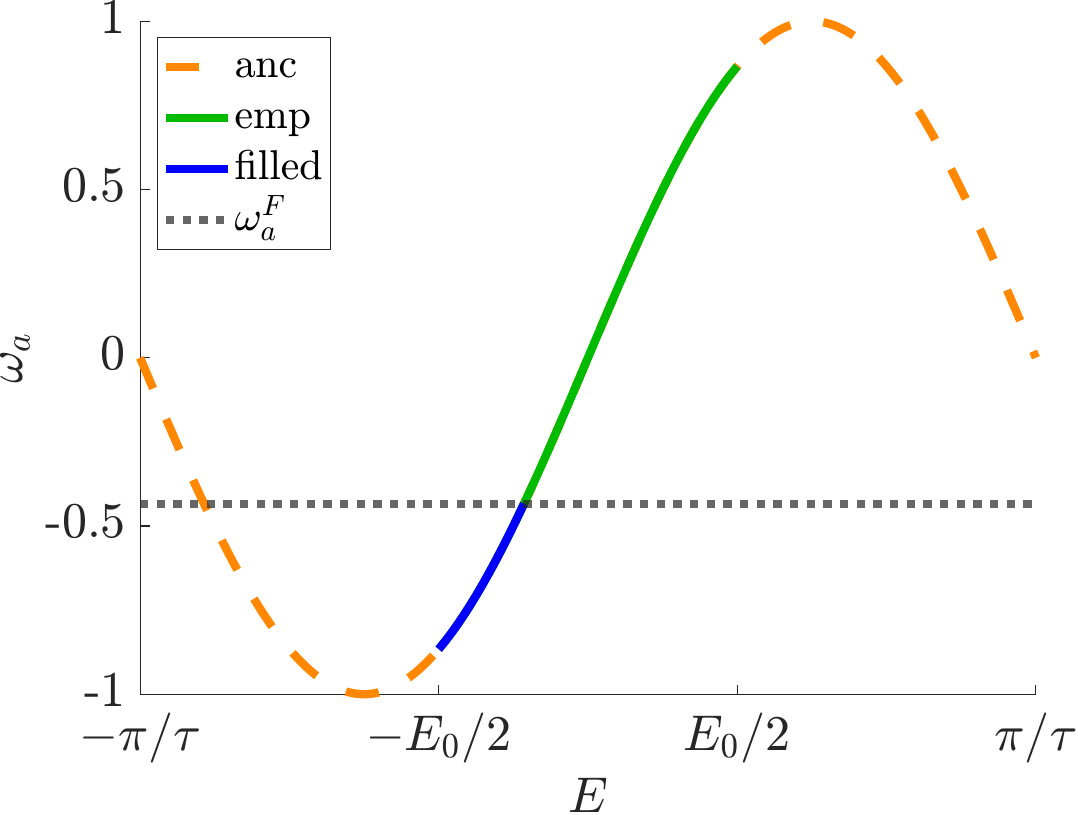}}
\put(-5,90){\footnotesize (a)}
\end{picture}}
\subfloat{\label{fig16b}
\begin{picture}(0.48\linewidth,0.48\linewidth)
\put(-5,90){\footnotesize (b)}
\put(0,45){\includegraphics[width=0.48\linewidth]{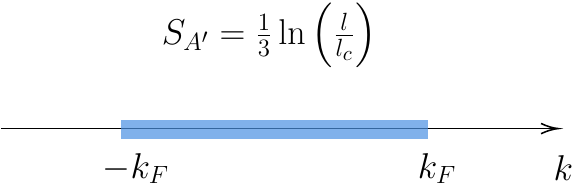}}
\put(-5,40){\footnotesize (c)}
\put(0,0){\includegraphics[width=0.48\linewidth]{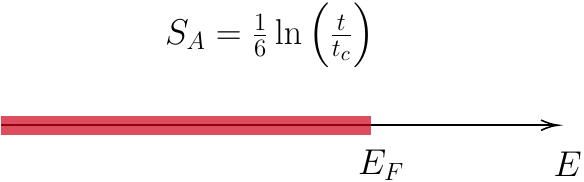}}
\end{picture}}
\caption{(a) The ancillary dispersion for explaining $S_A$ when $\tau<\tau_0$ (``anc" for ancillary, ``emp" for empty).  (b) For 1D gapless lattice fermion, the zero temperature Fermi sea state has two Fermi points at $\pm k_F$ in the momentum space. Accordingly, the Calabrese-Cardy formula for the spatial entanglement entropy is $S_{A'} =\frac{1}{3}\ln\left(\frac{l}{l_{c}}\right)$. (c) In the energy space, the zero temperature state has single-particle states with $E \leq E_F$ are filled, so there is only one Fermi point $E_F$ in the energy domain. Accordingly, the time-direction entanglement entropy in the continuum time limit is $S_{A} =\frac{1}{6}\ln\left(\frac{t}{t_{c}}\right)$. 
}
\label{fig16}
\end{figure}

To explain the numerical results in \cref{fig2d}, we imagine the discrete time points at site $0$ can be understood as an ``ancillary lattice fermion model" with lattice constant $\tau$, and the time direction is playing the role of ``spatial direction" of this ``ancillary model". In this picture, the original energy $E$ can be viewed as the ``quasi-momentum" of this ``ancillary model", which lives in the Brillouin zone of this ``ancillary model" which is an interval $[-\frac{\pi}{\tau},\frac{\pi}{\tau}]$. Assume this ``ancillary model" has an ancillary dispersion $\omega_a(E)$ as shown in \cref{fig16a}, and a ``Fermi level" $\omega_a^F=\omega_a(E_F)$, where $E_F$ is the physical Fermi energy, or the ``Fermi momentum" of this ``ancillary model". Then the physical Fermi sea state, which has levels $E<E_0$ fully occupied, maps to the ``Fermi sea state" of the ``ancillary model" with levels $\omega_a(E)<\omega_a^F$ fully occupied.

Meanwhile, physically, the energy $E$ can only take values in the energy range of the physical band $[-\frac{E_0}{2},\frac{E_0}{2}]$, which is much smaller than the Brillouin zone of the ``ancillary model". Therefore, the physical energy range $[-\frac{E_0}{2},\frac{E_0}{2}]$ plays the role of a ``quasi-momentum" cutoff near the right ``Fermi surface", as shown by the solid line in \cref{fig16a}, keeping only the right-moving chiral fermions of the ``ancillary model". Thus, the correlation matrix $D$ resembles that of one right-moving chiral fermion mode, and accordingly, we expect the time-direction entanglement entropy $S_A$ in the $\tau\ll\tau_0$ limit to stabilize into the Calabrese-Cardy formula for a chiral fermion mode with $c_L=0, c_R=1$ (in the limit $L\rightarrow\infty$), namely:
%if we enlarge the range of energy $E$ from $[-\frac{E_0}{2},\frac{E_0}{2}]$ to an ancillary range $[-\frac{\pi}{\tau},\frac{\pi}{\tau}]$, we can think of $E$ as the ``quasi-momentum" in the Brillouin zone $[-\frac{\pi}{\tau},\frac{\pi}{\tau}]$ of an ``ancillary lattice fermion model" with lattice constant $\tau$, which we assume has an ancillary dispersion $\omega_a(E)$ in \cref{fig16a}. Assume the ancilla model has a ``Fermi level" $\omega_a^F=\omega_a(E_F)$. Meanwhile, the physical energy range $[-\frac{E_0}{2},\frac{E_0}{2}]$ plays the role of a ``quasi-momentum" cutoff near the right ``Fermi surface" in \cref{fig16a}, keeping only the right-moving chiral fermions. As a result, the correlation matrix $D$ resembles that of one right-moving chiral fermion mode. Thus, $S_A$ stabilizes into the Calabrese-Cardy formula with $c_L=0, c_R=1$:
\begin{equation}\label{eq:S-largeK}
S_A \approx\frac{1}{6}\ln \left(\frac{t}{t_c}\right)\approx\frac{1}{6}\ln \left(\frac{K\tau}{t_{c}}\right).\quad (\tau\ll\tau_0, T=0)
\end{equation}
where again $t_c$ is some UV cutoff constant.

More intuitively, the result of \cref{eq:S-largeK} in the continuum time limit can be understood from the picture of \cref{fig16}b-c. In an infinite 1D lattice model, the ground state, when viewed in the momentum $k$ space, is a Fermi sea state occupying the momentum interval $[-k_F,k_F]$ (\cref{fig16}b), which has two Fermi points at $k_F$ and $-k_F$. Accordingly, the \emph{spatial} entanglement entropy is $S_{A'}=\frac{1}{3}\ln(\ell/\ell_c)$ for a spatial subregion $A'$ of length $\ell$, where the coefficient $\frac{1}{3}$ is given by $\frac{1}{6}$ times the total number of Fermi points in the momentum space. This is because each Fermi point contributes a chiral fermion degree of freedom, contributing chiral central charge $1$. In analogy, to understand the time-direction entanglement entropy, we should view the ground state in the energy $E$ space, which is a Fermi sea state occupying the energy range $E<E_F$ (\cref{fig16}c). In this case, there is only one Fermi point $E_F$ in the energy space (the lower bound $-\frac{E_0}{2}$ plays the role of the UV cutoff rather than a Fermi point, as explained by \cref{fig16}a). Accordingly, the \emph{time-direction} entanglement entropy $S_A=\frac{1}{6}\ln(t/t_c)$, and the coefficient $\frac{1}{6}$ is given by $\frac{1}{6}$ times the number of Fermi points in the energy space.

Therefore, we see that the time-direction entanglement entropy in the continuum time limit is probing the ``effective central charge" in the time direction, which is the number of Fermi points in the energy space. This piece of information is different from that given by the spatial entanglement entropy.

\subsection{Finite temperature} 
At temperature $T>0$, the total state is in the thermal mixed state. As shown in \cref{fig2c}, $S_A$ for fixed total time $t$ still shows a triangle or trapezoid like curve with respect to $K$ when $K<t/\tau_0$ ($\tau>\tau_0$), and stabilizes into a linear relation when $K>t/\tau_0$ ($\tau<\tau_0$). This shows that the time-direction entanglement entropy $S_A$ shows a volume law in $t$ (namely, linear in $t$) at finite temperature. The small $K$ (namely, $\tau\gg \tau_0$ for fixed total time $t$) slope obeys \cref{eq:S-smallK} with a similar reasoning.

We also calculate the mutual information $\I$ in \cref{eq:mutual}. For fixed total time $t$, $\I$ at temperature $T>0$ is shown in \cref{fig2e}, which shows stabilization at $K>t/\tau_0$ ($\tau<\tau_0$), similar to $S_A$ at $T=0$ (\cref{fig2a}). At $T=\infty$, it is straightforward to show that $\I\equiv 0$. For fixed $\tau<\tau_0$, we find $\I$ increases at small $t$, and saturates as $t\gtrsim \beta=T^{-1}$, as shown in \cref{fig2f}.

%{\color{blue}Although we calculated entanglement entropy and mutual information, neither is a good entanglement measure for mixed states.}

\subsection{Non-eigenstates} 
\begin{figure}[tbp]
\centering
\vspace{-12mm}
\subfloat{\label{fig3a}
\begin{picture}(0.48\linewidth,0.48\linewidth)
\put(0,0){\includegraphics[width=0.48\linewidth]{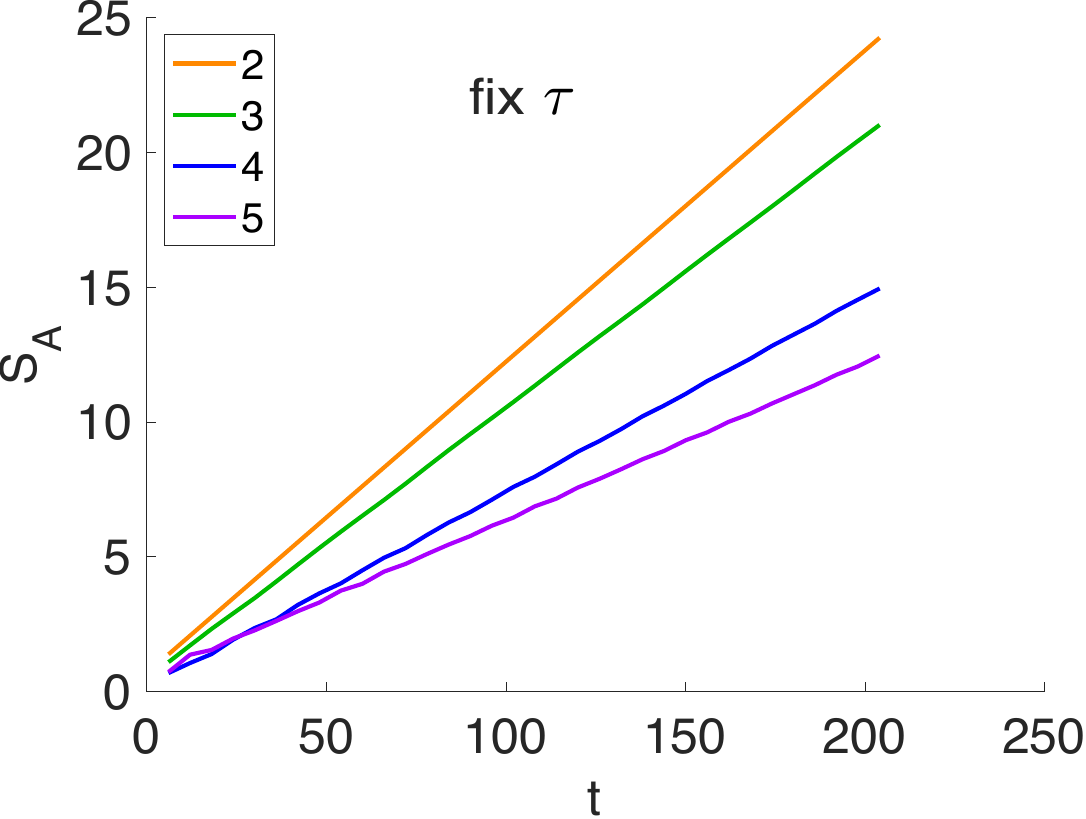}}
\put(-5,90){\footnotesize (a)}
\end{picture}}
\subfloat{\label{fig3b}
\begin{picture}(0.48\linewidth,0.48\linewidth)
\put(0,0){\includegraphics[width=0.48\linewidth]{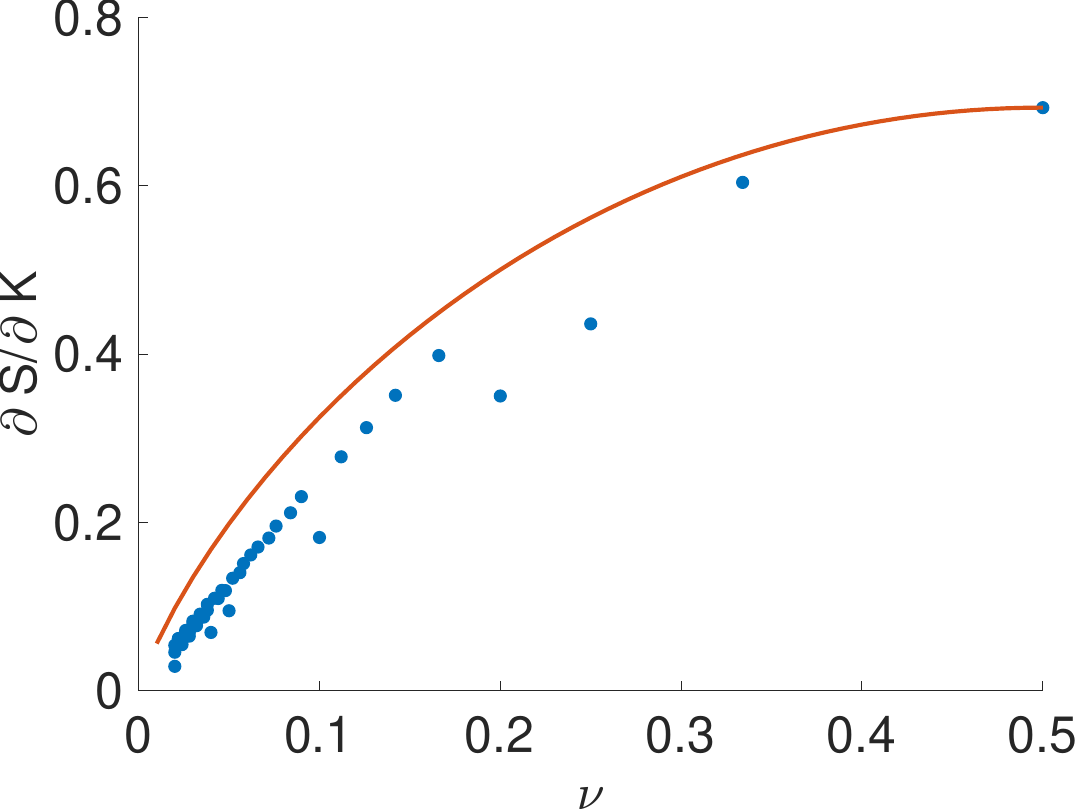}}
\put(-5,90){\footnotesize (b)}
\end{picture}}

\vspace{-10mm}
\subfloat{\label{fig3c}
\begin{picture}(0.48\linewidth,0.48\linewidth)
\put(0,0){\includegraphics[width=0.48\linewidth]{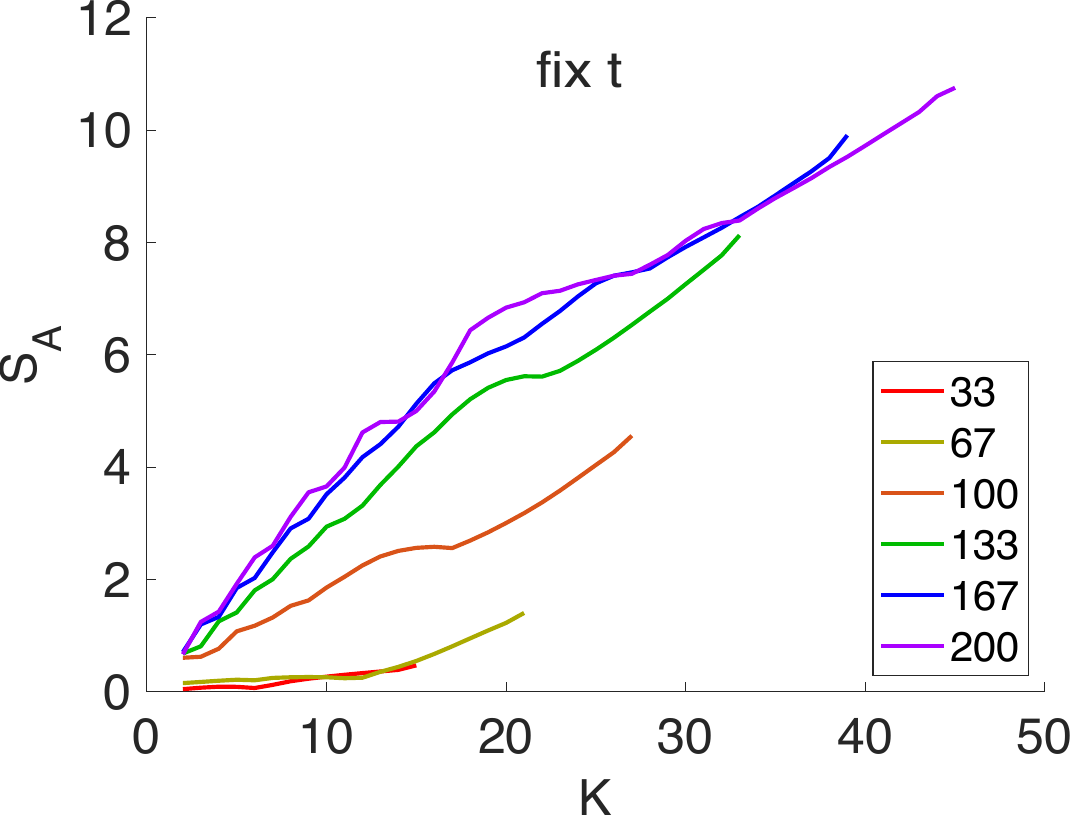}}
\put(-5,90){\footnotesize (c)}
\end{picture}}
\subfloat{\label{fig3d}
\begin{picture}(0.48\linewidth,0.48\linewidth)
\put(0,0){\includegraphics[width=0.48\linewidth]{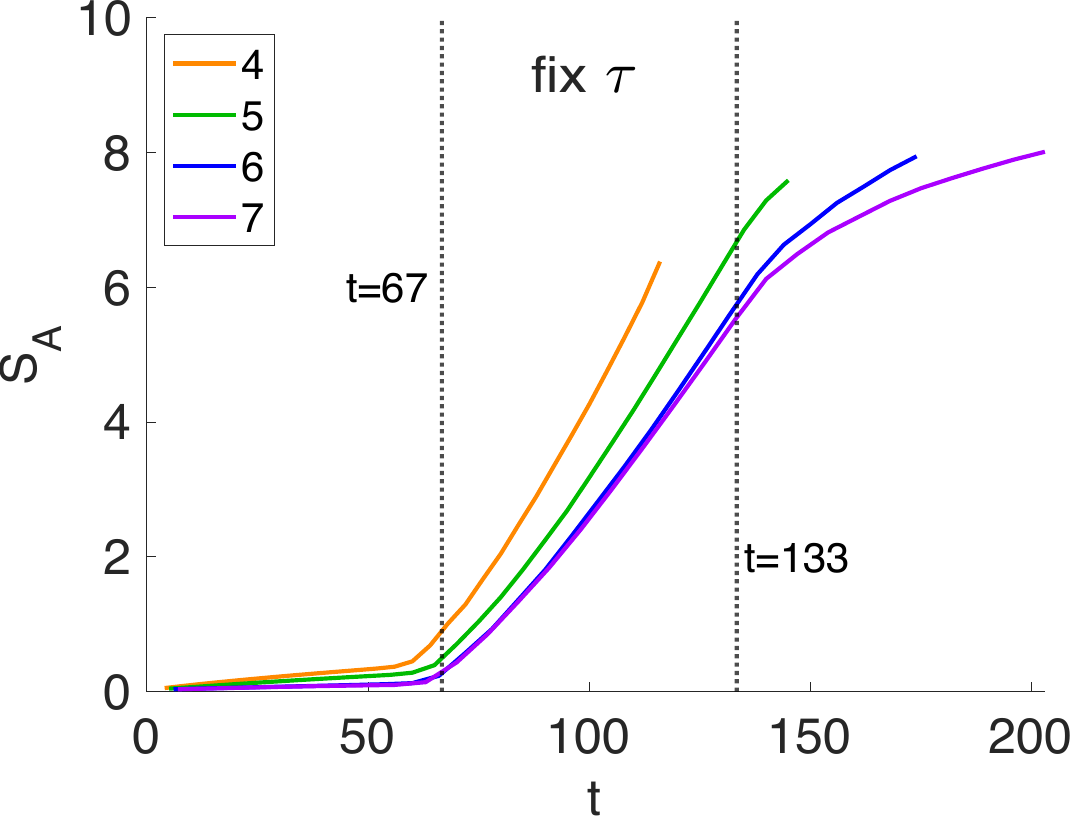}}
\put(-5,90){\footnotesize (d)}
\end{picture}}
\caption{Entanglement entropy $S_A$ of time-direction slice of $K$ points at site $m=0$ for non-eigenstates (a)-(b) $|\psi_q\rangle=\prod_{m} c_{q m}^\dag|0\rangle$ ($L=500$) with fixed $2 \pi \tau / \tau_0 = 6$, and (c)-(d) $|\psi\rangle=\prod_{-\frac{N_f}{3}<m<\frac{2N_f}{3}}c_m^\dag|0\rangle$ ($L=1000$) with $N_f=100$. $q$ is given in the legend of (a). In (b), the dots are the fitted $\left(\frac{\partial S_A}{\partial K}\right)_\tau$ and the red line is $-\nu\ln\nu-(1-\nu)\ln(1-\nu)$. (c) is calculated with fixed $2 \pi t / \tau_0$ given in the legend, while (d) has fixed $2 \pi \tau / \tau_0$ given in the legend.
}
\label{fig3}
\end{figure}

We now study the time-direction entanglement entropy $S_A$ in time direction slice $A$ of \cref{fig1b} at site $m=0$ for pure non-eigenstates. We first consider state 
\begin{equation}
|\psi_q\rangle=\prod_{m} c_{q m}^\dag|0\rangle\ ,
\end{equation}
with one occupied site per $q$ sites ($q\in\Z$), which has filling $\nu \approx \frac{1}{q}$ (exact if $q$ divides $L$). As shown in \cref{fig3a}, $S_A$ with fixed spacing $\tau$ shows a volume law $S_A \propto t$, which has a slope $\left(\frac{\partial S_A}{\partial K}\right)_\tau$ upper bounded by $-\nu\ln \nu - (1-\nu) \ln(1-\nu)$, as shown in \cref{fig3b}. This is similar to that of thermal states (\cite{suppl} Fig. S7), indicating the similarity between a partially filled pure state at finite energy density and a thermal state. Another state we calculate is 
\begin{equation}
|\psi\rangle=\prod_{-\frac{N_f}{3}<m<\frac{2N_f}{3}}c_m^\dag|0\rangle\ ,
\end{equation}
which has $N_f$ consecutive occupied sites. As shown in \cref{fig3c,fig3d} with $N_f=100$, the behavior of $S_A$ changes sharply as $t$ reaches two times $t_L=\frac{N_f a_0}{3 v_\text{max}}\approx67$ and $t_R=\frac{2N_f a_0}{3 v_\text{max}}\approx 133$, which correspond to the times for signals at two edges of the occupied interval to reach slice $A$. 

The above behaviors of $S_A$ can be understood as follows. For $t<\min(t_L,t_R)$, $S_A$ versus $K$ with fixed $t$ (\cref{fig3c}) or fixed $\tau$ (\cref{fig3d}) remains almost zero for $t<\min(t_L,t_R)$, resembling that of a fully occupied state. This is because the signals from the edges of the occupied interval has not yet propagated to the time direction slice $A$, and the slice $A$ effectively sees a fully occupied state with zero entropy. When $t>\text{min}(t_L,t_R)$, the time direction slice $A$ starts to see the state as partially filled, and $S_A$ starts to grow.

\section{LINEAR SPACETIME SLICES}\label{sec:linear}
\begin{figure}[tbp]
\centering
\vspace{-12mm}
\subfloat{\label{fig4a}
\begin{picture}(0.48\linewidth,0.48\linewidth)
\put(0,0){\includegraphics[width=0.48\linewidth]{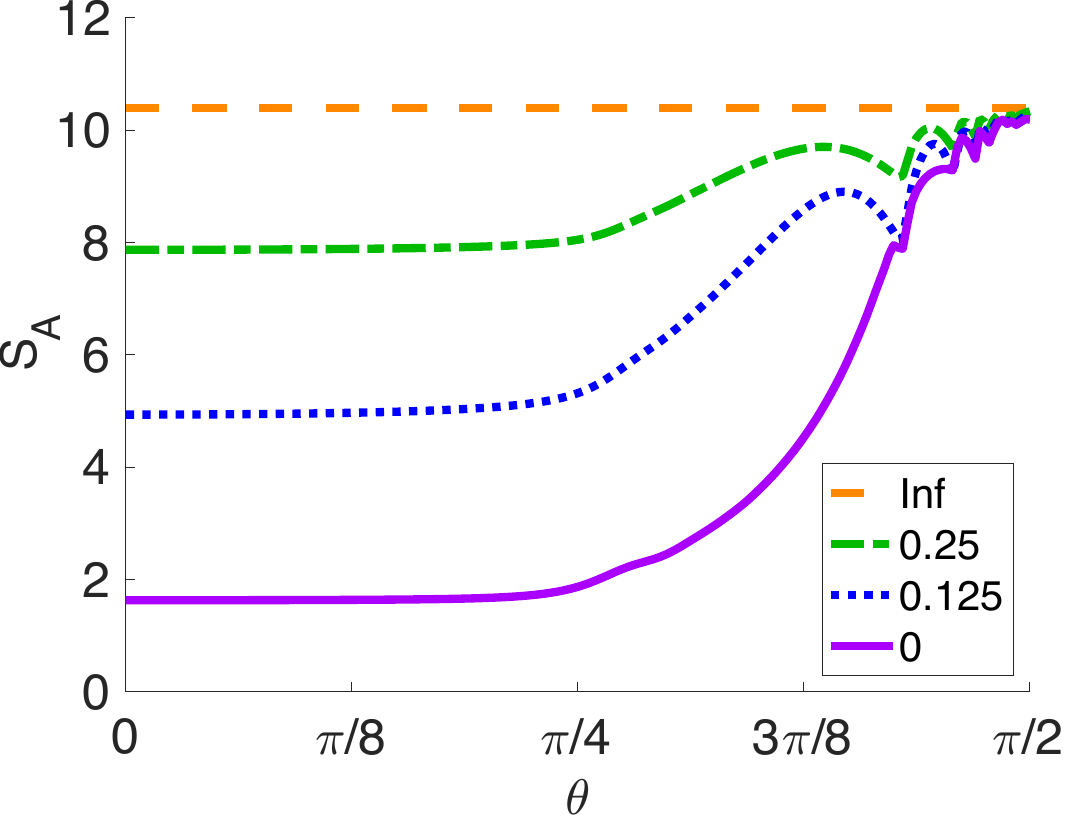}}
\put(-5,90){\footnotesize (a)}
\end{picture}}
\subfloat{\label{fig4b}
\begin{picture}(0.48\linewidth,0.48\linewidth)
\put(0,0){\includegraphics[width=0.48\linewidth]{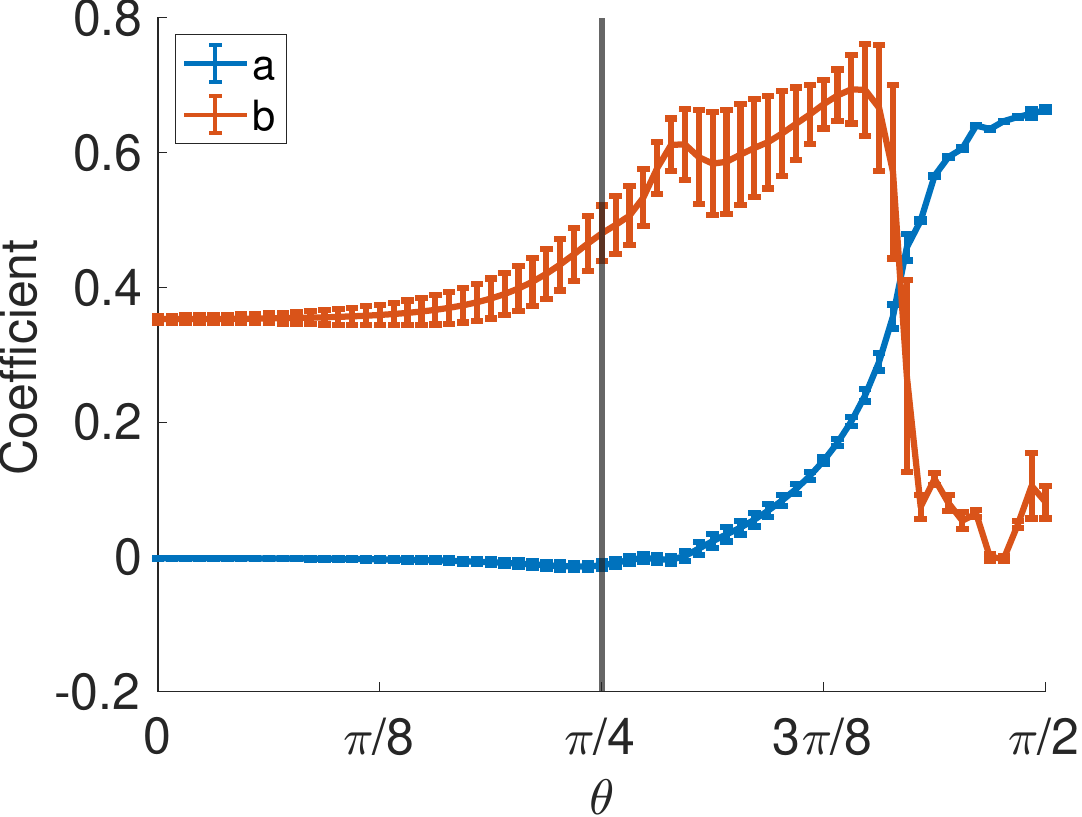}}
\put(-5,90){\footnotesize (b)}
\end{picture}}
\caption{(a) Entanglement entropy $S_A$ as a function of $\theta$ with fixed $\ell=15$ in the linear slice in \cref{fig1c} at $\nu=0.5$ and $T/E_0$ given in the legend. (b) For the linear slice, we fit $S_A = a\ell + b\ln(\ell) + c$ to the zero temperature entanglement entropy at $1\leq\ell\leq20$. Then we plot $a$ and $b$ against $\theta$. At $\theta=0$, $a\approx0$ and $b\approx0.35$, which agrees with the theoretical prediction $S_A \approx\frac{1}{3}\ln(\ell/\ell_c)$. The value of $a$ starts to increase significantly at around $\theta=0.25\pi\sim0.3\pi$, when the ln term becomes sub-leading. When the linear term begins to saturate at $\theta\approx0.4\pi$, $b$ sharply drops to below $0.1$. At $\theta=\frac{\pi}{2}$, $a\approx0.66$ and $b\approx0.08$, which also agrees with the theoretical prediction $S_A\approx\ln(2)\ell+c$.
}
\label{fig4}
\end{figure}

We further investigate the entanglement entropy $S_A$ in a linear spacetime slice at angle $\theta$ containing $\ell$ points at $(x_n,t_n)=na_0(1,v_{\text{max}}^{-1}\tan\theta)$, where $0\le n\le \ell-1$ (see \cref{fig1c}). Therefore, the slice is spacelike (non-causal)
when $\theta<\frac{\pi}{4}$, and timelike (causal) when $\theta>\frac{\pi}{4}$. As shown in \cref{fig4a}, $S_A$ for any temperature $T$ thermal state stays nearly constant for $\theta < \frac{\pi}{4}$, while increases when $\theta>\frac{\pi}{4}$. This suggests an approximate ``conformal symmetry" among all spacelike slices. We fit the zero temperature $S_A$ by 
\begin{equation}
S_A=a\ell +b \ln(\ell/\ell_c)\ , 
\end{equation}
and the coefficients $a$ and $b$ as a function of $\theta$ are shown in \cref{fig4b}.
We find $a\approx 0$ and $S_A \approx\frac{1}{3}\ln(\ell/\ell_c)$ for $\theta<\frac{\pi}{4}$, as expected from the Calabrese-Cardy formula \cite{Cardy_Peschel_1988,calabrese_qft_2004,calabrese_cft_2009}. For timelike slice with $\theta>\frac{\pi}{4}$, $S_A$ shows a volume law with $a>0$, and shows similarity to the time-direction slice case: as $\theta$ approaches $\frac{\pi}{2}$, the time separation $\tau=\frac{a_0}{v_{\text{max}}}\tan\theta\gg\tau_0$, and $S_A$ approaches \cref{eq:S-smallK} with $K=\ell$ points. To summarize, in this setup we find
\begin{equation}
S_A\approx \begin{cases}
& \frac{1}{3}\ln(\ell/\ell_c)\ ,\qquad (\text{spacelike})\ ,\\
& a\ell\ ,\quad (a>0), \quad (\text{timelike}) \ .
\end{cases}
\end{equation}

\section{DISCUSSION}\label{sec:discussion}
In this paper, we have defined a generalization of entanglement entropy in a generic spacetime slice for free fermion lattice models, as given by \cref{eq:S-Dexpression}. We also discussed calculating this spacetime entanglement entropy for free fermion lattice models with pairings. A future question is to generalize this spacetime entanglement entropy we defined to free boson systems, which should also be in principle calculable from two-point correlation functions due to the Wick's theorem \cite{peschel_calculation_2003}. 

For 1D free lattice fermion tight-binding model, we find the zero temperature time-direction entanglement entropy in a time-direction slice approaches $S_A=\frac{1}{6}\ln \left(t/t_c\right)$ (\cref{eq:S-largeK}) when $\tau< \tau_0$. This indicates that the zero temperature time-direction entanglement entropy $S_A$ of 1D free lattice fermions in the continuous time limit $\tau\rightarrow 0$ resembles the spatial entanglement entropy of a chiral fermion mode, given that the system size $L\gg K=\frac{t}{\tau}+1$. We showed that the coefficient $\frac{1}{6}$ is probing the number of Fermi points in the energy space. A continuous time limit also exists for mutual information $\I$ at finite temperature. Moreover, in the special case $\tau=\tau_0$, we find the time-direction entanglement entropy is given by $S_A=\frac{1}{3}\ln \left(t/t_c\right)$, which is dual to the spatial entanglement entropy. Lastly, we note that when $K\ge L$, the Hilbert space $h_A$ will saturates to $h_\text{tot}$ (unless constrained by symmetries), leading to $S_A=S_\text{tot}$ \cite{suppl}. 

Numerically, larger number of spacetime points $K$ in slice $A$ requires higher precision in the calculation of $S_A$. This can be simplified by defining a $\delta$-cutoff of Hilbert space $h_A$. We approximate the $K\times K$ matrix $B$ in \cref{eq:anti-comm-matrix} as $B\approx Q_\delta \Lambda_\delta Q_\delta^\dag$, where $\Lambda_\delta$ is a $K_\delta\times K_\delta$ diagonal matrix with diagonals being the $K_\delta$ eigenvalues larger than $\delta$ (thus $K_\delta\le K$), the eigenvectors of which are the columns of $Q_\delta$. We can then define a matrix $M=\Lambda_\delta^{-1/2}Q_\delta^\dag$ and an orthonormal basis of $K_\delta$ operators $d_m$ by \cref{eq:orthonormal}, which spans a sub-Hilbert space $h_A^\delta$. We can calculate the zero temperature $S_A$ and finite temperature $\I$ for $h_A^\delta$, which well approximate those for $h_A$ when the cutoff $\delta$ is small enough \cite{suppl}.

An interesting question is the relation between our spacetime entanglement entropy and the temporal entanglement entropy of influence matrix \cite{abanin_prx,abanin_1,abanin_2,abanin_3,feynman-vernon,Muller-Hermes_2012,Hasting_connecting_ent_2015}, which differ in definitions and scaling behaviors (see \cite{suppl} Sec. II for details). A key difference is as follows. Since $c_{r_m}(t_m),c_{r_m}(t_m)^\dag$ at different spacetime points $(r_m,t_m)$ in slice $A$ do not form an orthogonal fermion basis, the local Hilbert space at different points $(r_m,t_m)$ are not orthogonal to each other. In our treatment here, we linear transform $c_{r_m}(t_m),c_{r_m}(t_m)^\dag$ into an orthogonal fermion basis \cref{eq:orthonormal} (which requires free fermion Hamiltonian), and then define the tensor product of their Hilbert spaces as the sub-Hilbert space $h_A$ of slice $A$, which ensures $h_A$ to be a well-defined subspace of the physical Hilbert space. In temporal entanglement of tensor networks, such an orthogonalization of the local Hilbert spaces at different spacetime points $(r_m,t_m)$ is absent (thus not requiring free Hamiltonian), and the basis of spatial transfer matrix is regarded as the sub-Hilbert space of slice $A$, which is however not a subspace of the physical Hilbert space. This difference in definitions leads to different resulting entanglement entropy behaviors.

The entanglement spectrum \cite{topological_spectrum} and operators in entanglement Hamiltonian \cite{lian_conserved_2022} of the timelike reduced density matrix in our formalism await future studies. The generalization of spacetime entanglement entropy into interacting models is also a challenging question. A simple generalization is to define the spacetime entanglement entropy in the interaction picture in the method formulated in this paper. Namely, for Hamiltonian $H=H_0+H_I$ where $H_0$ is a free fermion Hamiltonian and $H_I$ is interaction, we can consider operators in the interaction picture
\begin{equation}
O^{(I)}_{\br,j}(t)=e^{iH_0t}O_{\br,j}e^{-iH_0t} 
\end{equation}
at coordinates $(\br,t)$ in spacetime slice $A$, and define the sub-Hilbert space $h_A$ as the minimal subspace in which $n$-point functions within slice $A$ are calculable (similar to \cref{eq:n-point}):
\begin{equation}
\langle \prod_{\alpha\in A}O^{(I)}_{\br_{n_\alpha},j_\alpha}(t_{n_\alpha})\rangle =\text{tr}\left[\rho_A\prod_{\alpha\in A}O^{(I)}_{\br_{n_\alpha},j_\alpha}(t_{n_\alpha})\right]\ , 
\end{equation}
where $\rho_A$ is the corresponding reduced density matrix. In this way, identification of the sub-Hilbert space $h_A$ of slice $A$ follows the same procedure as \cref{sec:def} with $H$ replaced by $H_0$. The only difference is that the quantum states are no longer time-independent, but is evolved by the interaction. Accordingly, the calculation of entanglement entropy becomes much more difficult. We leave the study of such interacting systems to the future.

\begin{acknowledgments}
\emph{Acknowledgments}. We thank Ying Zhao, J. Alexander Jacoby and Xiao-Liang Qi for helpful conversations. This work is supported by the Alfred P. Sloan Foundation, the National Science Foundation through Princeton University’s Materials Research Science and Engineering Center DMR-2011750, and the National Science Foundation under award DMR-2141966. Additional support is provided by the Gordon and Betty Moore Foundation through Grant GBMF8685 towards the Princeton theory program.
\end{acknowledgments}

\bibliography{main}

\pagebreak
\widetext
\clearpage
\begin{center}
\textbf{\large Supplemental Material for ``Entanglement Entropy of Free Fermions in Timelike Slices"}
\end{center}
\begin{center}
Bowei Liu, Hao Chen, and Biao Lian \\
%\today
\end{center}
% \tableofcontents
%%%%%%%%%% Prefix a "S" to all equations, figures, tables and reset the counter %%%%%%%%%%
\setcounter{equation}{0}
\setcounter{figure}{0}
\setcounter{table}{0}
\setcounter{page}{1}
\setcounter{section}{0}
\makeatletter
\renewcommand{\theequation}{S\arabic{equation}}
\renewcommand{\thefigure}{S\arabic{figure}}
\renewcommand{\bibnumfmt}[1]{[S#1]}
%\renewcommand{\citenumfont}[1]{S#1}
%%%%%%%%%% Prefix a "S" to all equations, figures, tables and reset the counter %%%%%%%%%%

\section{Fermion Operator Orthonormalization}\label{a1}
We give the details of orthonormalizing the free fermion creation and annihilation operators in an arbitrary spacetime slice $A$ in this section. 

\subsection{The case with fermion number conservation}\label{sec:a1-number}
Consider the full system with a total number of fermion modes $L$. Given $K$ single-particle annihilation operators $c_{\br_1}(t_1), c_{\br_2}(t_2), \cdots c_{\br_K}(t_K)$ in slice $A$, we can express them as 
\begin{equation}
c_{\br_j}(t_j) = e^{iHt_n}c_{\br_j}e^{-iHt_n}= \sum_l a_l \phi_{l, \br_j} e^{-\ri\varepsilon_l t_j}\ ,
\label{eq:c}\end{equation}
where $\phi_{l,\br_j}$ is the $l$-th ($1\le l\le L$) single-particle eigenstate wave function (normalized) of the full system, $a_l$ is the $l$-th single-particle eigenstate annihilation operator, and $\varepsilon_l$ 
are the energy of the $l$-th single-particle eigenstate. In the case that the lattice model has one fermion mode per site and has the translation symmetry (which is the case we study in this paper), the eigenfunctions are simply plane waves:
\begin{equation}
\phi_{l,\br_j} = \frac{1}{\sqrt{L}}e^{i \mathbf{k}_l \cdot \br_j}\ ,
\label{eq:phi}\end{equation}
where $\mathbf{k}_l$ is the quasi-momentum of the $l$-th single-particle eigenstate. Define the $K\times L$ matrix $\phi^A$ as 
\begin{equation}
\phi_{j l}^A = \phi_{l,\br_j} e^{-\ri\varepsilon_l t_j}\ ,\qquad 1\le j\le K,\quad 1\le l\le L\ .
\end{equation}
The anticommutator matrix in slice $A$ is then: 
\begin{equation}\label{B}
B_{m n} = \{c_{\br_m}(t_m), c_{\br_n}^\dagger(t_n) \} = \sum_l \phi_{l,\br_m} \phi_{l,\br_n}^* e^{-\ri\varepsilon_l (t_m - t_n)} = (\phi^A \phi^{A\dagger})_{m n}\ , \quad 
1\le m\le K,\quad 1\le n\le K\ .
\end{equation}
The $\{c_{\br_j}(t_j)\}$ operators no longer serve as an orthonormal fermion operator basis because $B_{m n}$ is not the identity matrix. To find an orthonormal fermion operator basis, we notice that $B$ is Hermitian and positive semi-definite. \textit{Assuming $B$ is full rank} (if not full rank, one can work with the full rank part of $B$, see the section ``Definition of $\delta$ cutoff''), there exists $K \times K$ matrix $M$ such that
\begin{equation}\label{M}
M^\dagger M = B^{-1}\ .
\end{equation}
In general, $M$ is not unique. Here we choose to diagonalize $B$ as $B=Q \Lambda Q^\dag$, where $\Lambda$ is diagonal and $Q$ is unitary, and we then define the matrix $M=\Lambda^{-1/2}Q^\dag$. 
Then we can introduce new operators $d_m=\sum_{n\in A}M_{mn}c_n(t_n)$ which are canonically orthonormalized: 
\begin{equation}
\{d_m, d_n^\dagger\} = \sum_{j, l} M_{m j} M^{\dagger}_{l n} \{c_{\br_j}(t_j), c_{\br_l}^\dagger(t_l) \} =  \sum_{j, l} M_{m j} (M^\dagger M)^{-1}_{j l} M^{\dagger}_{l n} = \delta_{m n}\ .
\end{equation}
By defining
\begin{equation}
c = \begin{bmatrix}
c_{\br_1}(t_1) \\
c_{\br_2}(t_2) \\
\vdots \\
c_{\br_K}(t_K)
\end{bmatrix}\ ,\qquad 
a = \begin{bmatrix}
a_{1} \\
a_{2} \\
\vdots \\
a_{L}
\end{bmatrix}\ ,\qquad
d = \begin{bmatrix}
d_{1} \\
d_{2} \\
\vdots \\
d_{K}
\end{bmatrix}\ ,
\end{equation}
we can rewrite the transformation in matrix form as
\begin{equation}\label{d}
c = \phi^A a\ , 
\quad
d = Mc= M \phi^A a\ .
\end{equation}
The correlation matrices satisfy
\begin{equation}
D_{mn}=\text{tr}(\rho_{\text{tot}}d^\dag_m d_n)=\sum_{j,l} M_{n l} M^{*}_{m j} \operatorname{tr} (\rho_{\text{tot}} c_{\br_j}^\dag(t_j) c_{\br_l}(t_l))
=(M^{*} C M^{T})_{mn}\ , \quad m,n\in A\ .
\end{equation}

For a time-direction slice shown in the main text Fig. 1b, every operator sits at site $0$ but are at different time $t_n$. So we only need to substitute
\begin{equation}
c = \begin{bmatrix}
c_{\br_0}(t_1) \\
c_{\br_0}(t_2) \\
\vdots \\
c_{\br_0}(t_K)
\end{bmatrix},
\quad
\phi_{j l}^A = \phi_{l,0} e^{-\ri\varepsilon_l t_j}\ ,
\end{equation}
into \cref{B,M,d} to define fermion basis $d$.

\subsection{Derivation for the case with pairing}\label{sec:a1-pairing}

Here we show how to diagonalize the non-negative Hermitian matrix $R$ defined in \cref{eq:anti-comm-Nambu} in the Nambu basis $\Phi_A=(c_{\br_1} (t_1),\cdots,c_{\br_K} (t_K),c_{\br_1}^\dag (t_1),\cdots, c_{\br_K}^\dag (t_K))^T$:
\begin{equation}\label{eq:anti-comm-Nambu-S}
\{\Phi_A,\Phi_A^\dag\}=\left(
\begin{array}{cc}
B&V\\
V^* &B^*
\end{array}
\right)=R\ ,
\end{equation}
where $B=B^\dag$ is Hermitian, and $V=V^T$ is symmetric (not necessarily real). We mentioned in the main text that the system has a particle-hole symmetry $P=I_K\otimes \sigma_x$ due to the property of the Nambu basis: 
\begin{equation}
P\Phi_A=(\Phi_A^\dag)^T\ ,\qquad PRP^{-1}=R^*\ .
\end{equation}

We now perform transformation to find a simplified basis. First, as we noted in the previous subsection, there exists a matrix $M$ satisfying $M^\dag M=B^{-1}$, which indicates $M BM^\dag=I$. So we have
\begin{equation}
S_1 R S_1^\dag=\left(
\begin{array}{cc}
M&0\\
0 &M^*
\end{array}
\right)R
\left(
\begin{array}{cc}
M^\dag&0\\
0 &M^T
\end{array}
\right)=
\left(
\begin{array}{cc}
I&MVM^T\\
M^*V^*M^\dag & I
\end{array}
\right)\ .
\end{equation}
Note that $MVM^T$ is still a symmetric matrix. Therefore, according to the Autonne-Takagi factorization, there exists a unitary matrix $U_V$ such that
\begin{equation}
U_V(MVM^T)U_V^T=\Lambda_V
\end{equation}
is a real diagonal matrix with non-negative entries. We then define a matrix
\begin{equation}
S_2=\left(
\begin{array}{cc}
U_V&0\\
0& U_V^*
\end{array}
\right)\ ,
%S=\left(
%\begin{array}{cc}
%aU_V&b^*U_V^*\\
%bU_V& a^*U_V^*
%\end{array}
%\right)\ ,\qquad \qquad |a|^2+|b|^2=1\ .
\end{equation}
It is easy to see that
\begin{equation}
S_2\left(
\begin{array}{cc}
I&MVM^T\\
M^*V^*M^\dag & I
\end{array}
\right)S_2^\dag=
\left(
\begin{array}{cc}
I&\Lambda_V\\
\Lambda_V & I
\end{array}
\right)\ .
\end{equation}
At this stage, if we define a new Nambu basis
\begin{equation}
\Phi'_A=S_2S_1\Phi_A=\left(
\begin{array}{cc}
U_VM&0\\
0 & U_V^*M^*
\end{array}
\right)\Phi_A\ ,
\end{equation}
it will still satisfy the particle-hole symmetry $P\Phi'_A=(\Phi_A'^\dag)^T$, since $PS_2P^{-1}=S_2^*$, $PS_1P^{-1}=S_1^*$. Therefore, we have
\begin{equation}
\Phi'_A=(f_1,f_2,\cdots, f_K,f_1^\dag,f_2^\dag,\cdots, f_K^\dag)^T\ ,
\end{equation}
where $f_j$ are fermionic linear operators (not yet fermion creation/annihilation operators) satisfying
\begin{equation}
\{f_i,f_j^\dag\}=\delta_{ij}\ ,\qquad 
\{f_i,f_j\}=\delta_{ij}\lambda_j\ ,
\end{equation}
where we have assumed the real diagonal matrix $\Lambda_V$ has elements $\Lambda_V=\text{diag}(\lambda_1,\lambda_2,\cdots,\lambda_K)$, where $\lambda_j\ge0$. Moreover, since
\begin{equation}
2-2\lambda_j=2\{f_j,f_j^\dag\}-\{f_j,f_j\}-\{f_j^\dag,f_j^\dag\}=2A_j^\dag A_j\ge 0\ ,\qquad \text{with }A_j=f_j-f_j^\dag\ ,
\end{equation}
%{\color{red}I don't understand the last inequality $2(f_j-f_j^\dag)^\dag (f_j-f_j^\dag)\ge 0$. Does it hold for any linear combinations of creation and annihilation operators? } {\color{cyan}BL: This is more general than that. Any operator of the form $A^\dag A$ is non-negative. I made it more clear above.}
(we have used the fact that any operator $A^\dag A$ is non-negative) we know
\begin{equation}
0\le \lambda_j\le 1\ ,\qquad \rightarrow \qquad 0\le \Lambda_V\le 1\ .
\end{equation}
We can then define fermion opeartors $d_j$, $d_j^\dag$ as follows:
\begin{equation}
f_j=\sqrt{\frac{1+\sqrt{1-\lambda_j^2}}{2}}d_j+\sqrt{\frac{1-\sqrt{1-\lambda_j^2}}{2}}d_j^\dag\ ,\qquad f_j^\dag=\sqrt{\frac{1-\sqrt{1-\lambda_j^2}}{2}}d_j+\sqrt{\frac{1+\sqrt{1-\lambda_j^2}}{2}}d_j^\dag\ ,
\end{equation}
or more explicitly after solving the above equation:
\begin{equation}
d_j=\sqrt{\frac{1+\sqrt{1-\lambda_j^2}}{2(1-\lambda_j^2)}}f_j-\sqrt{\frac{1-\sqrt{1-\lambda_j^2}}{2(1-\lambda_j^2)}}f_j^\dag\ ,\qquad d_j^\dag=\sqrt{\frac{1+\sqrt{1-\lambda_j^2}}{2(1-\lambda_j^2)}}f_j^\dag-\sqrt{\frac{1-\sqrt{1-\lambda_j^2}}{2(1-\lambda_j^2)}}f_j\ ,
\end{equation}
which satisfies
\begin{equation}
\{d_i,d_j^\dag\}=\delta_{ij}\ ,\qquad \{d_i,d_j\}=\{d_i^\dag,d_j^\dag\}=0\ ,
\end{equation}
namely, they are the usual orthogonal fermion annihilation and creation operators. More symbolically, we define the following matrix
\begin{equation}
S_3=\left(
\begin{array}{cc}
s_+&-s_-\\
-s_- &s_+
\end{array}\right)\ ,\qquad s_\pm=\sqrt{\frac{1\pm\sqrt{1-\Lambda_V^2}}{2(1-\Lambda_V^2)}}\ .
\end{equation}
We then have
\begin{equation}\label{seq:transR}
S_3S_2S_1RS_1^\dag S_2^\dag S_3^\dag=L RL^\dag=\left(
\begin{array}{cc}
I&0\\
0 &I
\end{array}\right)\ ,\qquad L=S_3S_2S_1\ ,\qquad PLP^{-1}=L^*\ ,
\end{equation}
namely, $R$ can be diagonalized by a particle-hole preserving transformation (note that $L$ is not unitary). We can rewrite the orthonormal Nambu basis \begin{equation}
\Psi_A=(d_1,d_2,\cdots, d_K,d_1^\dag,d_2^\dag,\cdots, d_K^\dag)^T\ ,
\end{equation}
which is related to the original basis via
\begin{equation}
\Psi_A=L\Phi_A\ .
\end{equation}
Note that \cref{seq:transR} implies that the transformation matrix $L$ satisfies
\begin{equation}
L^\dag L=R^{-1}\ , \qquad P L P^{-1}=L^*\ .
\end{equation}

\section{Comparison with the Temporal Entanglement Entropy for Influence Matrix}

Here we comment on the differences between the spacetime slice entanglement entropy we defined in this paper and the temporal entanglement entropy for influence matrix studied in \cite{abanin_prx,abanin_1,abanin_2,abanin_3,feynman-vernon}.

The influence matrix setup considers quantum systems (circuits, etc.) at discrete times $t_n=n\tau$. In the example of fermions, the influence matrix on a fixed site $m=0$ takes the form
\begin{equation}\label{eq:IF}
\mathcal{F}(\{\sigma_{0},\sigma_{\tau},\cdots,\sigma_{n\tau},\cdots;\overline{\sigma}_{0},\overline{\sigma}_{\tau},\cdots,\overline{\sigma}_{n\tau},\cdots\})\ ,\qquad \sigma_{n\tau},\overline{\sigma}_{n\tau}\in\{0,1\}\ ,
\end{equation}
which is understood as the quantum amplitude that the density matrix $\rho_{\text{tot}}(n\tau)=e^{-iH n\tau}\rho_{\text{tot}}e^{iH n\tau}$ at time $t=n\tau$ is locally in the state $|\sigma_{n\tau}\rangle\langle \overline{\sigma}_{n\tau}|_{m=0}$, where $|0\rangle_{m}$ and $|1\rangle_{m}$ are the unoccupied and occupied fermion state on site $m$ at time $n\tau$. Or in the notations of our paper, $c^\dag_m(n\tau)c_m(n\tau)|\sigma_{n\tau}\rangle_{m}=\sigma_{n\tau}|\sigma_{n\tau}\rangle_{m}$ for $\sigma_{n\tau}=0$ or $1$, where $c_m(n\tau)=e^{-iHt}c_m e^{iHt}$ and $c_m^\dag(n\tau)=e^{-iHt}c_m^\dag e^{iHt}$. 
More explicitly, the influence matrix satisfies
\begin{equation}
\prod_{n'\neq n}\sum_{\sigma_{n'\tau}=\overline{\sigma}_{n'\tau}} \mathcal{F}(\{\sigma_{n\tau};\overline{\sigma}_{n\tau}\})=\text{tr}(\rho_{\text{tot}}(n\tau)|\sigma_{n\tau}\rangle\langle \overline{\sigma}_{n\tau}|_{m=0})\ .
\end{equation}
The temporal entanglement entropy for influence matrix $\mathcal{F}(\{\sigma_{n\tau};\overline{\sigma}_{n\tau}\})$ is defined by treating $\mathcal{F}(\{\sigma_{n\tau};\overline{\sigma}_{n\tau}\})$ in \cref{eq:IF} as a ``wavefunction" in a doubled time ``Fock basis":
\begin{equation}\label{eq:IFbasis}
|\sigma_{0},\sigma_{\tau},\cdots,\sigma_{n\tau},\cdots,\overline{\sigma}_{0},\overline{\sigma}_{\tau},\cdots,\overline{\sigma}_{n\tau},\cdots\rangle_{m=0}\ ,
\end{equation}
and calculate the entanglement entropy $S_A^\text{temp}$ of this ``wavefunction" in a time interval $t$ in the usual sense.

This temporal entanglement entropy $S_A^\text{temp}$ above is by definition a different concept from our time direction entanglement entropy $S_A$ in this paper, which we explain below.

1) Since the influence matrix is an operator, $S_A^\text{temp}$ is calculated by mapping it into a ``state" in the operator space, and thus $S_A^\text{temp}$ is a characterization of operator entanglement. In contrast, our time-direction entanglement entropy $S_A$ is calculated for a sub-Hilbert space $h_A$ of the physical quantum state, which is a characterization of state entanglement. For a slice $A$ of time interval $t$ containing $K$ points, $S_A^\text{temp}$ is calculated in an operator sub-Hilbert space of dimension $2^{2K}$, while our $S_A$ is calculated in a state sub-Hilbert space $h_A$ of dimension $2^K$.

2) In the definition of $S_A^\text{temp}$, an effective doubled Fock basis in \cref{eq:IFbasis} is defined. However, two states $|\sigma_{n\tau}\rangle_{m}$ and $|\sigma_{n'\tau}\rangle_{m}$ at different times $n\tau$ and $n'\tau$ ($n\neq n'$) are not orthogonal in the physical Hilbert space, since $\{c_m(n\tau),c_m^\dag(n'\tau)\}\neq 0$ as we explained in the main text. Therefore, the doubled Fock basis, defined in \cref{eq:IFbasis} as direct product of states $|\sigma_{n\tau}\rangle_{m}$ at different times, do not correspond to a physical state in the physical doubled Hilbert space. In contrast, in our definition of time-direction entanglement entropy $S_A$, an orthogonal physical Fock basis is defined by transforming the fermion operators $c_m(n\tau)$ into an orthogonal fermion basis $d_m$ as defined in main text Eq. (3), so $S_A$ does correspond to the entanglement entropy of a physical state in a physical sub-Hilbert space.

3) Given the above differences in the definitions of $S_A^\text{temp}$ and $S_A$, they do show different behaviors numerically. As shown in \cite{abanin_3}, in both non-interacting fermion models and interacting models, the influence matrix temporal entanglement entropy $S_A^\text{temp}$ at infinite temperature (or for high energy states) is sublinear in time $t$, and in many cases shows area law in $t$. In contrast, in our paper, we showed that the time-direction entanglement entropy $S_A$ is linear in time $t$ (namely, volume law) at any finite temperature when $\tau<\tau_0$ (\cref{fig:s_tau_temp}). These two results are thus clearly different. This is because the temporal entanglement entropy $S_A^\text{temp}$ detects the entanglement of operators, which is low (less correlated) at high temperatures; while our time-direction entanglement entropy $S_A$ detects the entanglement of states, which show volume law at high temperatures/energies.

An entropy along the time direction has also been defined for tensor networks by transverse contractions and folding algorithms \cite{Muller-Hermes_2012,Hasting_connecting_ent_2015}, similar to that of the influence matrix. Similarly, it (at least) has the significant distinction from our entanglement entropy that, its basis at different times are implicitly regarded as orthogonal and used to form a direct product Fock basis (similar to the second point above). We do note that for fixed $\tau\gg \tau_0$, the main text Eq. (9) indicates that our zero temperature time-direction entanglement entropy $S_A=K[-\nu\ln \nu - (1-\nu) \ln(1-\nu)]$ is linear in time $t=(K-1)\tau$, which is similar to the linear-in-$t$ temporal entanglement entropy without folding in \cite{Muller-Hermes_2012,Hasting_connecting_ent_2015}. We conjecture this is because, in the limit $\tau\gg \tau_0$, the fermion basis $c_m(n\tau)$ at different times $n\tau$ become approximately orthogonal, and our Fock basis approximately agrees with that used in \cite{Muller-Hermes_2012,Hasting_connecting_ent_2015}.

We leave the understanding of the underlying connection between the influence matrix $S_A^\text{temp}$ and our $S_A$ to future studies.

\section{Alternative Fermion Dispersions}
\begin{figure}[H]
\centering
\subfloat{
\begin{picture}(0.48\columnwidth,0.48\columnwidth)
\put(0,0){\includegraphics[width=0.48\columnwidth]{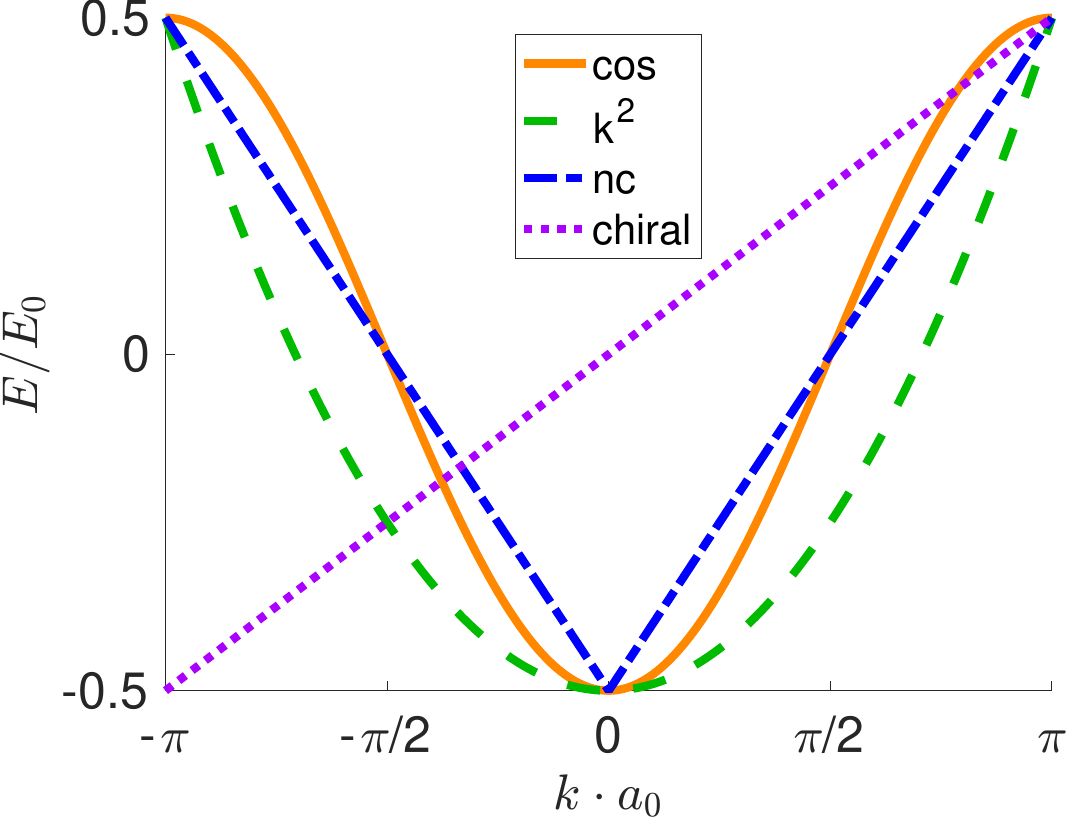}}
\put(-4,190){\footnotesize (a)}
\label{fig:dispersiona}
\end{picture}}
\subfloat{
\begin{picture}(0.48\columnwidth,0.48\columnwidth)
\put(0,0){\includegraphics[width=0.48\columnwidth]{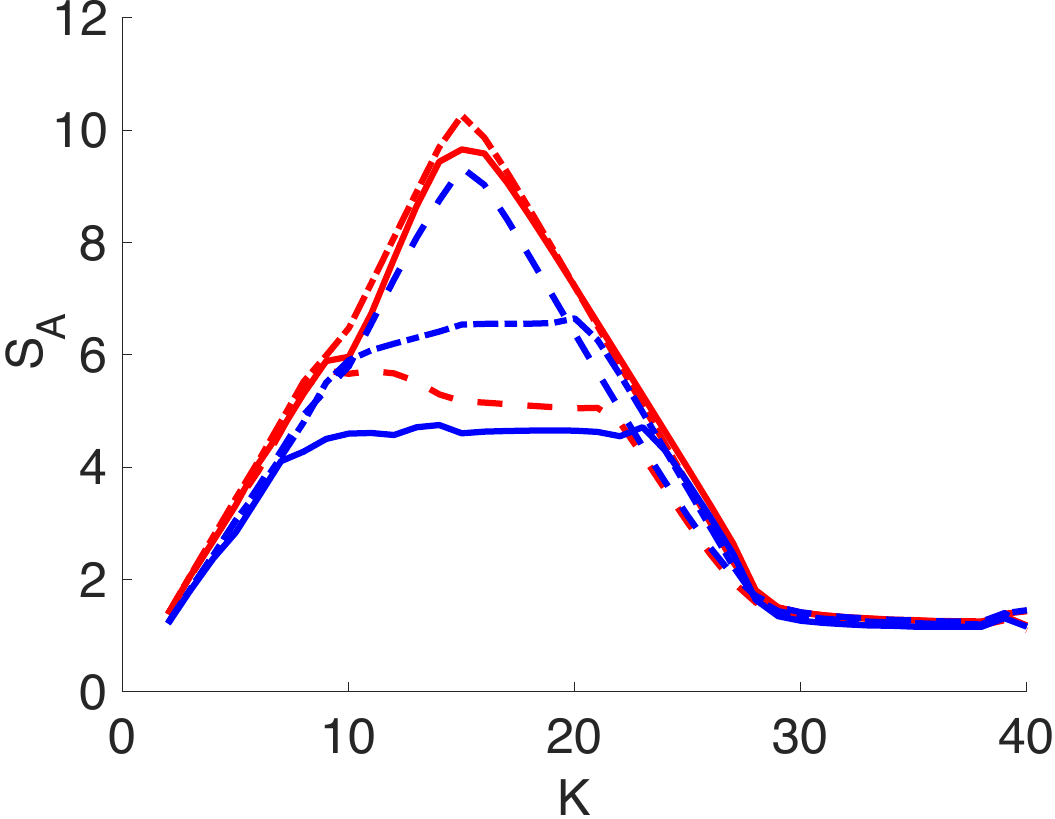}}
\put(-4,190){\footnotesize (b)}
\label{fig:dispersionb}
\end{picture}}
\caption{(a) Cosine dispersion \cref{cos}, quadratic ($k^2$) dispersion \cref{k2}, nonchiral (nc) linear dispersion \cref{nonchiral}, and chiral linear dispersion \cref{chiral}.
(b) Zero temperature $S_A$ in time-direction slice with fixed $2 \pi t / \tau_0=168$ at $\nu=0.5$ (red) and $\nu=0.7$ (blue) for different fermion dispersions in (a): the cosine (solid lines), quadratic (dashed lines), nonchiral
linear and chiral linear (which have equal $S_A$, dashed-dotted lines).}
\label{fig:dispersion}
\end{figure}

We consider the following four dispersion relations of 1D free fermions:
\begin{equation}
\text{cosine dispersion (the tight-binding model in main text):}\quad E(k) = -\frac{E_0}{2} \cos (k a_0) 
\label{cos}\end{equation}
\begin{equation}
\text{quadratic dispersion }(k^2):\quad E(k) = -\frac{E_0}{2} \left(1-\frac{2 (k a_0)^{2}}{\pi^2}\right)
\label{k2}\end{equation}
\begin{equation}
\text{nonchiral linear dispersion:}\quad E(k) = -\frac{E_0}{2} \left(1-\frac{2|k a_0|}{\pi}\right)
\label{nonchiral}\end{equation}
\begin{equation}
\text{chiral linear dispersion:}\quad E(k) = \frac{E_0}{2} \frac{k a_0}{\pi}
\label{chiral}\end{equation}
In particular, the results presented in the main text are all for the cosine dispersion in \cref{cos}, or the model in main text Eq. (7). 
The range of energy spectrum is $E_0$ for all the four dispersion relations, such that they have the same characteristic time $\tau_0=2\pi/E_0$, see \cref{fig:dispersiona}. \cref{fig:dispersionb} shows the fixed $t$ zero-temperature entanglement entropy $S_A$ calculated for these energy dispersions, which are similar except for the trapezoid plateaus of the curves.

\section{Scaling Behavior and Oscillation Period at $\tau\le\tau_0$}
\begin{figure}[H]
\centering
\vspace{-8.5mm}
\subfloat{
\begin{picture}(0.32\columnwidth,0.32\columnwidth)
\put(0,0){\includegraphics[width=0.32\columnwidth]{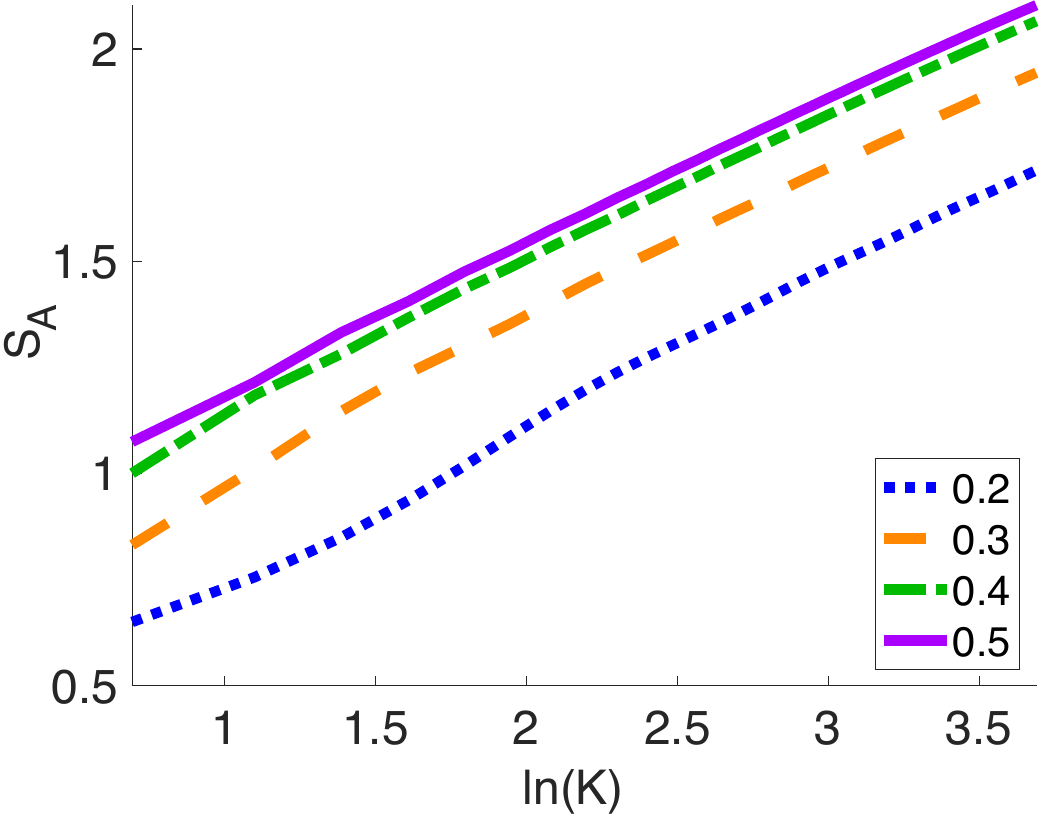}}
\put(-4,122){\footnotesize (a)}
\end{picture}}%
\subfloat{
\begin{picture}(0.32\columnwidth,0.32\columnwidth)
\put(0,0){\includegraphics[width=0.32\columnwidth]{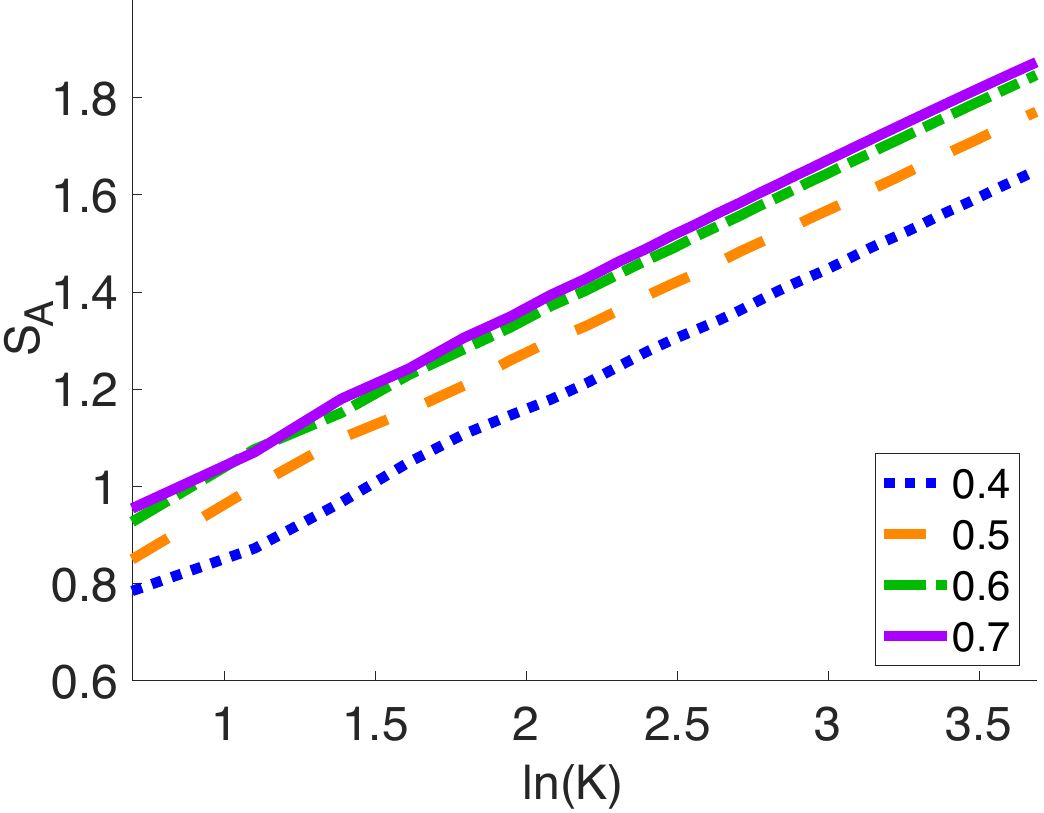}}
\put(-4,122){\footnotesize (b)}
\end{picture}}%
\subfloat{
\begin{picture}(0.32\columnwidth,0.32\columnwidth)
\put(0,0){\includegraphics[width=0.32\columnwidth]{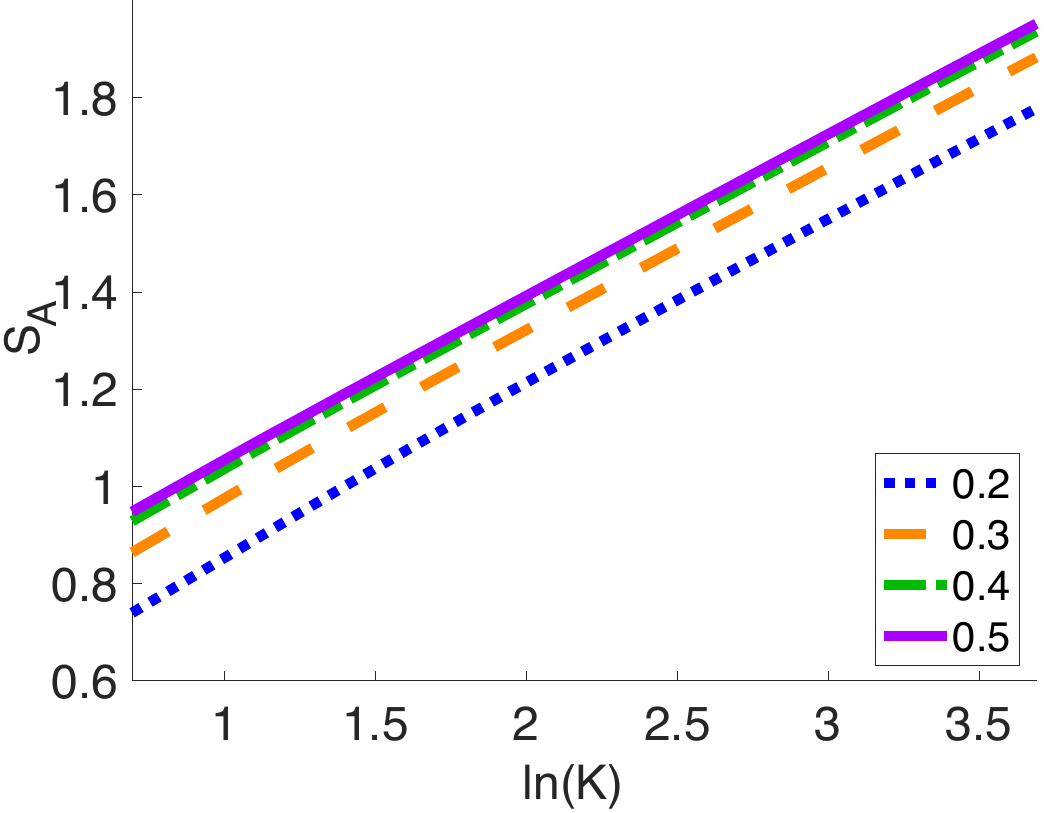}}
\put(-4,122){\footnotesize (c)}
\end{picture}}
\caption{Zero temperature entanglement entropy $S_A$ versus $\ln K$ in time direction slice for different fillings $\nu$ (see legend) at fixed $\tau=\tau_0$ for (a) cosine dispersion \cref{cos}, (b) quadratic dispersion \cref{k2}, and (c) chiral linear dispersion \cref{chiral}.} 
\label{fig:s1}
\end{figure}

\begin{figure}[H]
\centering
\vspace{-8.5mm}
\subfloat{
\begin{picture}(0.32\columnwidth,0.32\columnwidth)
\put(0,0){\includegraphics[width=0.32\columnwidth]{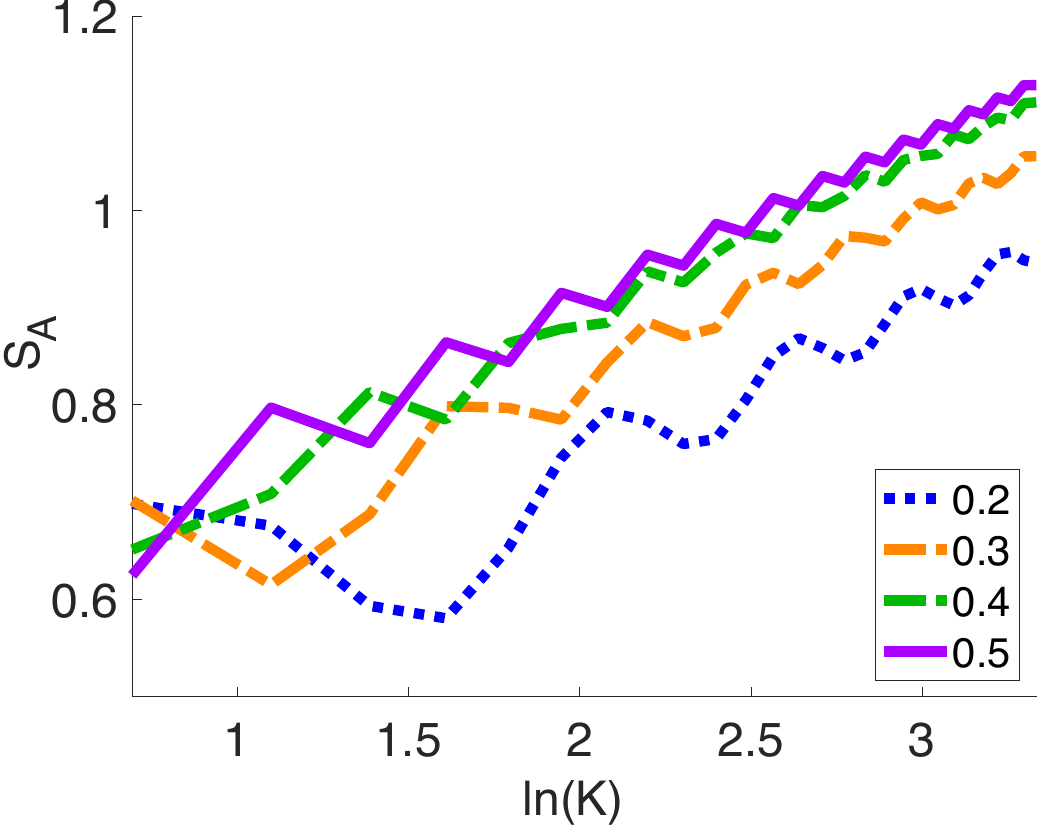}}
\put(-4,122){\footnotesize (a)}
\end{picture}}%
\subfloat{
\begin{picture}(0.32\columnwidth,0.32\columnwidth)
\put(0,0){\includegraphics[width=0.32\columnwidth]{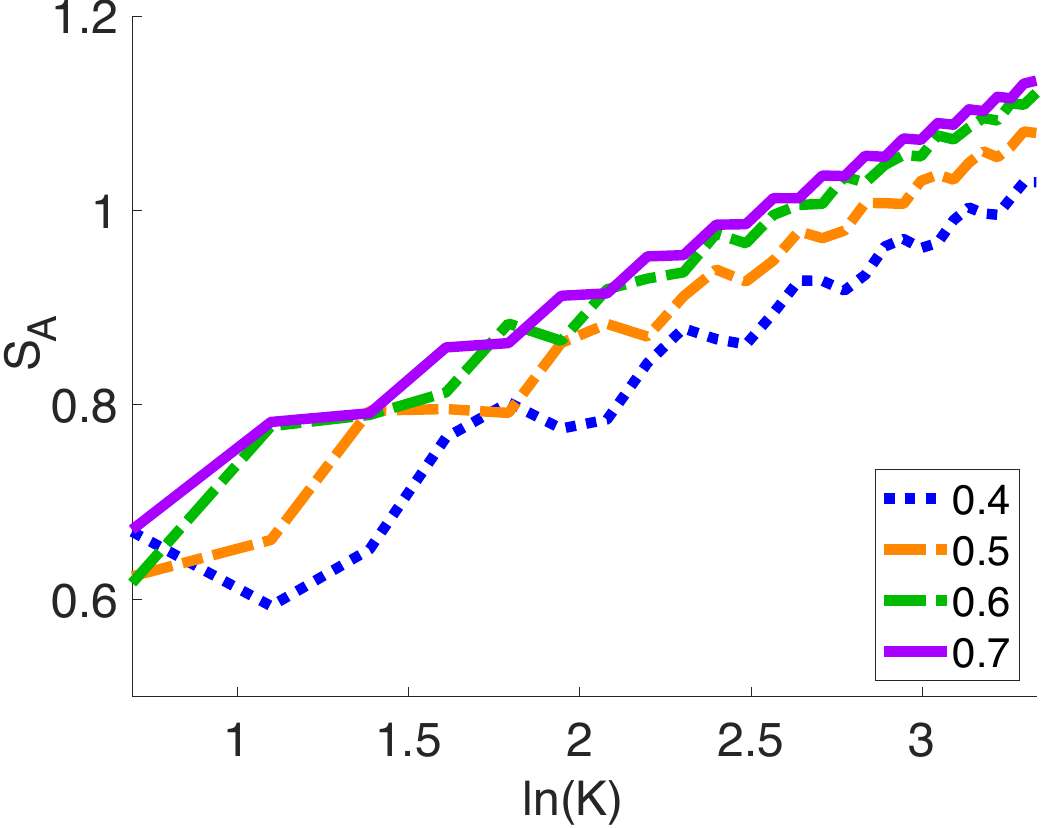}}
\put(-4,122){\footnotesize (b)}
\end{picture}}%
\subfloat{
\begin{picture}(0.32\columnwidth,0.32\columnwidth)
\put(0,0){\includegraphics[width=0.32\columnwidth]{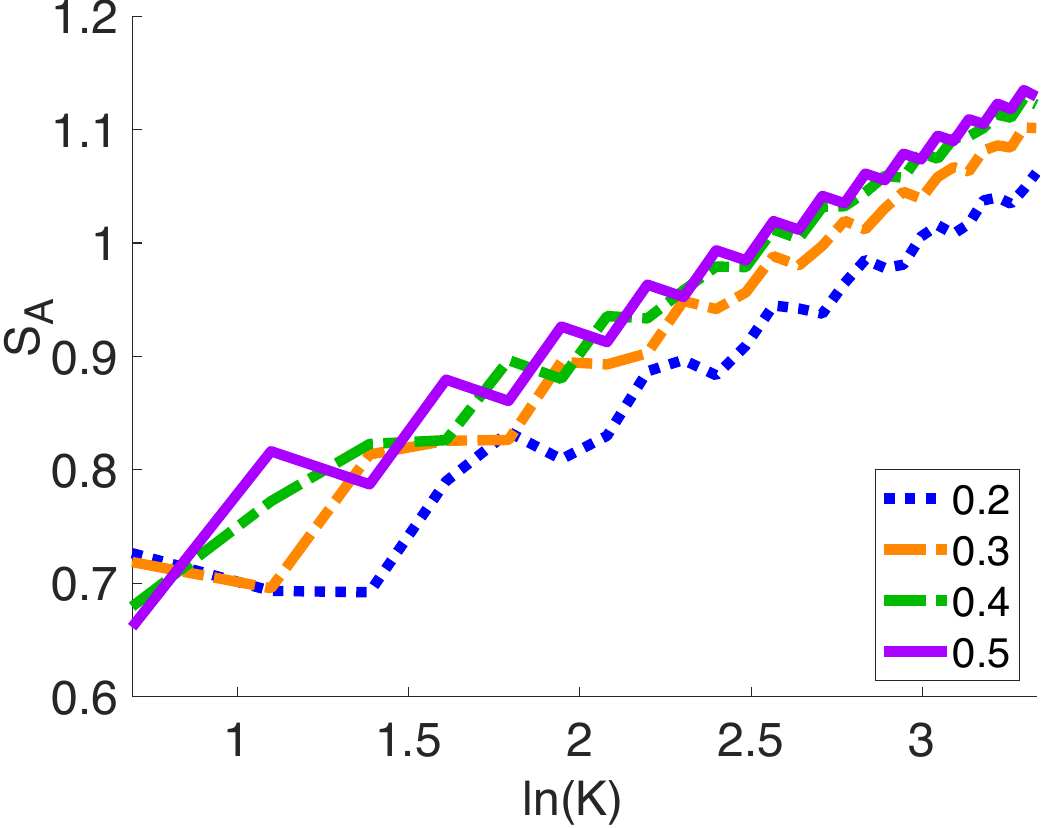}}
\put(-4,122){\footnotesize (c)}
\end{picture}}
\caption{Zero temperature entanglement entropy $S_A$ versus $\ln K$ in time direction slice for different fillings $\nu$ (see legend) at fixed $\tau=\frac{2}{\pi} \tau_0$ for (a) cosine dispersion \cref{cos}, (b) quadratic dispersion \cref{k2}, and (c) chiral linear dispersion \cref{chiral}.} 
\label{fig:s2}
\end{figure}

\begin{figure}[H]
\centering
\vspace{-8.5mm}
\subfloat{
\begin{picture}(0.32\columnwidth,0.32\columnwidth)
\put(0,0){\includegraphics[width=0.32\columnwidth]{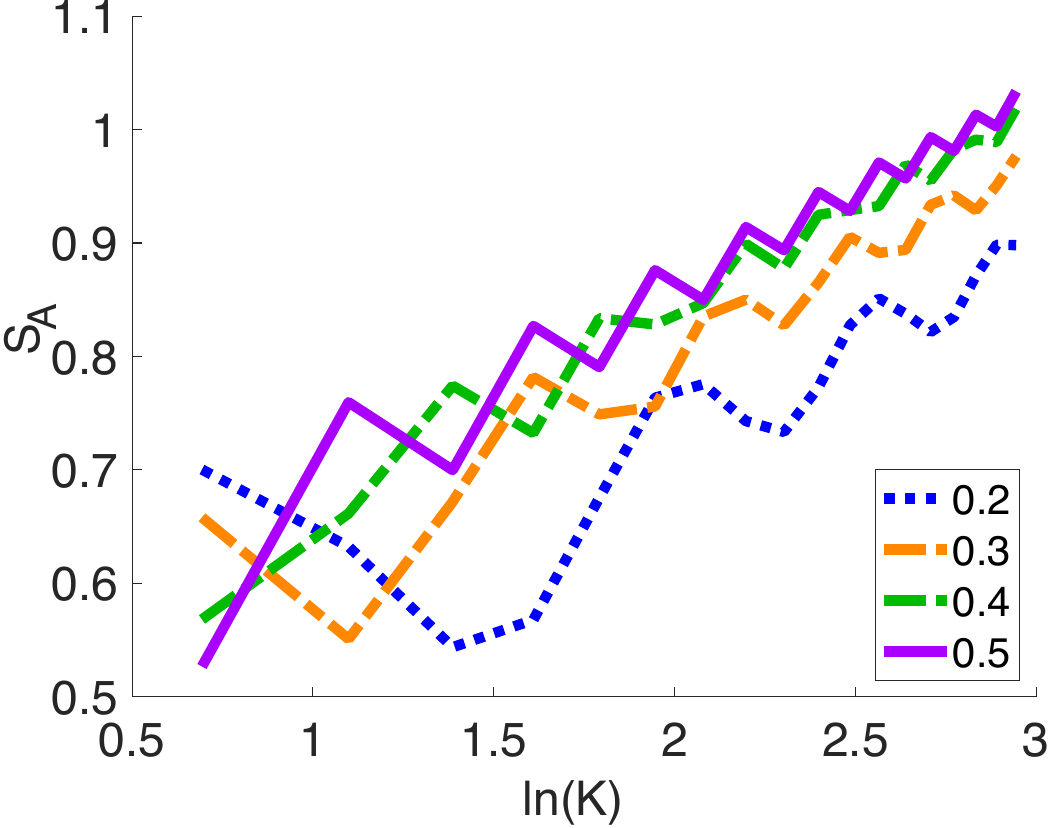}}
\put(-4,122){\footnotesize (a)}
\end{picture}}%
\subfloat{
\begin{picture}(0.32\columnwidth,0.32\columnwidth)
\put(0,0){\includegraphics[width=0.32\columnwidth]{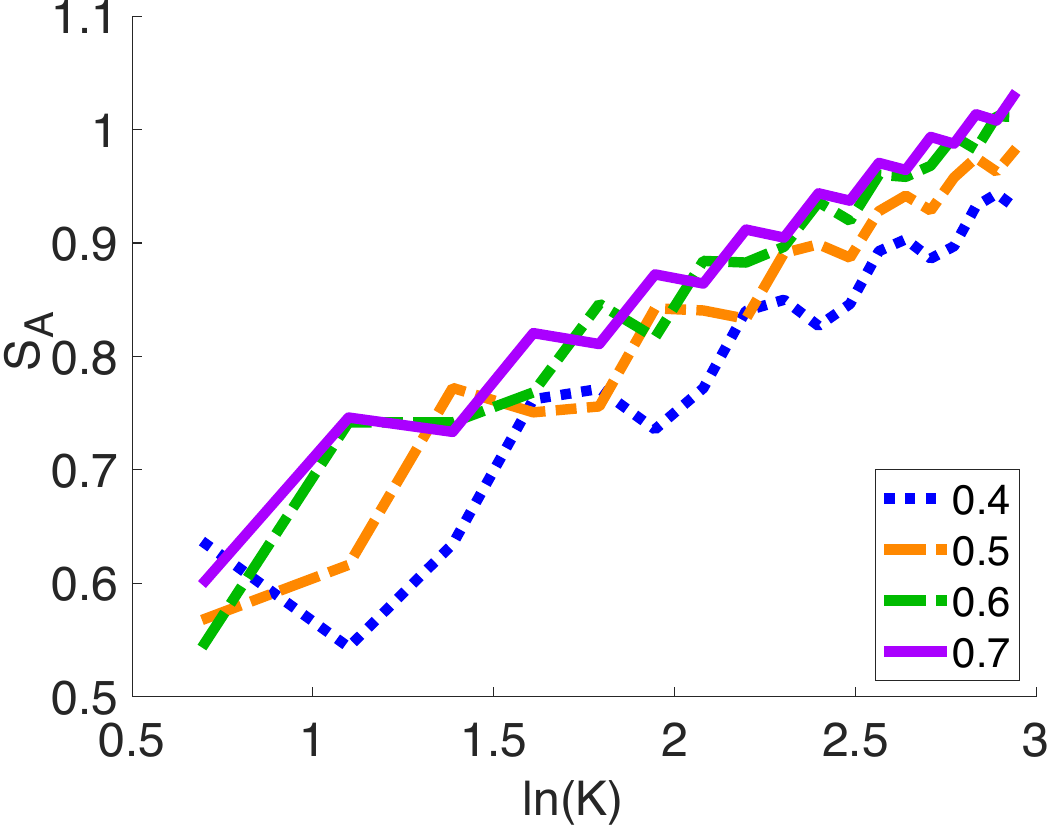}}
\put(-4,122){\footnotesize (b)}
\end{picture}}%
\subfloat{
\begin{picture}(0.32\columnwidth,0.32\columnwidth)
\put(0,0){\includegraphics[width=0.32\columnwidth]{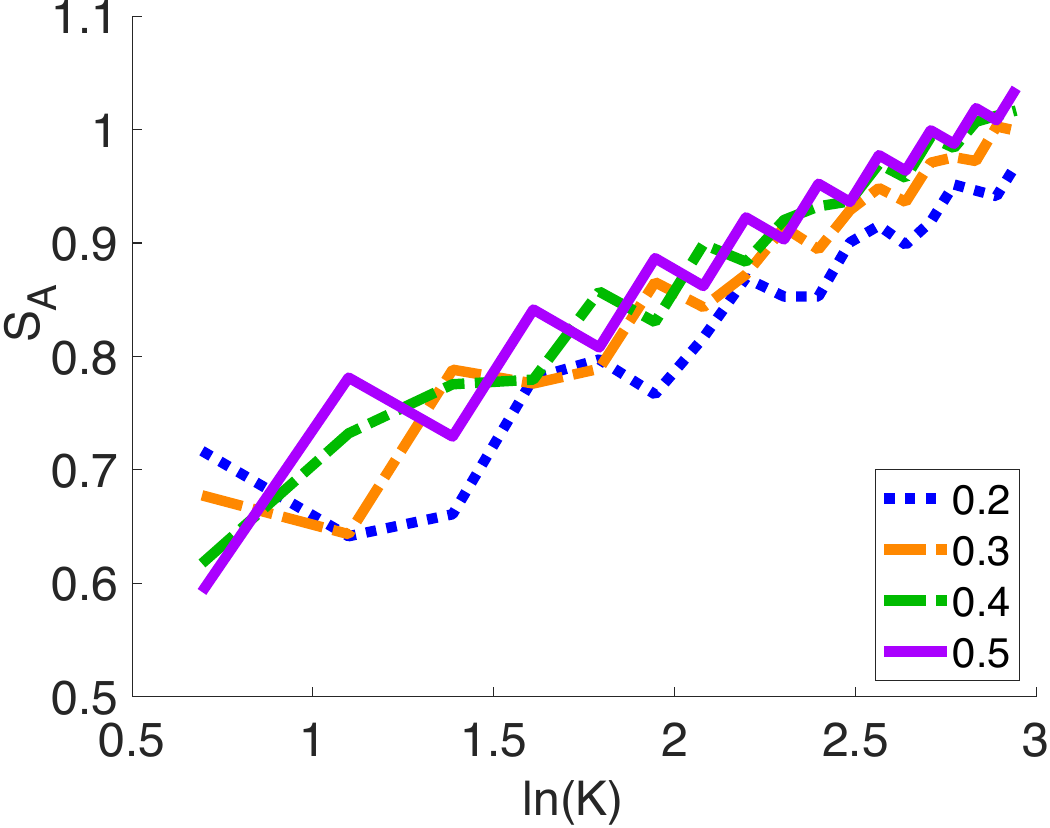}}
\put(-4,122){\footnotesize (c)}
\end{picture}}
\caption{Zero temperature entanglement entropy $S_A$ versus $\ln K$ in time direction slice for different fillings $\nu$ (see legend) at fixed $\tau=\tau_0/2$ for (a) cosine dispersion \cref{cos}, (b) quadratic dispersion \cref{k2}, and (c) chiral linear dispersion \cref{chiral}.} 
\label{fig:s3}
\end{figure}

We now show the scaling behavior and oscillation period of the zero-temperature entanglement entropy $S_A$ in the time-direction slice with fixed time separation $\tau=\tau_0, \frac{2}{\pi}\tau_0$ or $\frac{1}{2}\tau_0$. For $\tau\le\tau_0$, the zero temperature $S_A$ is in general a linear function in $\ln K$ plus an oscillation in $K$. The oscillation amplitude is small when $\tau=\tau_0$ (\cref{fig:s1}), but is large when $\tau\ll\tau_0$ (\cref{fig:s2,fig:s3}). We first fit a linear function $S_{\text{fit}}=a\cdot\ln(K)+b$ to extract $a=\left(\frac{\partial S_A }{\partial\ln(K)}\right)_\tau$. Then we find the oscillation period of $S_A-S_{\text{fit}}$ (using the fitted parameters $a$ and $b$) against $K$. The oscillation is periodic in $K$. To mitigate the effect of oscillation to the fitting, we have excluded the entropy at small $K$, see captions of \cref{tab: s1,tab: s2,tab: s3}. The number of filled single-fermion eigenstates are $101,151,201,251,301,351$ for $\nu=0.2,0.3,0.4,0.5,0.6,0.7$ respectively (total size $L=500$), to avoid breaking the double degeneracy in the cases of cosine, $k^2$, and nonchiral spectrum in \cref{cos,k2,nonchiral}. The nonchiral spectrum in \cref{nonchiral} gives almost the same entanglement entropy as chiral spectrum in \cref{chiral} due to their nearly identical density of states $\Omega(E)$, so we have only plotted the entanglement entropy of the cosine, $k^2$ and chiral spectra in \cref{fig:s1,fig:s2,fig:s3}. 

As shown in \cref{tab: s1,tab: s2,tab: s3}, $\left(\frac{\partial S_A}{\partial\ln(K)}\right)_\tau$ is about $\frac{1}{3}$ at $\tau=\tau_0$, whereas $\left(\frac{\partial S_A}{\partial\ln(K)}\right)_\tau$ is around $\frac{1}{6}$ at $\tau=\frac{2}{\pi}\tau_0$ or $\tau=\frac{1}{2}\tau_0$. This result is independent of filling $\nu$ or dispersion relations. 


\begin{table}[H]
\centering
\begin{tabular}{c|c|c|c|c|c|c}
$\nu$ & $\left(\frac{\partial S_A}{\partial\ln(K)}\right)_\tau$ at $\tau=\tau_0$ & period at $\tau=\tau_0$ &  $\left(\frac{\partial S_A}{\partial\ln(K)}\right)_\tau$ at $\tau=\frac{2}{\pi} \tau_0$ & period at $\tau=\frac{2}{\pi} \tau_0$ & $\left(\frac{\partial S_A}{\partial\ln(K)}\right)_\tau$ at $\tau=\frac{1}{2}\tau_0$ & period at $\tau=\frac{1}{2}\tau_0$ \\
\hline
0.2 & 0.3423 $\pm$ 0.0022 & 9.0 & 0.1606 $\pm$ 0.0310 & 5.7 & 0.1979 $\pm$ 0.0483 & 5.5 \\
0.3 & 0.3317 $\pm$ 0.0017 & 4.5 & 0.1638 $\pm$ 0.0179 & 3.7 & 0.1649 $\pm$ 0.0310 & 3.3 \\
0.4 & 0.3273 $\pm$ 0.0018 & 3.0 & 0.1667 $\pm$ 0.0133 & 2.6 & 0.1843 $\pm$ 0.0247 & 2.4 \\
0.5 & 0.3261 $\pm$ 0.0020 & 2.0 & 0.1697 $\pm$ 0.0158 & 2.0 & 0.1677 $\pm$ 0.0291 & 2.0
\end{tabular}
\caption{In time direction slice with cosine dispersion \cref{cos}, $95\%$ confidence bound of $\left(\frac{\partial S_A}{\partial\ln(K)}\right)_\tau$ is fitted from the zero temperature entanglement entropy $S_A$ at $K \in [11, 40], [12, 28], [5, 19]$ when fixing $\tau=\tau_0, \frac{2}{\pi}\tau_0, \frac{1}{2}\tau_0$, respectively.}
\label{tab: s1}
\end{table}
\begin{table}[H]
\centering
\begin{tabular}{c|c|c|c|c|c|c}
$\nu$ & $\left(\frac{\partial S_A}{\partial\ln(K)}\right)_\tau$ at $\tau=\tau_0$ & period at $\tau=\tau_0$ & $\left(\frac{\partial S_A}{\partial\ln(K)}\right)_\tau$ at $\tau=\frac{2}{\pi} \tau_0$ & period at $\tau=\frac{2}{\pi} \tau_0$ & $\left(\frac{\partial S_A}{\partial\ln(K)}\right)_\tau$ at $\tau=\frac{1}{2}\tau_0$ & period at $\tau=\frac{1}{2}\tau_0$ \\
\hline
0.2 & 0.3411 $\pm$ 0.0024 & $\infty$ & 0.1511 $\pm$ 0.0136 & 3.6 & 0.1438 $\pm$ 0.0275 & 3.7 \\
0.3 & 0.3362 $\pm$ 0.0015 & $\infty$ & 0.1594 $\pm$ 0.0094 & 2.9 & 0.1594 $\pm$ 0.0194 & 2.8 \\
0.4 & 0.3340 $\pm$ 0.0009 & $\infty$ & 0.1644 $\pm$ 0.0059 & 2.3 & 0.1685 $\pm$ 0.0131 & 2.3 \\
0.5 & 0.3335 $\pm$ 0.0007 & $\infty$ & 0.1660 $\pm$ 0.0102 & 2.0 & 0.1724 $\pm$ 0.0232 & 2.0
\end{tabular}
\caption{In time direction slice with chiral linear dispersion \cref{chiral}, $95\%$ confidence bound of $\left(\frac{\partial S_A}{\partial\ln(K)}\right)_\tau$ is fitted from the zero temperature entanglement entropy $S_A$ at $K \in [2, 40], [2, 28], [2, 19]$ when fixing $\tau=\tau_0, \frac{2}{\pi}\tau_0, \frac{1}{2}\tau_0$, respectively.}
\label{tab: s2}
\end{table}
\begin{table}[H]
\centering
\begin{tabular}{c|c|c|c|c|c|c}
$\nu$ & $\left(\frac{\partial S_A}{\partial\ln(K)}\right)_\tau$ at $\tau=\tau_0$ & period at $\tau=\tau_0$ & $\left(\frac{\partial S_A}{\partial\ln(K)}\right)_\tau$ at $\tau=\frac{2}{\pi} \tau_0$ & period at $\tau=\frac{2}{\pi} \tau_0$ & $\left(\frac{\partial S_A}{\partial\ln(K)}\right)_\tau$ at $\tau=\frac{1}{2}\tau_0$ & period at $\tau=\frac{1}{2}\tau_0$ \\
\hline
0.4 & 0.2926 $\pm$ 0.0021 & 5.8 & 0.1638 $\pm$ 0.0156 & 4.2 & 0.1909 $\pm$ 0.0294 & 4.3 \\
0.5 & 0.2992 $\pm$ 0.0025 & 3.8 & 0.1720 $\pm$ 0.0091 & 3.3 & 0.1751 $\pm$ 0.0242 & 3.3 \\
0.6 & 0.3015 $\pm$ 0.0019 & 2.7 & 0.1728 $\pm$ 0.0092 & 2.6 & 0.1618 $\pm$ 0.0171 & 2.5 \\
0.7 & 0.3054 $\pm$ 0.0022 & 2.0 & 0.1684 $\pm$ 0.0055 & 2.0 & 0.1651 $\pm$ 0.0160 & 2.0
\end{tabular}
\caption{In time direction slice with quadratic dispersion relation \cref{k2}, $95\%$ confidence bound of $\left(\frac{\partial S_A}{\partial\ln(K)}\right)_\tau$ is fitted from the zero temperature entanglement entropy $S_A$ at $K \in [2, 40], [2, 28], [3, 19]$ when fixing $\tau=\tau_0, \frac{2}{\pi}\tau_0, \frac{1}{2}\tau_0$, respectively.}
\label{tab: s3}
\end{table}

\section{Entanglement Entropy in the Plateau}
\begin{figure}[H]
\centering
\vspace{-8.5mm}
\subfloat{
\begin{picture}(0.32\columnwidth,0.32\columnwidth)
\put(0,0){\includegraphics[width=0.32\columnwidth]{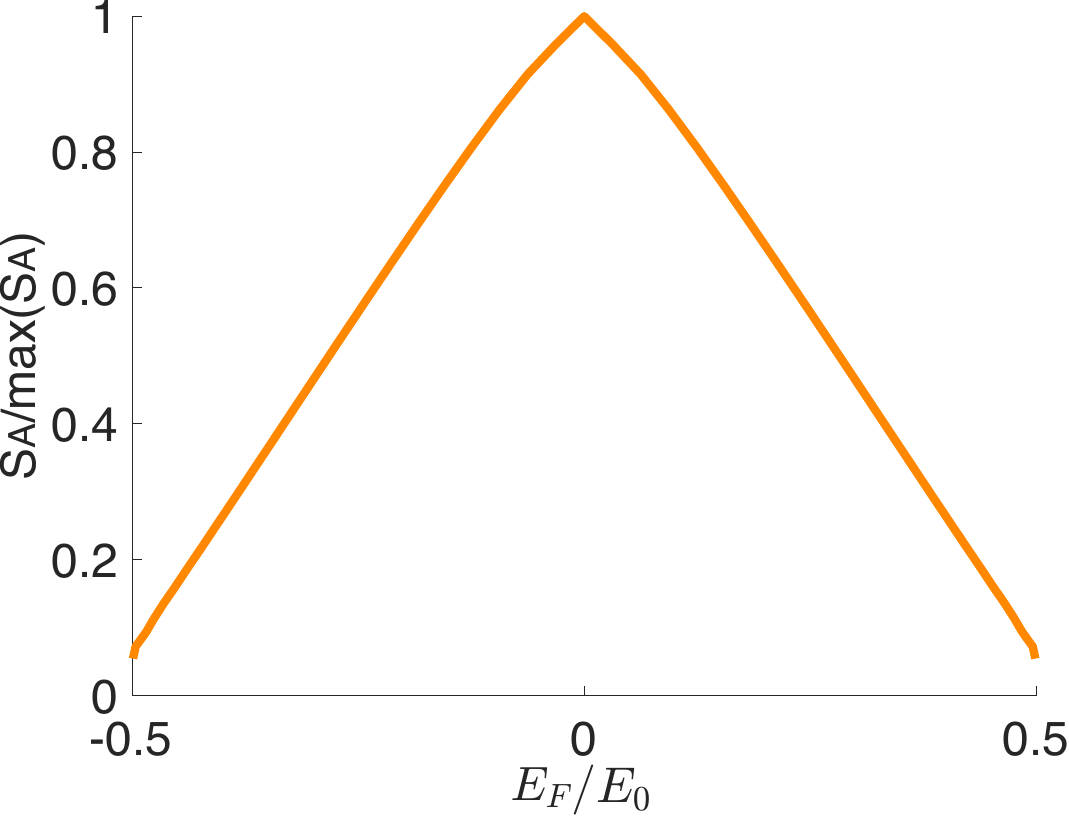}}
\put(-4,122){\footnotesize (a)}
\end{picture}}%
\subfloat{
\begin{picture}(0.32\columnwidth,0.32\columnwidth)
\put(0,0){\includegraphics[width=0.32\columnwidth]{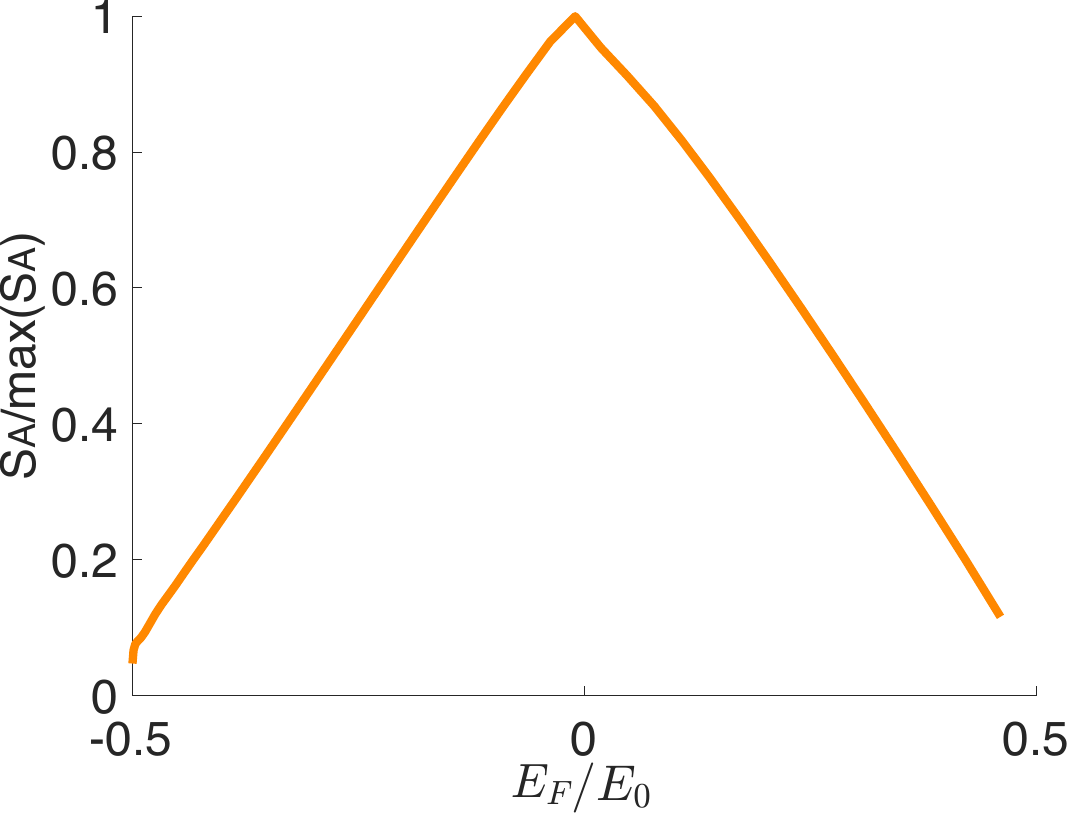}}
\put(-4,122){\footnotesize (b)}
\end{picture}}%
\subfloat{
\begin{picture}(0.32\columnwidth,0.32\columnwidth)
\put(0,0){\includegraphics[width=0.32\columnwidth]{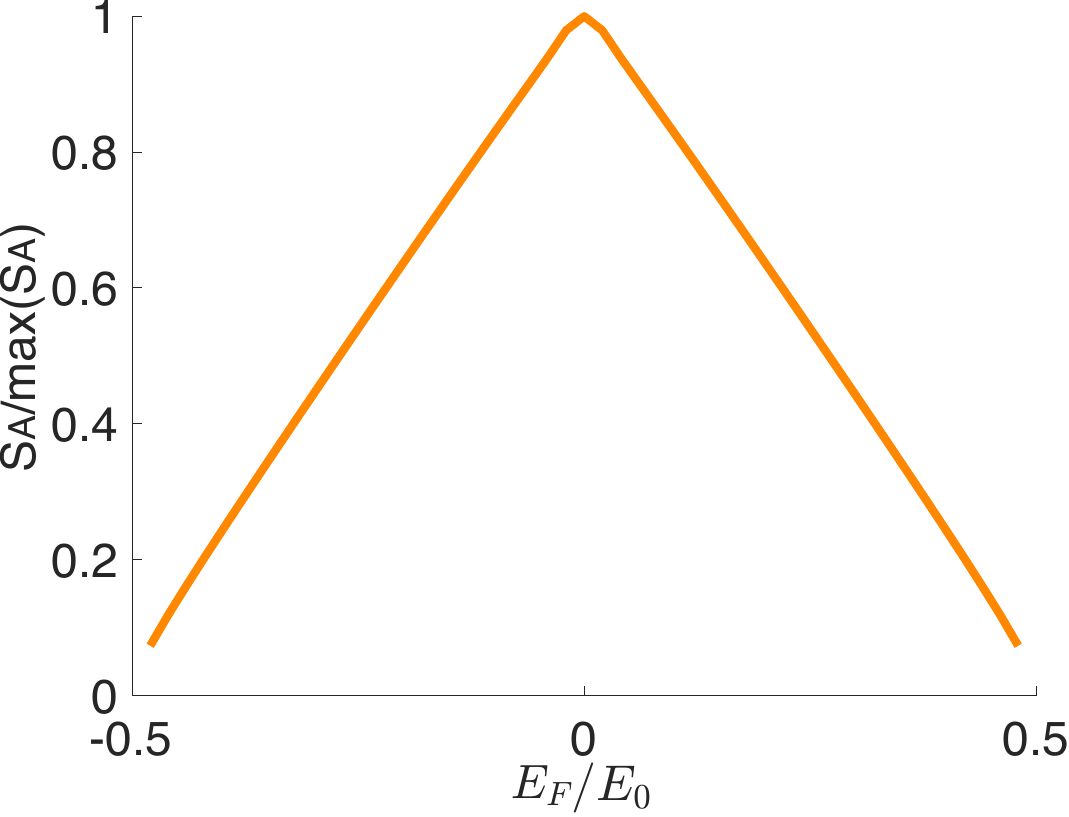}}
\put(-4,122){\footnotesize (c)}
\end{picture}}
\caption{At fixed $2\pi t/\tau_0 = 168$ and $K=15$ ($\tau\approx2\tau_0$) in a time direction slice, zero temperature entanglement entropy $S_A $ is at the tip of the triangle when $E_F=0$ or middle of the plateau when $E_F\neq0$. This plateau entanglement entropy $S_A $ normalized by the tip of the triangle $\text{max}(S_A )$ (entropy $S_A $ when $E_F=0$) is given for (a) cosine dispersion \cref{cos}, (b) quadratic dispersion \cref{k2}, and (c) chiral linear dispersion \cref{chiral} (which is the same as nonchiral linear dispersion \cref{nonchiral}), respectively. Note that $E_F=0$ corresponds to filling $\nu=0.5$ for cosine and chiral/nonchiral linear dispersions, and $\nu\approx 0.7$ for quadratic dispersion. }
\label{fig:s5}
\end{figure}

In the main text Fig. 2a, we have shown the time-direction entanglement entropy shows a trapezoid shape with a plateau or triangular shape when $K<t/\tau_0$, depending on filling $\nu$.  \cref{fig:s5} here shows the maximal entanglement entropy $S_A $ (occurring in the trapezoid plateau or the triangle top vertex) in the range of $K<t/\tau_0$ in a fixed $t$ zero temperature time-direction slice (normalized by the maximal possible $S_A $), with respect to the Fermi energy $E_F\in[-\frac{E_0}{2},\frac{E_0}{2}]$. The curve of $S_A $ forms a triangle with respect to Fermi energy $E_F$, which is almost independent of dispersion relation, suggesting the plateau entropy only depends on $E_F$ (instead of filling $\nu$).

\section{$t$ Dependence at Finite Temperature}
\begin{figure}[H]
\centering
\vspace{-8.5mm}
\subfloat{
\begin{picture}(0.32\columnwidth,0.32\columnwidth)
\put(0,0){\includegraphics[width=0.32\columnwidth]{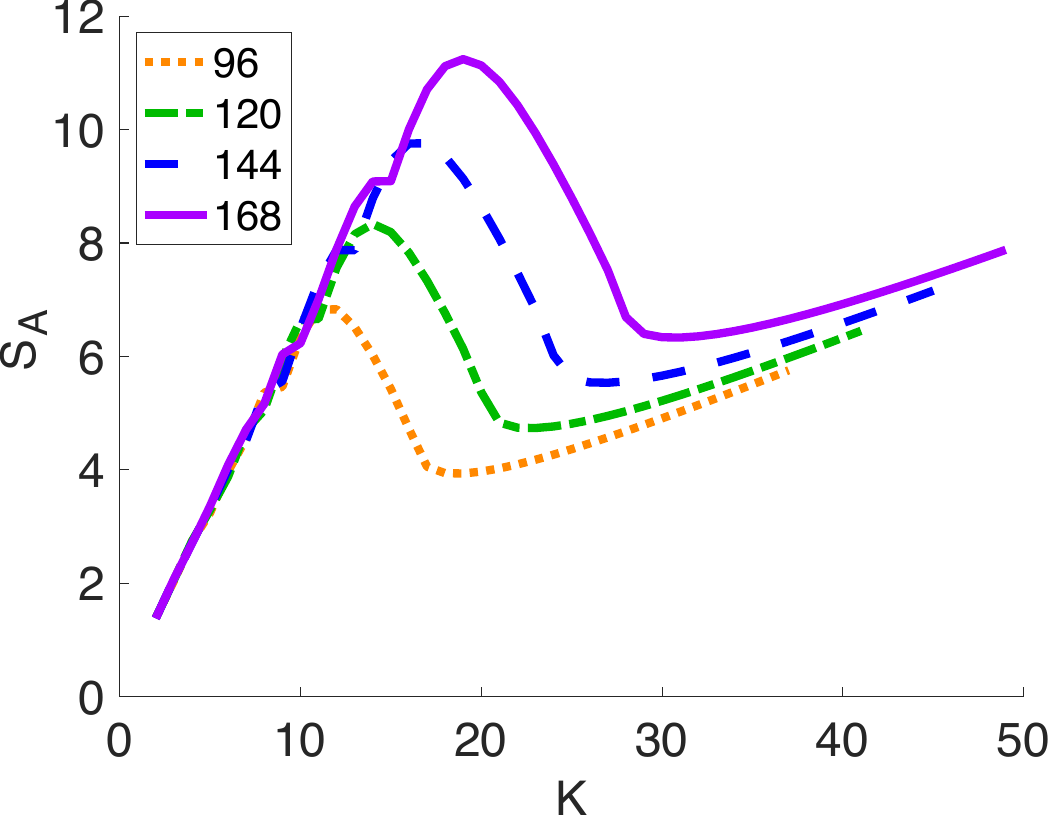}}
\put(-4,122){\footnotesize (a)}
\end{picture}}%
\subfloat{
\begin{picture}(0.32\columnwidth,0.32\columnwidth)
\put(0,0){\includegraphics[width=0.32\columnwidth]{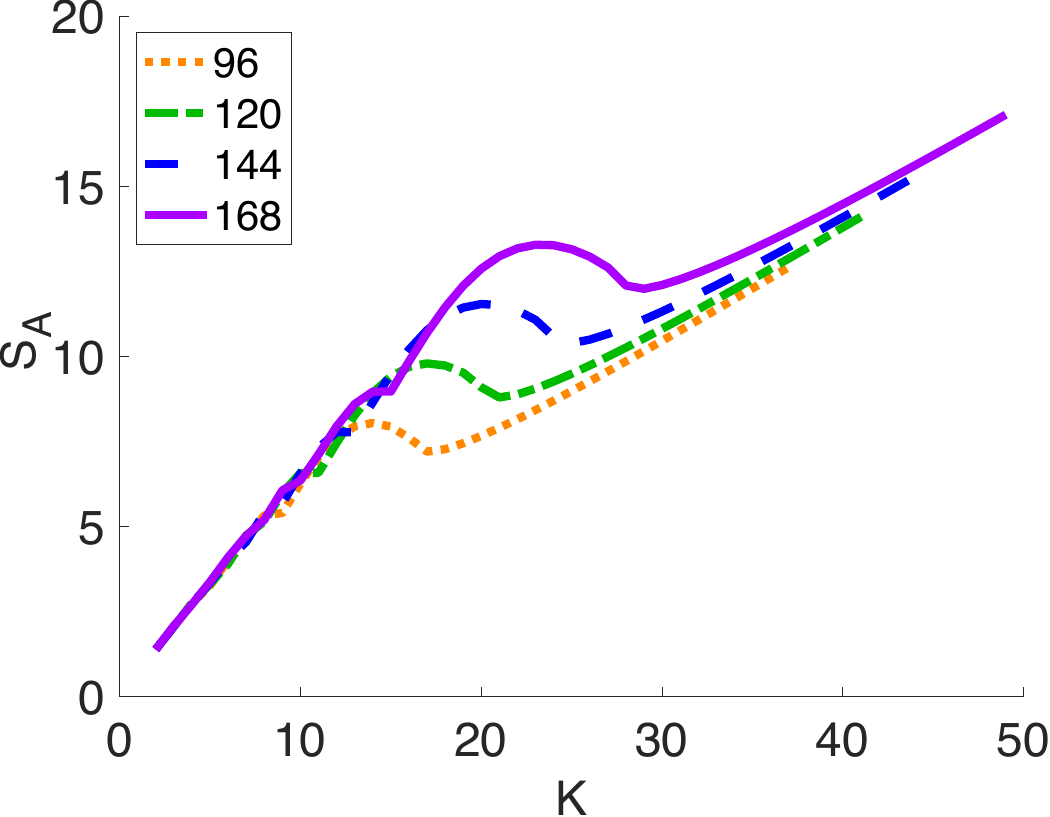}}
\put(-4,122){\footnotesize (b)}
\end{picture}}%
\subfloat{
\begin{picture}(0.32\columnwidth,0.32\columnwidth)
\put(0,0){\includegraphics[width=0.32\columnwidth]{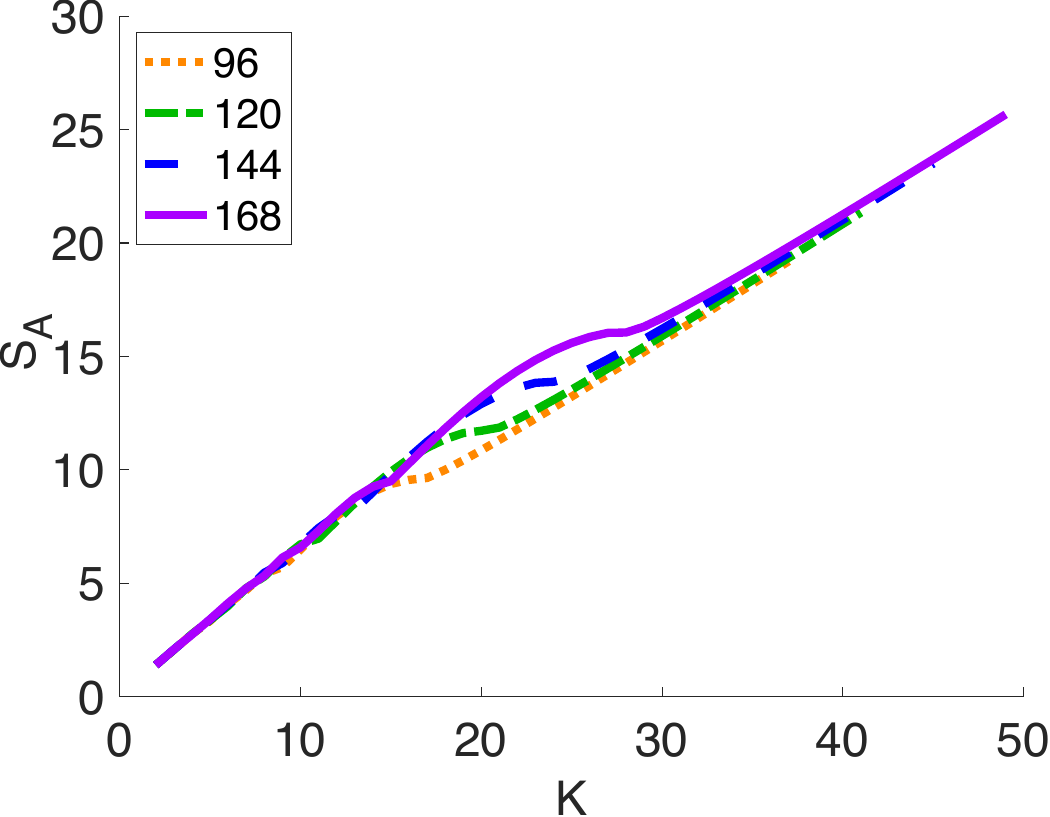}}
\put(-4,122){\footnotesize (c)}
\end{picture}}
\caption{Time-direction slice entanglement entropy $S_A$ of model \cref{cos} at fixed filling $\nu=0.5$ and finite temperature $T/E_0 = 0.0625, 0.1375, 0.25$ is shown in (a), (b), (c), respectively. In each panel, different values of $2\pi t/\tau_0$ is given in the legend. Each curve is kept up to a reliable value of $K$, since numerical accuracy becomes insufficient for larger $K$. }
\label{fig:s6}
\end{figure}

\cref{fig:s6} shows that in a time direction slice, $\left(\frac{\partial S_A }{\partial K}\right)_t$ at $K\gg t/\tau_0$ ($\tau\ll\tau_0$) with fixed $t$ is almost independent of $t$, regardless of temperature $T$.

\begin{figure}[H]
\centering
\vspace{-8.5mm}
\subfloat{
\begin{picture}(0.32\columnwidth,0.32\columnwidth)
\put(0,0){\includegraphics[width=0.32\columnwidth]{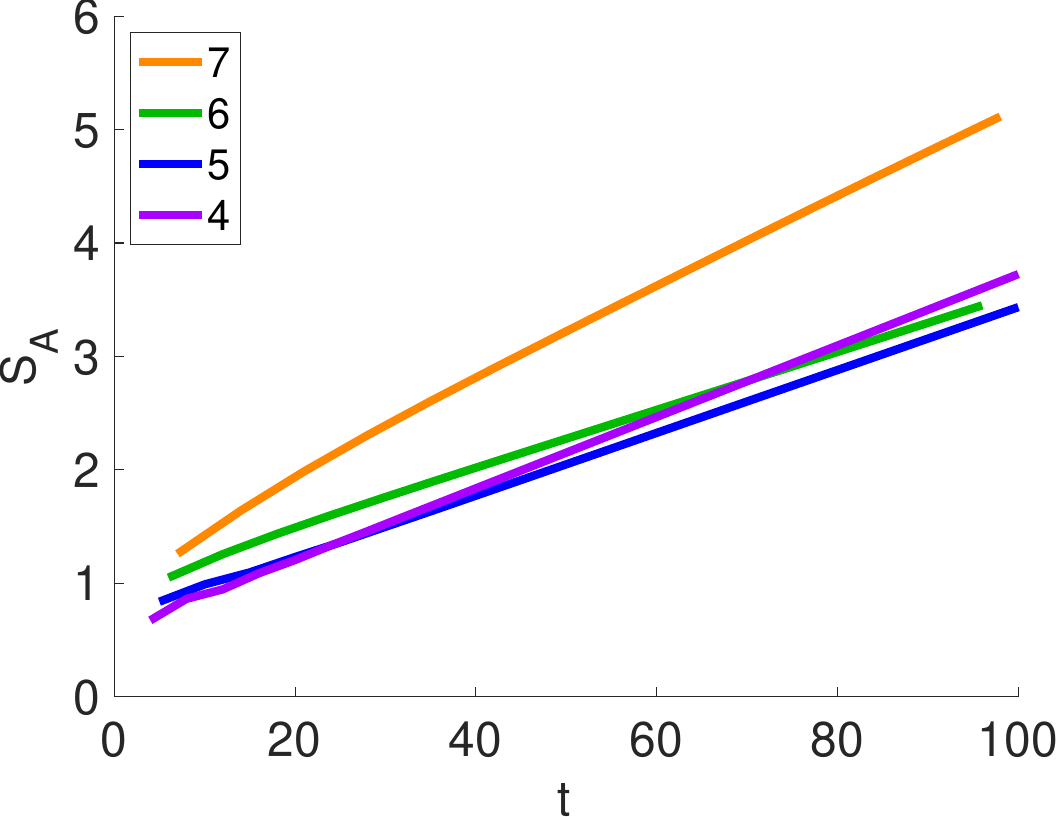}}
\put(-4,122){\footnotesize (a)}
\end{picture}}%
\subfloat{
\begin{picture}(0.32\columnwidth,0.32\columnwidth)
\put(0,0){\includegraphics[width=0.32\columnwidth]{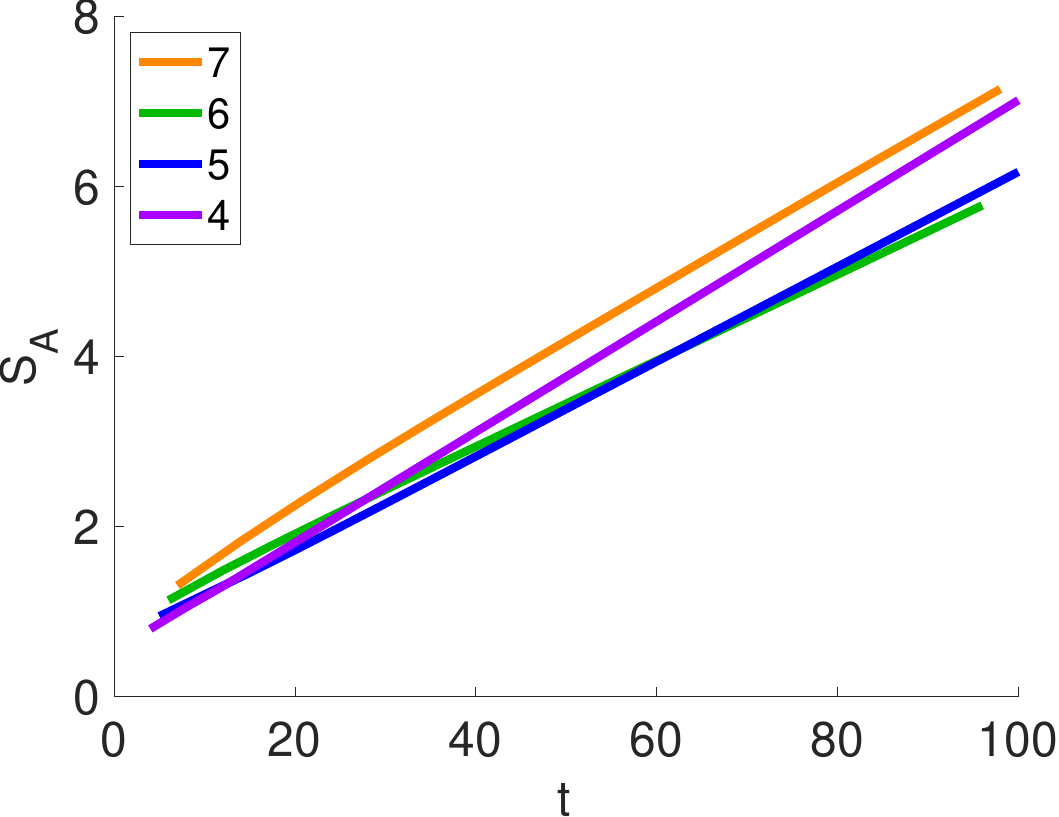}}
\put(-4,122){\footnotesize (b)}
\end{picture}}%
\subfloat{
\begin{picture}(0.32\columnwidth,0.32\columnwidth)
\put(0,0){\includegraphics[width=0.32\columnwidth]{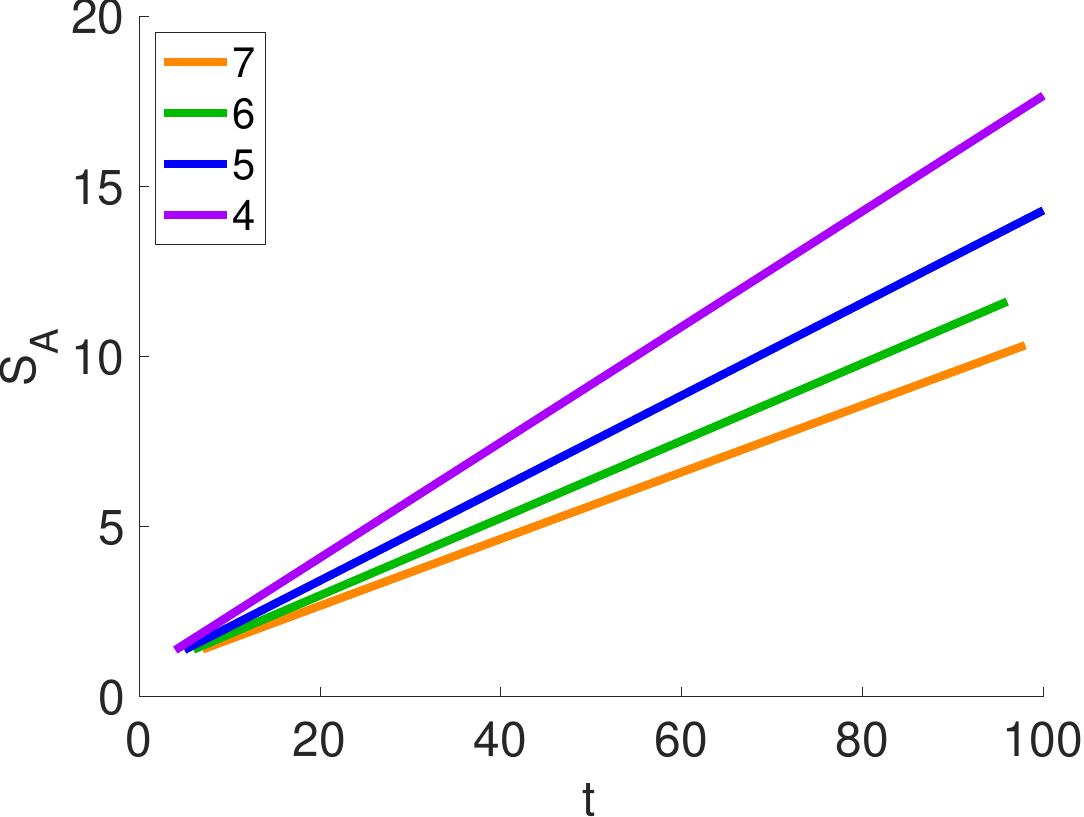}}
\put(-4,122){\footnotesize (c)}
\end{picture}}
\caption{Time-direction slice entanglement entropy $S_A$ of model \cref{cos} with fixed time spacing $\tau$ at fixed filling $\nu=0.5$, $\delta=10^{-18}$, and finite temperature $T/E_0 = 0.05, 0.1, 1$ is shown in (a), (b), (c), respectively. In each panel, different values of $2\pi \tau/\tau_0$ is given in the legend. }
\label{fig:s_tau_temp}
\end{figure}

When we fix time separation $\tau<\tau_0$, we get a volume law $S_A \propto t$ at finite temperature, see \cref{fig:s_tau_temp}.

\section{Entanglement Entropy in Linear Slice with Large $\theta$}

\begin{figure}[H]
\centering
\vspace{-8.5mm}
\subfloat{
\begin{picture}(0.32\columnwidth,0.32\columnwidth)
\put(0,0){\includegraphics[width=0.32\columnwidth]{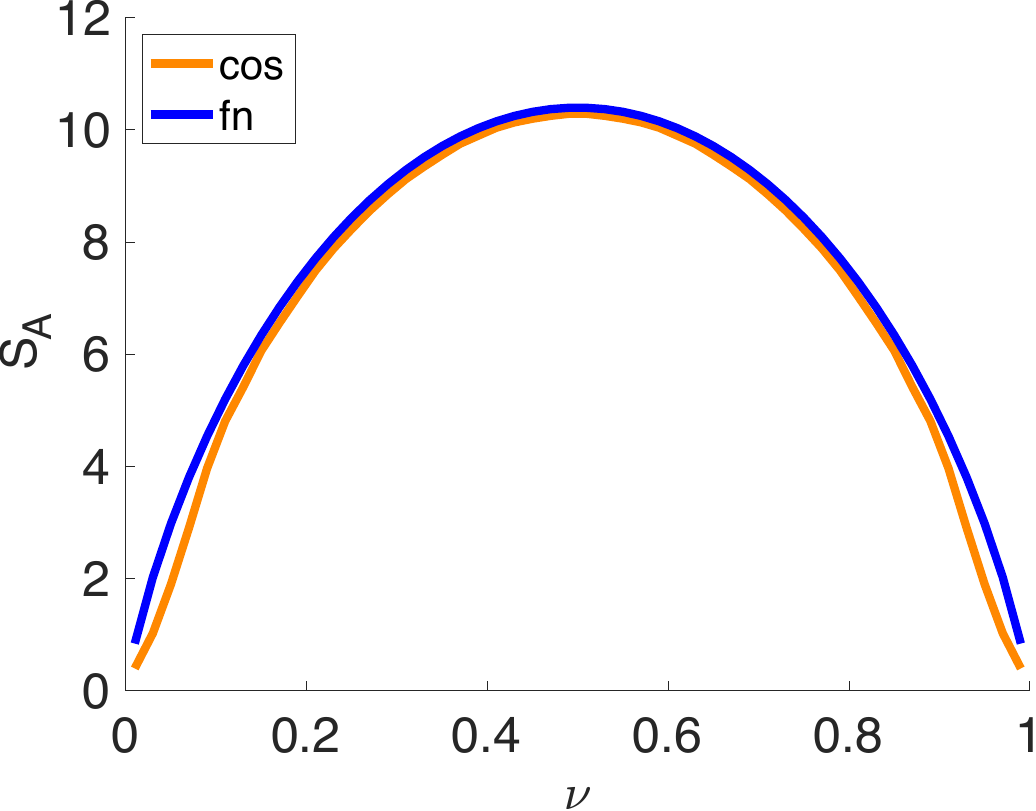}}
\put(-4,122){\footnotesize (a)}
\end{picture}}%
\subfloat{
\begin{picture}(0.32\columnwidth,0.32\columnwidth)
\put(0,0){\includegraphics[width=0.32\columnwidth]{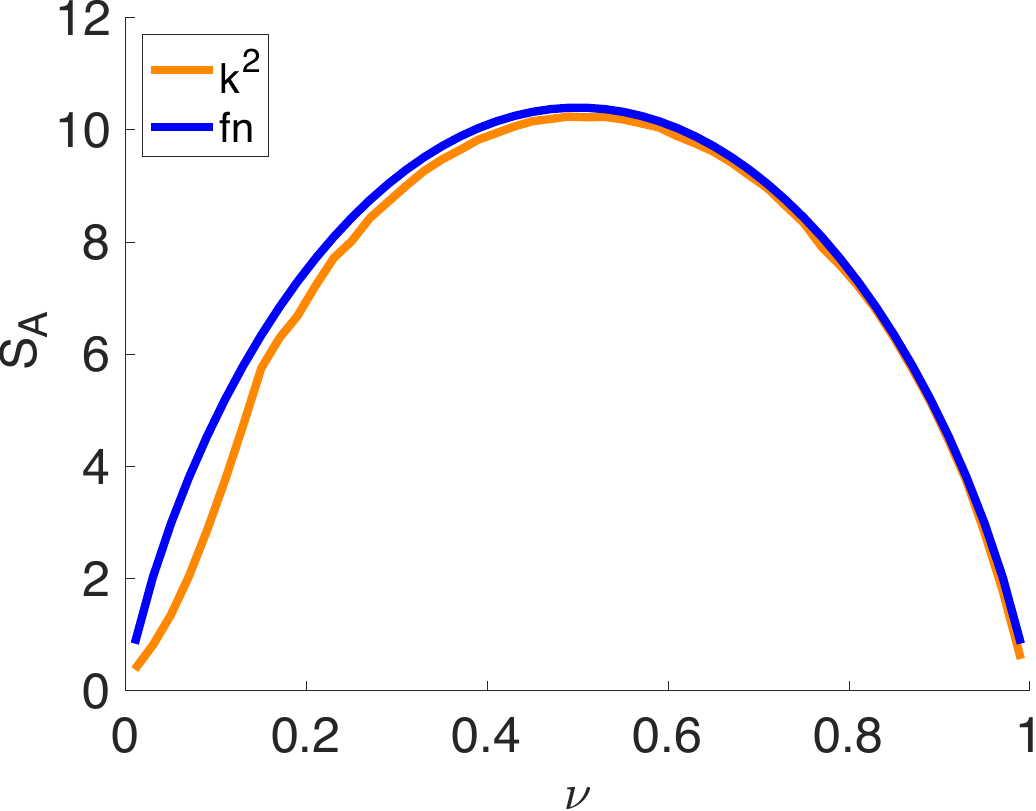}}
\put(-4,122){\footnotesize (b)}
\end{picture}}%
\subfloat{
\begin{picture}(0.32\columnwidth,0.32\columnwidth)
\put(0,0){\includegraphics[width=0.32\columnwidth]{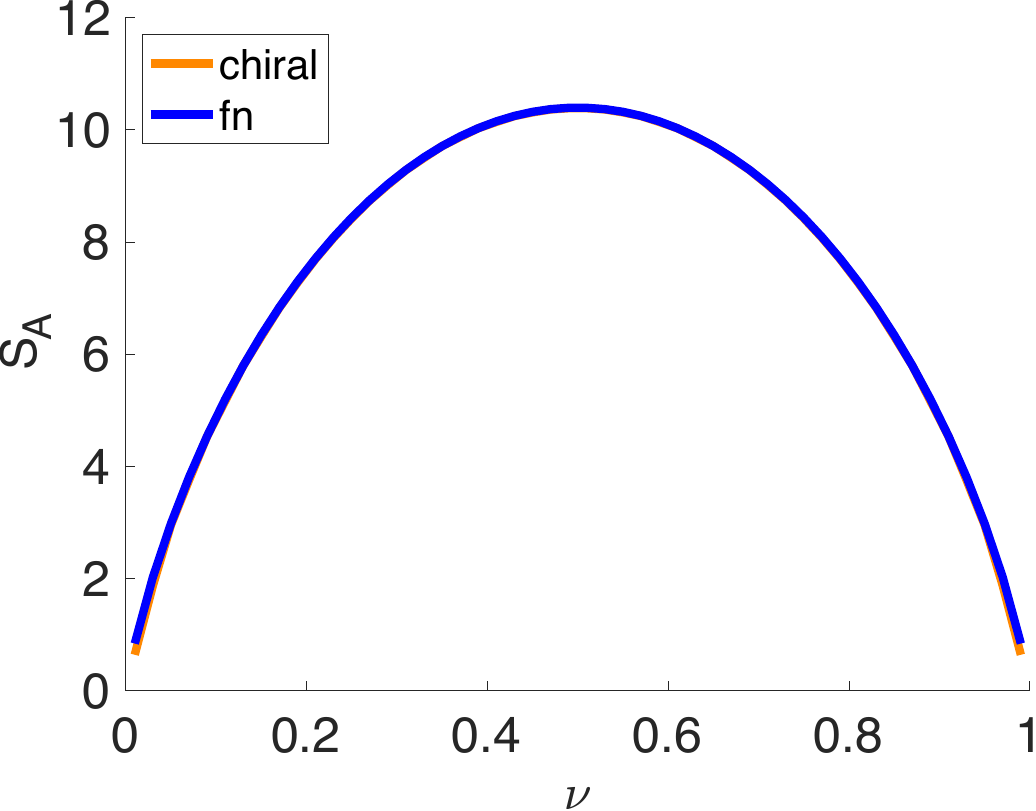}}
\put(-4,122){\footnotesize (c)}
\end{picture}}
\caption{Zero temperature entanglement entropy $S_A$ versus filling $\nu$ in a linear slice with $\ell = 15$ and (a) ${t_n E_0}/n = a_0 E_0 \tan(\theta) / v_{\text{max}} = 240$ for cosine dispersion \cref{cos}, (b) $a_0 E_0 \tan(\theta) / v_{\text{max}} = 240$ for quadratic dispersion \cref{k2}, (c) $a_0 E_0 \tan(\theta) / v_{\text{max}} = 150$ for chiral linear dispersion \cref{chiral}, where $v_{\text{max}}$ is the maximum Fermi velocity of the corresponding dispersion relation. ``fn'' stands for the analytic formula $S_A = \ell (-\nu\ln(\nu)-(1-\nu)\ln(1-\nu))$. }
\label{fig:s4}
\end{figure}

\cref{fig:s4} shows that the zero temperature entanglement entropy $S_A \approx \ell (-\nu\ln(\nu)-(1-\nu)\ln(1-\nu))$ when $\theta\rightarrow\frac{\pi}{2}$ in a linear slice. This is almost independent of dispersion relation.

\section{Quantum Mutual Information}
When the total Hilbert space decomposes as $h_{\text{tot}}=h_A\otimes h_{A^c}$, the quantum mutual information is 
\begin{equation}
\mathcal{I} = S_A + S_{A^c} - S_{\text{tot}}
\end{equation}
where $S_A\left(S_{A^c}\right)$ is the von Neumann entropy of the reduced density matrix $\rho_A\left(\rho_{A^c}\right)$, and $S_{\text{tot}}$ is the von Neumann entropy of the state $\rho_{\text{tot}}$.

In order to calculate $S_{A^c}$, we want to find a orthonormal basis $c_B$ of single fermion annihilation operators that generate $h_{A^c}$. For the subsystem $A$ we have (\cref{d})
\begin{equation}
c=\phi^A a, \quad
\phi_{j l}^A=\phi_{l, \mathbf{r}_j} e^{-\mathrm{i} \varepsilon_l t_j}, \quad 1 \leq j \leq K, \quad 1 \leq l \leq L
\end{equation}
If $\phi^A$ has rank $r$, suppose
\begin{equation}
c_B=\phi^B a \ ,
\end{equation}
where $\phi^B$ is a complex matrix with dimension $(L-r) \times L$. Because $a$ is the set of single-particle eigenstate annihilation operators, the \textit{rows} of the matrices $\phi^B$ and $\phi^A$ together has rank $L$, the total system size. The rows of $\phi^B$ can be found by solving for a column vector $x$ of dimension $L$ such that
\begin{equation}
\phi^A x^* = 0
\end{equation}
Numerically, this can be achieved by a QR decomposition of $\phi_A^T$:
\begin{equation}
\phi_A^T = Q R
=\begin{bmatrix}
Q_1 & Q_2
\end{bmatrix}
\begin{bmatrix}
R_1 \\
0
\end{bmatrix}, \quad
\phi_A = R^T Q^T = R_1^T Q_1^T
\end{equation}
where $Q$ is a $L \times L$ unitary matrix, $R$ is a $L\times K$ upper triangular matrix, and $R_1$ is a $r \times K$ upper triangular matrix. $Q_1, Q_2$ have dimensions $L\times r$, $L\times (L-r)$ respectively.
By the unitarity of $Q$, 
\begin{equation}
Q^\dag Q = I \implies Q_1^\dag Q_2 = 0
\end{equation}
Define
\begin{equation}
\phi_B = Q_2^T \implies
\phi_A \phi_B^\dag 
= R_1^T Q_1^T Q_2^*
= 0
\end{equation}
so $c_B$ is orthogonal to $c$. Then we can calculate $S_{A^c}$ by replacing $\phi_A$ by $\phi_B$ in \cref{d} and the rest of the calculation follows exactly like the calculation of $S_A$.

\section{Definition of $\delta$ Cutoff}
To improve numerical precision, we introduce a $\delta$ cutoff to the Hilbert space $h_A$.
Suppose the spectrum decomposition of $B$ is
\begin{equation}
B = Q \Lambda Q^\dag
\end{equation}
where $\Lambda$ is a positive semi-definite diagonal matrix and its diagonal elements are monotonically decreasing: $\Lambda_{nn} \geq \Lambda_{n+1,n+1}$. Because we need to invert $\Lambda$ to calculate the normalization matrix $M$, numerical errors on small eigenvalues of $\Lambda$ will result in large errors on $M$. To mitigate these errors, we can introduce a cutoff $\delta \geq 0$ to take only the eigenvalues that are bigger than $\delta$. Suppose $\Lambda_{K_\delta, K_\delta}$ is the last eigenvalue that is not smaller than $\delta$
\begin{equation}
\Lambda_{K_\delta, K_\delta} \geq \delta > \Lambda_{K_\delta+1, K_\delta+1}
\end{equation}
we can define a $K_\delta\times K_\delta$ matrix
\begin{equation}
\Lambda_\delta \equiv \Lambda_{1:K_\delta, 1:K_\delta}
\label{eq: delta}
\end{equation}
where $1:n$ stands for indices $\{1, 2, \cdots n\}$ and
\begin{equation}
B \approx Q_\delta \Lambda_\delta Q_\delta^\dag, \quad
Q_\delta \equiv Q_{1:K, 1:K_\delta}
\end{equation}
Therefore
\begin{equation}
M = \Lambda_\delta^{-\frac{1}{2}} Q_\delta^\dag
\end{equation}
and everything else follows as in \cref{d}. In practice, $B$ is never singular unless the number of points $K$ is on the order of $L$. In this paper, $\delta$ is set to $0$ unless otherwise mentioned.

\section{Total System Size $L$ Dependence}
\begin{figure}[H]
\centering
\subfloat{
\begin{picture}(0.48\columnwidth,0.48\columnwidth)
\put(0,0){\includegraphics[width=0.48\columnwidth]{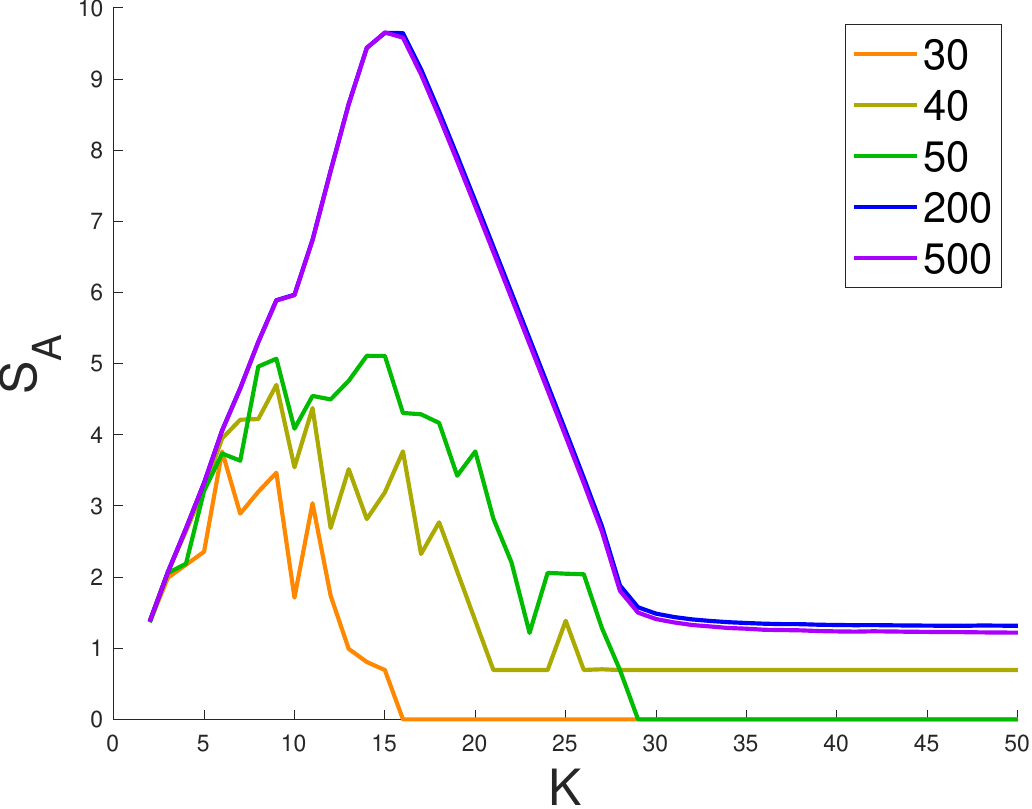}}
\put(-4,190){\footnotesize (a)}
\end{picture}}
\subfloat{
\begin{picture}(0.48\columnwidth,0.48\columnwidth)
\put(0,0){\includegraphics[width=0.48\columnwidth]{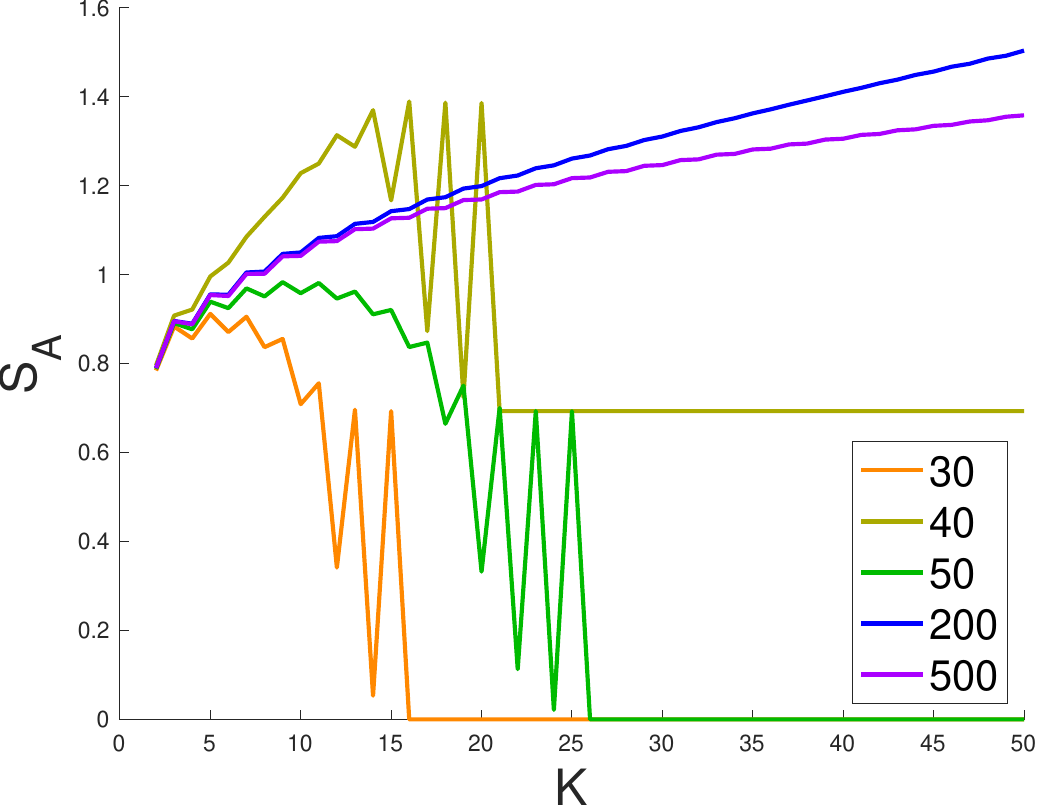}}
\put(-4,190){\footnotesize (b)}
\end{picture}}
\caption{Zero temperature entanglement entropy $S_A$ of model \cref{cos} in the time-direction slice with $K$ points for different values of total system size $L$ given in the legend with filling $\nu=\frac{1}{2}$ and $\delta=10^{-13}$ for (a) fixed $2 \pi t / \tau_0=168$ and (b) fixing $2 \pi \tau / \tau_0=5$. 
}
\label{fig:l1}
\end{figure}

As the total system size $L$ is gradually increased, the zero temperature $S_A$ in the time-direction slice increases from $0$ to a stabilizing value, see \cref{fig:l1}.

Now we focus on the $K > L$ regime. For the dispersion relation \cref{cos} and filling $\nu=0.5$, $S_A=0$ when $L/2$ is odd and $S_A=\log(2)=0.6931$ when $L/2$ is even. This is because of the particular discretization of the spectrum in \cref{cos}, which satisfies $E(k) = E(-k)$. Our spectrum is non-degenerate for the lowest and highest energies, but is doubly degenerate for all energies in between: $-0.5 E_0, -0.49 E_0, -0.49 E_0, \cdots, 0.49 E_0, 0.49 E_0, 0.5 E_0$.

Because $\nu=0.5$, this introduces an even-odd effect with respect to the filling number $L/2$: when $L/2$ is odd, the many-body ground state is a product state $a_{\text {sym}}^{\dagger} a_{\text {anti}}^{\dagger}|0\rangle$ of symmetric ($a^\dag_{sym} = \frac{1}{\sqrt{2}} (a_k^\dag + a_{-k}^\dag)$) and antisymmetric ($a^\dag_{anti} = \frac{1}{\sqrt{2}} (a_k^\dag - a_{-k}^\dag)$) combinations. When $L/2$ is even, the last pair of $a_k, a_{-k}$ is broken up in the ground state, so we have a superposition $(a^\dag_{sym} + a^\dag_{anti}) |0\rangle \propto a^\dag_k |0\rangle$.

As the sub-Hilbert space $h_A$ for the time direction slice at $\mathbf{r}_j=0$ only contains symmetric combinations ($c_{\mathbf{r}_j}\left(t_j\right)=\sum_l a_l \frac{1}{\sqrt{L}} e^{-\mathrm{i} \varepsilon_l t_j}$), tracing out the antisymmetric part gives $S_A=0$ when $L/2$ is odd and $S_A=\log(2)$ when $L/2$ is even.

This intuitive picture also implies that $S_A$ drops to $0$ or $\log(2)$ around $K = \frac{L}{2}$, because $h_A$ is contained in the symmetric sub-Hilbert space while the symmetric subspace is $2^{\frac{L}{2}}$ dimensional. For more generic dispersion relations and spacetime region $A$, we expect $S_A=0$ when $K > L$: because the total Hilbert space $h_{\text{tot}}$ is $L$ dimensional while we now have $K > L$ operators to build the sub-Hilbert space $h_A$, $h_A$ has the potential to be equal to $h_{\text{tot}}$ and $S_A=0$. We also point out that the above analysis is only intuitive: there are special parameters for which $S_A $ is not $0$ or $\log(2)$ at $K \gtrsim L/2$.

\section{$\delta$ Dependence}
\begin{figure}[H]
\centering
\subfloat{
\begin{picture}(0.48\columnwidth,0.48\columnwidth)
\put(0,0){\includegraphics[width=0.48\columnwidth]{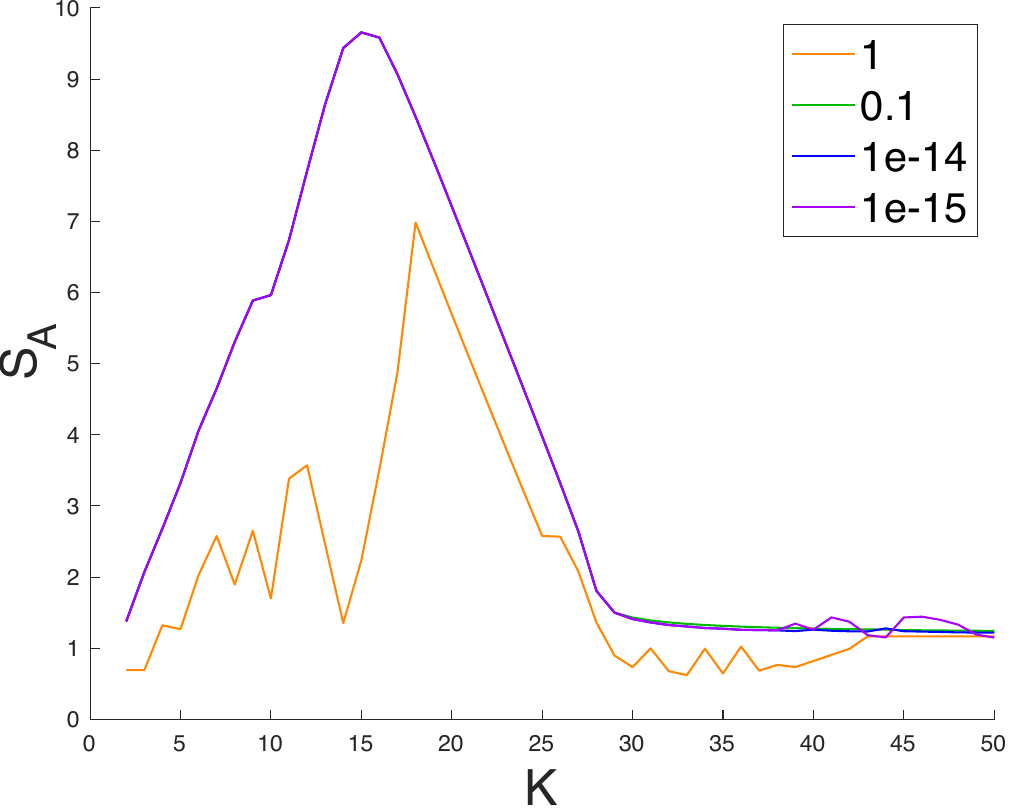}}
\put(-4,190){\footnotesize (a)}
\label{fig:delta1a}
\end{picture}}
\subfloat{
\begin{picture}(0.48\columnwidth,0.48\columnwidth)
\put(0,0){\includegraphics[width=0.48\columnwidth]{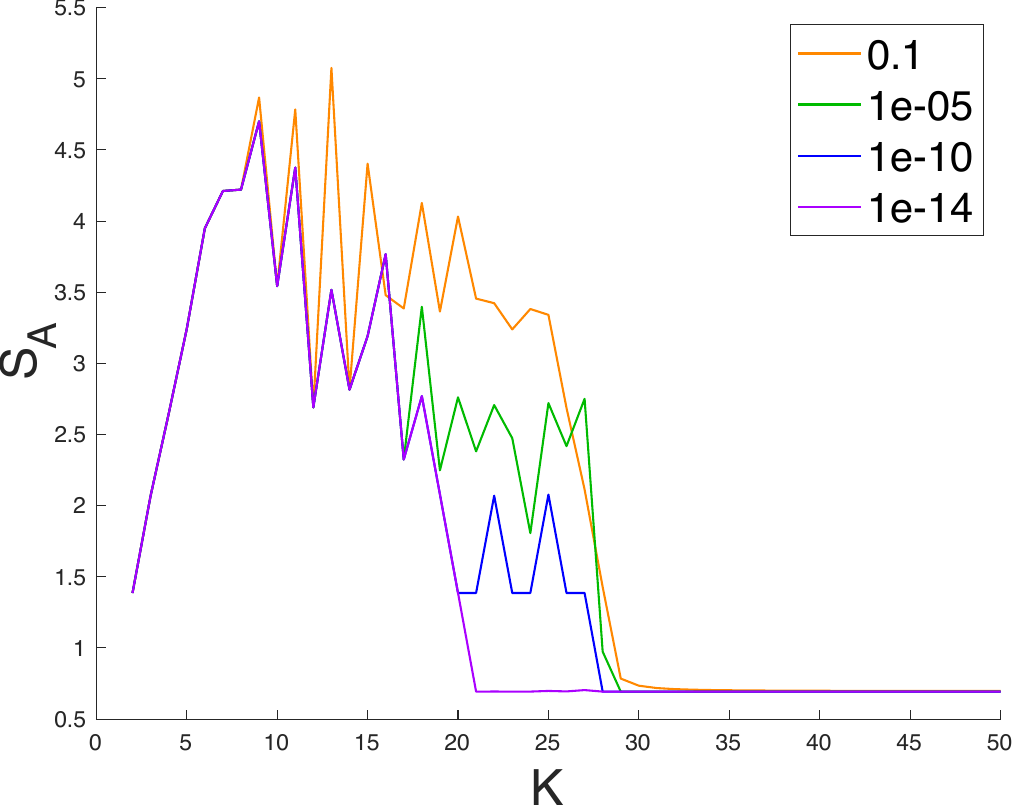}}
\put(-4,190){\footnotesize (b)}
\label{fig:delta1b}
\end{picture}}

\vspace{-17mm}
\subfloat{
\begin{picture}(0.48\columnwidth,0.48\columnwidth)
\put(0,0){\includegraphics[width=0.48\columnwidth]{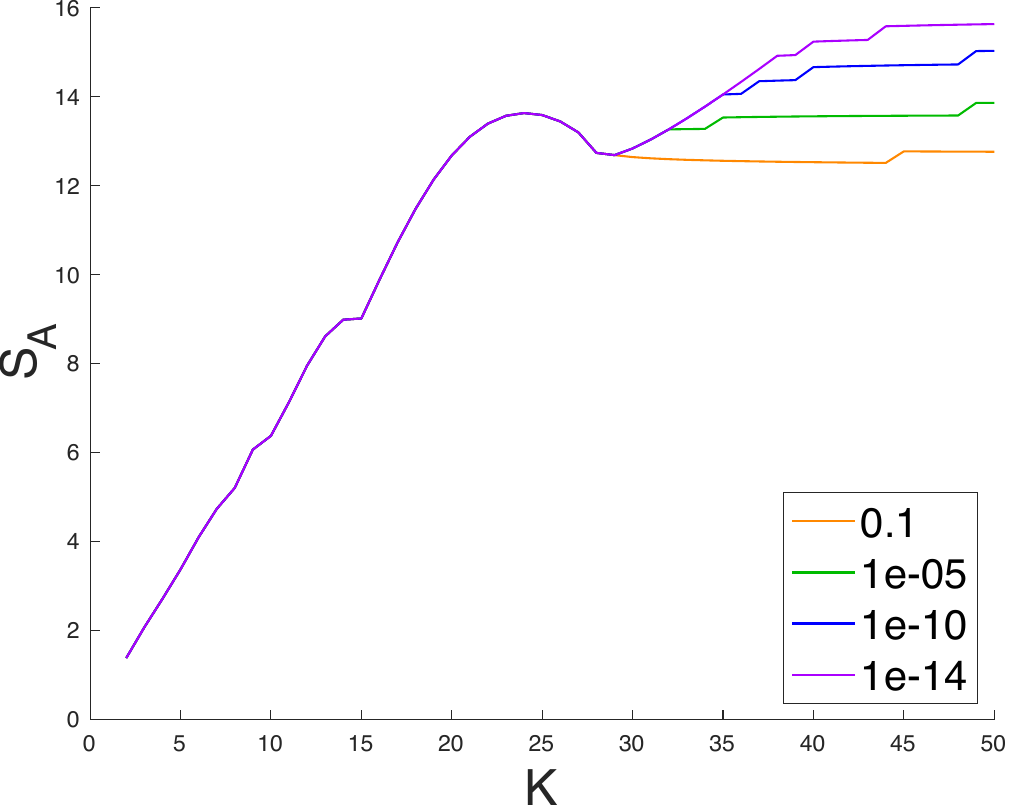}}
\put(-4,190){\footnotesize (c)}
\label{fig:delta1c}
\end{picture}}
\subfloat{
\begin{picture}(0.48\columnwidth,0.48\columnwidth)
\put(0,0){\includegraphics[width=0.48\columnwidth]{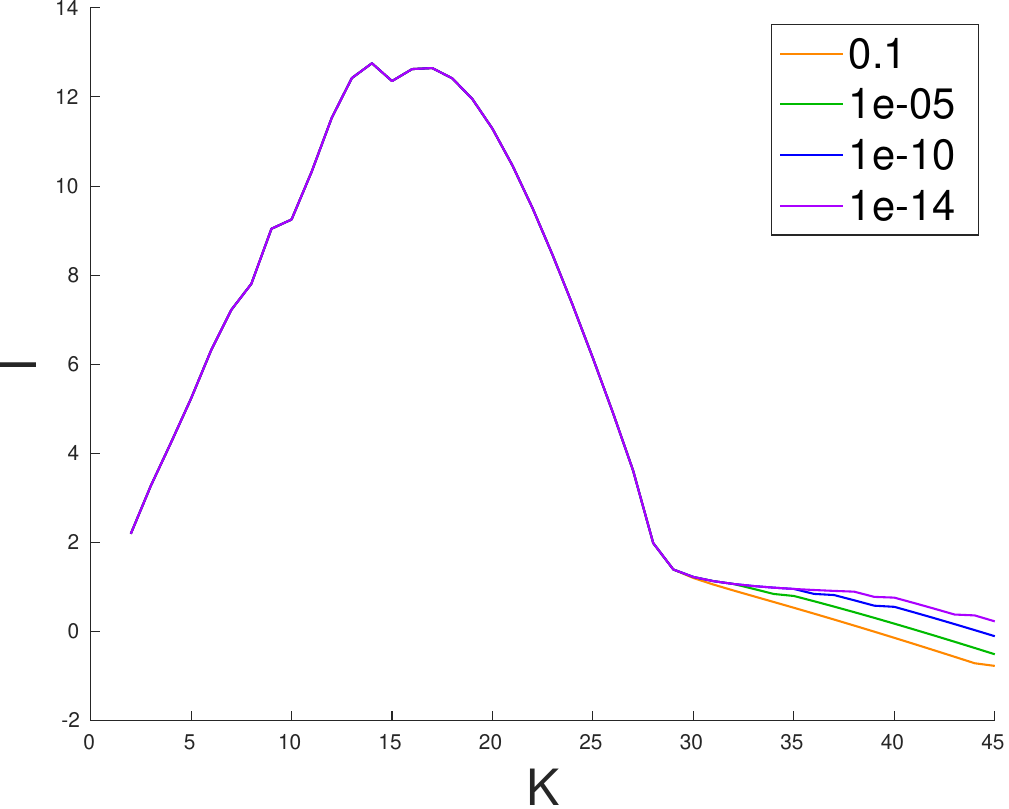}}
\put(-4,190){\footnotesize (d)}
\label{fig:delta1d}
\end{picture}}
\caption{Entanglement entropy $S_A$ (a-c) and quantum mutual information $\I$ (d) of model \cref{cos} in the time-direction slice at half-filling $\nu=0.5$ while fixing $2 \pi t / \tau_0=168$. Legend is the cutoff $\delta$ defined in \cref{eq: delta}, where 1e-n stands for $10^{-n}$. (a) $T=0$ and $L=500$. (b) $T=0$ and $L=40$. (c) $L=500$, $T/E_0 = 0.15$. (d) $L=200$ and $T/E_0=0.075$. }
\label{fig:delta1}
\end{figure}

In this section we explore the effect of $\delta$ defined in \cref{eq: delta}. As shown in \cref{fig:delta1a,fig:delta1b}, when $\delta\approx1$ the time-direction zero temperature entanglement entropy significantly deviates from the entropy when $\delta=10^{-14}$. Nonetheless, when $L=500$, entropy is very robust to changes in $\delta$ for all $\delta\leq0.1$. At large $K > t/\tau_0$, $\delta$ has a smoothing effect as shown in \cref{fig:delta1a}: the curve is very rigged when $\delta=10^{-15}$ but becomes flat when $\delta\geq10^{-14}$.

At finite temperature, a large $\delta$ flattens out the entropy's linear increase at large $K > t/\tau_0$ (\cref{fig:delta1c}). This is because $S_A$ is nearly thermal and linear in the number of points $K$; a large $\delta$ reduces the size of the matrices $M$ and $D$, so the effective number of points decreases and $S_A $ flattens out. A similar effect is seen for quantum mutual information $\I$ in \cref{fig:delta1d}: As $\delta$ increases, $\I$ at large $K > t/\tau_0$ decreases.

\section{Slope of Time-direction Entanglement Entropy vs. $\nu$}
The fitted ascending slope $\left(\frac{\partial S_A}{\partial K}\right)_t$ at small $K$ (orange line) for the model in \cref{cos} is close to $-\nu\ln \nu - (1-\nu) \ln(1-\nu)$ (the dashed line). The descending slope (green) is nearly independent of filling $\nu$.
\begin{figure}[H]
\centering
\includegraphics[width=0.3\linewidth]{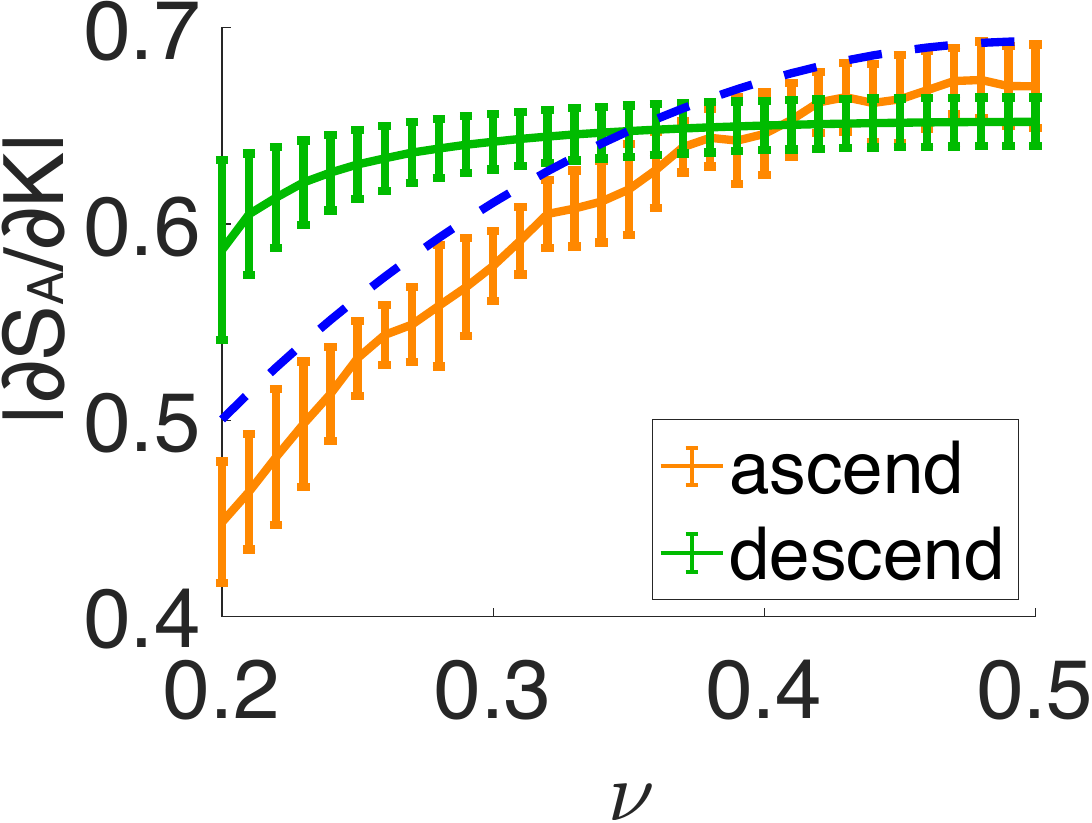}
\caption{For the zero temperature entanglement entropy $S_A$ of model \cref{cos} in the time-direction slice, we plot the absolute value of the slope $\left|\left(\frac{\partial S_A}{\partial K}\right)_t\right|$ with fixed $2\pi t/\tau_0=400$ fitted from the ascending region $K\in[2,6]$ and descending region $K\in[60,64]$, versus $\nu$ (solid lines, error bar $95\%$ confidence). The dashed curve is the function $-\nu\ln(\nu)-(1-\nu)\ln(1-\nu)$.}
\label{fig:slope}
\end{figure}

\section{Non-Eigenstate}
\begin{figure}[H]
\centering
\subfloat{
\begin{picture}(0.48\columnwidth,0.48\columnwidth)
\put(0,0){\includegraphics[width=0.48\columnwidth]{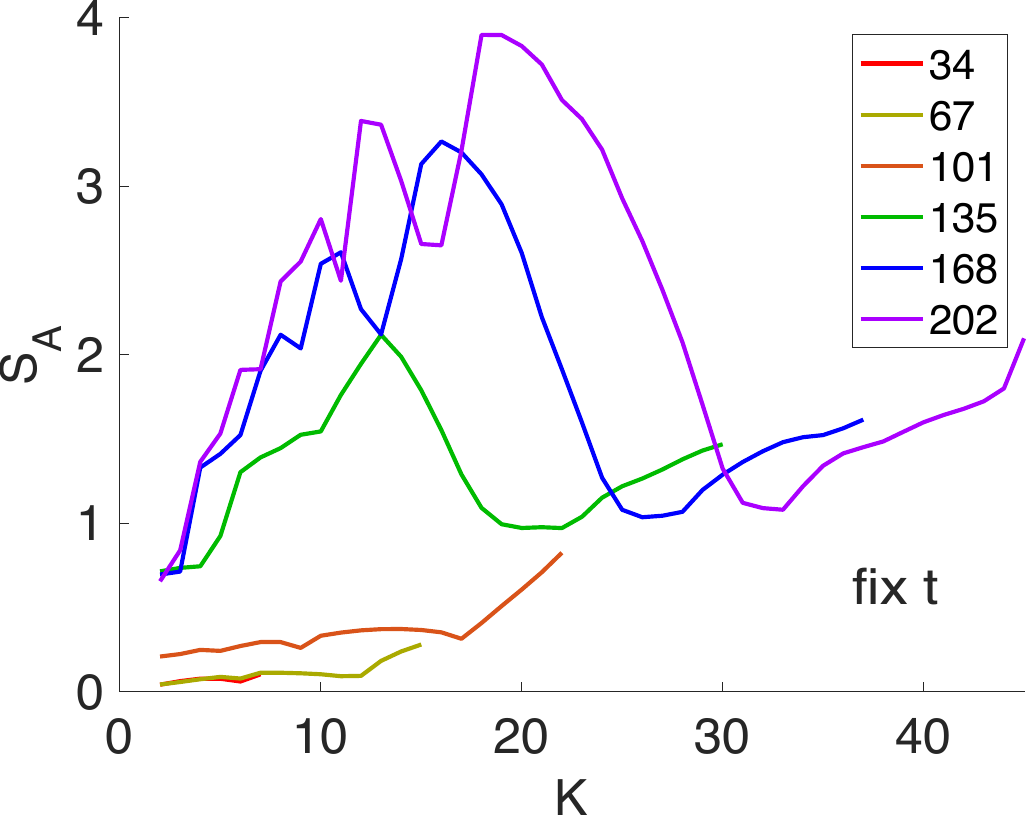}}
\put(-4,190){\footnotesize (a)}
\label{fig:ca}
\end{picture}}
\subfloat{
\begin{picture}(0.48\columnwidth,0.48\columnwidth)
\put(0,0){\includegraphics[width=0.48\columnwidth]{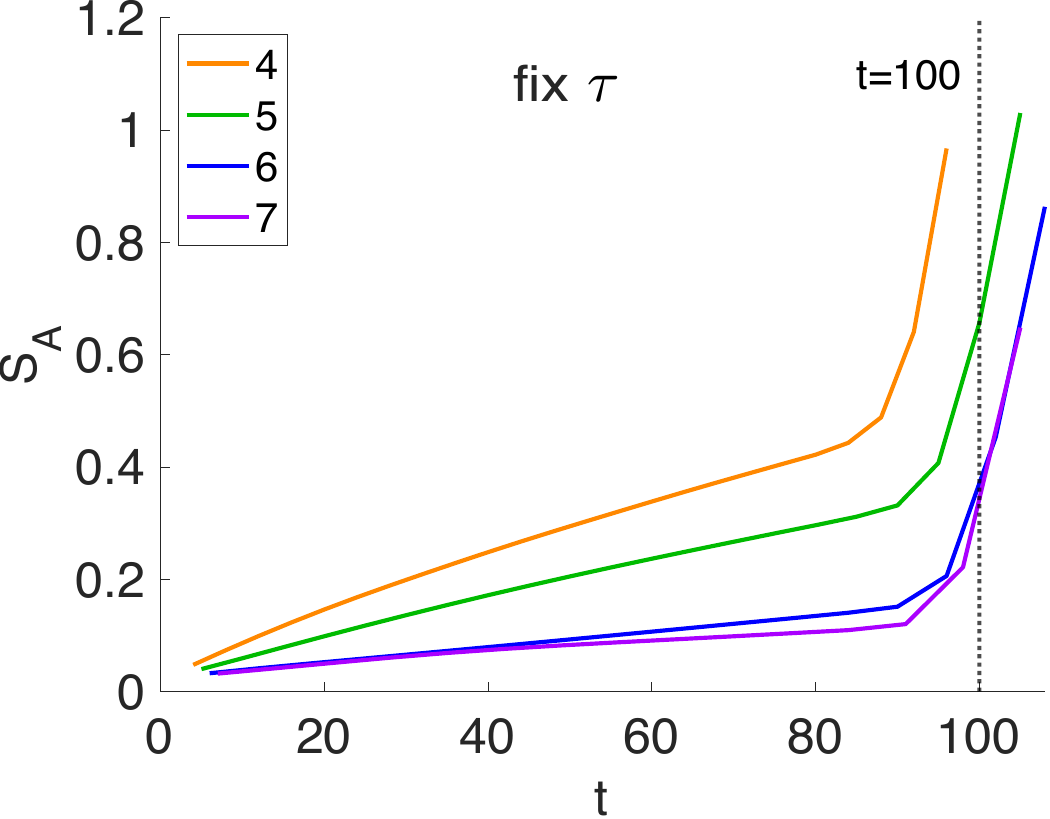}}
\put(-4,190){\footnotesize (b)}
\label{fig:cb}
\end{picture}}
\caption{Entanglement entropy $S_A$ of time-direction slice at site $m=0$ for non-eigenstates $|\psi\rangle=\prod_{m\in [m_I,m_I+N_f)}c_m^\dag|0\rangle$ ($L=1000$), with $N_f=101$, $m_I=-(N_f-1)/2$. (a) is calculated with fixed $2 \pi t / \tau_0$ given in the legend, while (b) is calculated with fixed $2 \pi \tau / \tau_0$ in the legend.
}
\label{fig:c}
\end{figure}

Generically, we can consider the non-eigenstate $|\psi\rangle=\prod_{m\in [m_I,m_I+N_f)}c_m^\dag|0\rangle$ in main text Fig. 3c and 3d. This time we put the time-direction slice at the center of the filling positions with $N_f=101$ and $m_I=-\frac{N_f-1}{2}$, in which case $t_L=t_R= 100$. \cref{fig:ca} shows $S_A$ with respect to $K$ at fixed total time $t$. When $t<t_L$, $S_A$ remains almost zero, while for $t>t_L$, $S_A$ shows a similar curve (with fluctuations) as that of the finite temperature state in main text Fig. 2c. The crossover of $S_A$ as a function of $t$ at $t=t_L$ can be seen more clearly in \cref{fig:cb}, where time spacing $\tau$ is fixed. 

\end{document}